\pdfoutput=1

\documentclass[11pt,a4paper]{article}

\usepackage[normalem]{ulem}
\usepackage{jheppub}
\usepackage{amsmath}
\usepackage{amssymb}
\usepackage{appendix}
\usepackage{multirow}
\usepackage{rotating}
\usepackage{diagbox}
\usepackage{float}
\usepackage{xcolor}
\usepackage{cancel}
\usepackage{tabularx}
\usepackage{color}

\newcommand{\bea}{\begin{eqnarray}}
\newcommand{\eea}{\end{eqnarray}}

\usepackage{color}
\definecolor{niceblue}{rgb}{0,0,1}
\definecolor{nicered}{rgb}{0.7,0.1,0.1}
\definecolor{nicegreen}{rgb}{0.1,0.5,.1}

\usepackage{graphicx}
\usepackage{subcaption}
\graphicspath{{figures/}{./}}
\title{A new puzzle in non-leptonic $B$ decays}

\author[a]{Aritra Biswas,}
\author[b]{S\'ebastien Descotes-Genon,}
\author[a]{Joaquim Matias.}
\author[b,c]{Gilberto Tetlalmatzi-Xolocotzi}

\affiliation[a]{Universitat Aut\`onoma de Barcelona, 08193 Bellaterra, Barcelona,\\
Institut de F\'{i}sica d'Altes Energies (IFAE), The Barcelona Institute of Science and Technology, Campus UAB, 08193 Bellaterra (Barcelona)}
\affiliation[b]{Universit\'e Paris-Saclay, CNRS/IN2P3, IJCLab, 91405 Orsay, France}
\affiliation[c]{Theoretische Physik 1, Center for Particle Physics Siegen (CPPS), Universit\"at Siegen, Walter-Flex-Str. 3, 57068 Siegen, Germany}

\abstract{We build a set of 
new observables 
using closely related non-leptonic penguin-mediated $B_d$ and $B_s$ decays: 
${\bar B}_{d,s} \to K^{*0} \bar{K}^{*0}$, 
${\bar B}_{d,s} \to K^{0} \bar{K}^{0}$, ${\bar B}_{d,s} \to K^{0} \bar{K}^{*0}$ and ${\bar B}_{d,s} \to \bar{K}^{0} {K^{*0}}$ together with their CP conjugate partners.
These optimised observables are designed to reduce hadronic uncertainties, mainly coming from form factors and power-suppressed infrared divergences, and thus maximize their sensitivity to New Physics (NP). The deviations observed with respect to the Standard Model (SM)
in the ratios of branching ratios of ${\bar B}_{d,s} \to K^{*0} \bar{K}^{*0}$ ($2.6\sigma$) and
${\bar B}_{d,s} \to K^{0} \bar{K}^{0}$ ($2.4\sigma$) can be explained by simple NP scenarios 
involving the Wilson coefficients ${\cal C}_4$ and ${\cal C}_6$ associated with QCD penguin operators and the coefficient ${\cal C}_{8g}$ of the chromomagnetic operator.
The optimised observables for ${\bar B}_{d,s} \to K^{0} \bar{K}^{*0}$ and ${\bar B}_{d,s} \to \bar{K}^{0} {K^{*0}}$ show distinctive patterns of deviations with respect to their SM predictions under these NP scenarios. The pattern of deviations of individual branching ratios, though affected by significant hadronic uncertainties, suggests that NP is needed both in $b\to d$ and $b\to s$ transitions.
We provide the regions for the Wilson coefficients consistent with both optimised observables and individual branching ratios.
The NP scenarios considered to explain the deviations of ${\bar B}_{d,s} \to K^{*0} \bar{K}^{*0}$ and
${\bar B}_{d,s} \to K^{0} \bar{K}^{0}$
can yield deviations up to an order of magnitude among the observables that we introduced for 
 ${\bar B}_{d,s} \to K^{0} \bar{K}^{*0}$ and ${\bar B}_{d,s} \to \bar{K}^{0} {K^{*0}}$.
Probing these new observables experimentally by improving the measurements of individual branching ratios of penguin-mediated decays would confirm the consistency of the deviations already observed and provide a highly valuable hint of NP in the non-leptonic sector.
}

\emailAdd{abiswas@ifae.es}
\emailAdd{sebastien.descotes-genon@ijclab.in2p3.fr}
\emailAdd{matias@ifae.es}
\emailAdd{gtx@physik.uni-siegen.de}

\arxivnumber{2301.yyyyy}

\begin{document}
%\pacs{}
\begin{flushright}

SI-HEP-2023-01\\
P3H-23-003

\end{flushright}

%%%%%%%%%%%%%%%%%%%%%%%%%%%%%%%%%
%%%%%%%%%%%%%%%%%%%%%%%%%%%%%%%%%
\maketitle
%%%%%%%%%%%
\section{Introduction}
\label{intro}
%%%%%%%%%%%

Given the possible hints of New Physics (NP) observed in semileptonic $B$-meson decays (the so-called $b$-quark anomalies~\cite{Albrecht:2021tul,London:2021lfn}), one could expect that deviations also arise in other rare $B$-meson decays, such as non-leptonic $B$ decays associated with flavour-changing neutral currents (FCNC). Such penguin-mediated decays can be expected to exhibit a particular sensitivity to NP, given that the SM contribution occurs only at the loop level. 
However, it is well known that going from semileptonic $B$ decays to  non-leptonic ones
involves an increased degree of difficulty to control the hadronic uncertainties. It proves interesting to choose FCNC non-leptonic $B$ decays where theoretical tools are available to reduce these uncertainties, either by computing hadronic matrix elements or by relating them in different modes. 
These considerations promote the study of $\bar{B}_{d,s}\to K^{(*)0}\bar{K}^{(*)0}$, which can be analysed relying on different approaches:
$SU(3)$ symmetry (see Ref.~\cite{Amhis:2022hpm} for a recent example of this type of analysis  combining $\bar{B}_{s}\to K^{0}\bar{K}^{0}$ with other isospin and U-spin related  modes   and Ref.~\cite{Fleischer:1999pa,Fleischer:2007hj,Fleischer:2010ib,Fleischer:2016jbf,Bhattacharya:2022akr} for other attempts), but also QCD Factorisation (QCDF)~\cite{Beneke:2000ry,Beneke:2001ev,Beneke:2003zv,Beneke:2006hg,Bartsch:2008ps}, or even a combination of the two approaches~\cite{Descotes-Genon:2006spp,Descotes-Genon:2007iri,Descotes-Genon:2011rgs,Alguero:2020xca}.

In a recent article~\cite{Alguero:2020xca} we studied the observable $L_{{K}^* \bar{K}^*}$ defined as the ratio of the longitudinal branching ratios of $\bar{B}_s \to K^{*0} \bar{K}^{*0}$ versus $\bar{B}_d \to K^{*0} \bar{K}^{*0}$. This observable exhibits a tension of 2.6$\sigma$ between its SM prediction and data. We performed a model-independent analysis of this tension within the Weak Effective Theory at the $b$-quark mass scale (keeping the analysis at the level of the operators generated in the SM and their chirally-flipped counterparts). We ended up with an 
explanation relying on NP contributions to the Wilson coefficients of two operators: a)  the QCD penguin operator $O_{4s}=(\bar{b}_i s_j)_{V-A} \sum_q (\bar{q}_j q_i)_{V-A}$ and b) the chromomagnetic operator $O_{8g}=- \frac{g_s}{8\pi^2} m_b \bar{s} \sigma_{\mu\nu} (1+\gamma_5) G^{\mu\nu}b$ (see ref.~\cite{Alguero:2020xca} and App.~\ref{app:WET} for the basis of operators of the Weak Effective Theory).  

We need further information to understand this deviation and confirm its NP origin. This can be done if 
\begin{itemize}
\item[i)] we identify other modes that exhibit a similar sensitivity to the same NP,
\item[ii)] we can design quantities that combine observables for these modes with a reduced sensitivity to hadronic uncertainties, 
\item[iii)] these additional quantities yield clear patterns of consistent deviations for different modes depending on the NP scenario considered. 
\end{itemize}
Under these conditions, we will be able to confirm the NP origin of the deviations observed with a reasonable certainty, given the robustness of our framework. As usual in these kind of global analyses, these robust observables can be complemented by additional less clean observables that can favour one scenario with respect to another in a more qualitative way.

In this article, we thus extend the discussion to a larger set of 
decays, with the same underlying quark transitions and thus a similar potential sensitivity to NP, but with different 
final states. In practice we consider non-leptonic $\bar{B}_d$ and $\bar{B}_s$ meson decays not only to two vectors ($VV$), but also two pseudoscalars ($PP$) and a vector and pseudoscalar ($PV$ or $VP$), with $V=K^{*0}$ and $P=K^0$. For these decays, we will identify and construct observables with reduced hadronic uncertainties following the same strategy as for $L_{K^*\bar{K}^*}$. Thanks to this larger set of decays,
we can probe some of the simple NP scenarios suggested to explain 
$L_{K^*\bar{K}^*}$ as they may yield distinct and consistent patterns 
of deviations for the other modes.
The additional observables for these other modes have very similar values in the SM but they can differ by one order of magnitude in some NP scenarios. Such hierarchy among observables, which can hardly be attributed to residual hadronic uncertainties, may be 
 tested at LHCb and Belle II. Interestingly, the individual branching ratios show patterns of deviations that will lead us to lift the ambiguity between NP explanations favouring $b\to d$ or $b\to s$ transitions, in the sense that NP is needed in both types of transitions to explain the deviations currently observed in the individual branching ratios, in addition to the one observed in the $L$ observables.

The structure of the article is as follows. In section~\ref{sec:theory} we introduce the theoretical framework that will be used for the construction of the observables in section~\ref{sec:observables}. In the latter section we construct the set of observables of interest and we provide their SM prediction as well as their current experimental value (when available). Some of the observables designed in this section will require an LHCb upgrade to be accessible. In Section~\ref{sec:modelindep} a model-independent analysis of $L_{K\bar{K}}$ is performed to check the shifts in SM Wilson coefficients able to describe the anomaly. 
We also study the sensitivity of the different observables to NP contributions to the Wilson coefficients.
Section~\ref{sec:combined} shows the domain allowed by $L_{K\bar{K}}$
and  $L_{K^*\bar{K}^*}$  for the Wilson coefficients assuming NP only in $b\to s$ first and provides the prediction for the observables associated with pseudoscalar-vector modes. This defines the pattern of deviations to be tested by LHCb. In Sec.~\ref{sec:indivBR}, we discuss individual branching ratios which are affected by larger hadronic uncertainties but still provide interesting glimpses of potential NP scenarios to be tested. They will point to NP scenarios in both $b \to s$ and $b \to d$ transitions for which we will discuss the allowed ranges for the relevant Wilson coefficients. We draw our conclusions in section~\ref{sec:conclusions}. 
The appendices are devoted to details for the computation of these modes within QCDF, some aspects of statistics concerning the distributions of the observables in the SM, and the results for the observables discussed in this article at some illustrative benchmark NP scenarios.

\section{Theoretical considerations} \label{sec:theory}

Following the discussion in the introduction, we focus on pairs of penguin-mediated non-leptonic decays which are related by $U$-spin symmetry. This selects naturally decays to $K^0$ and/or $K^{*0}$ mesons, and we will discuss each mode in turn.

\subsection{$\bar{B}_{d,s} \to K^{*0} \bar{K}^{*0}$}\label{sec:theory-VV}

\begin{figure} 
\begin{center}
\includegraphics[width=6.5 cm]{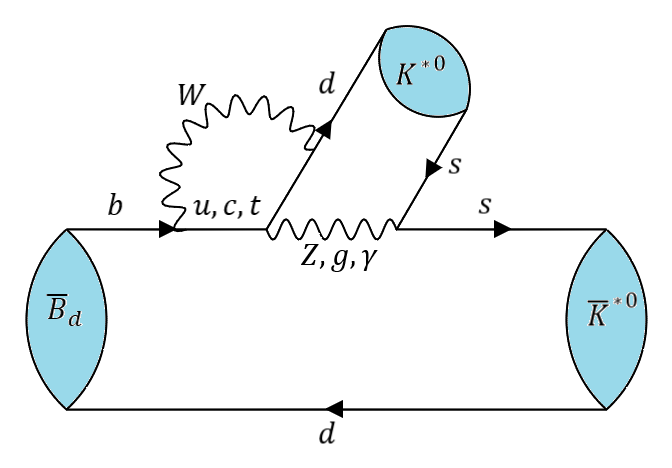} \qquad
\includegraphics[width=6.5 cm]{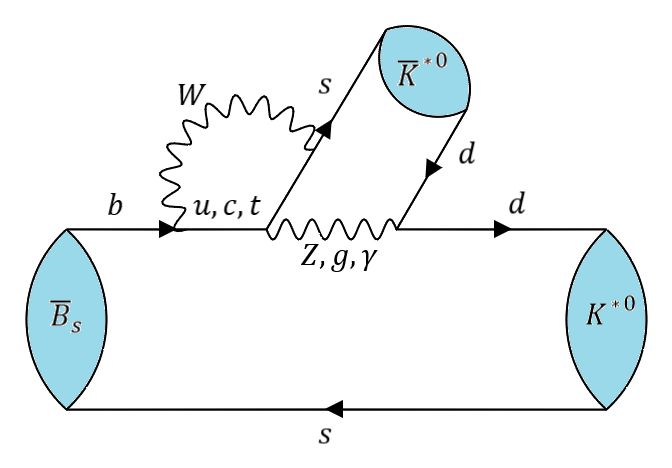}
   \end{center}  \caption{SM contributions to the non-leptonic decays $\bar{B}_{d,s}\to K^{*0} \bar{K}^{*0}$. The internal quark lines consist in a $u$-type quark, the curved wavy line is a $W$ boson, and the horizontal wavy line may be a gluon, a photon or a $Z$ boson, leading to different types of penguins. }
   \label{fig:kstarkstar}   \end{figure}

We recall briefly a few elements of our detailed discussion in Ref.~\cite{Alguero:2020xca}. The final state can be in three different polarisations states.
For a $\bar B_q$ meson decaying through a $b\to q$  process to a $K^{*0}\bar{K}^{*0}$ of given polarisation, the decompositions
\begin{equation}
\bar{A}_f\equiv A(\bar{B}_q\to K^{*0} \bar{K}^{*0})
  =\lambda_u^{(q)} T_q + \lambda_c^{(q)} P_q
  =\lambda_u^{(q)}\, \Delta_q - \lambda_t^{(q)} P_q 
\label{dec}
\end{equation}
are always possible, with the CKM factors $\lambda_U^{(q)}=V_{Ub} V_{Uq}^*$~\footnote{The weak phase in $\lambda_t^{(q)}$ is the angle $\beta_q$, defined as
$
    \beta_q\equiv \arg \left(- \frac{V_{tb} V_{tq}^*}{V_{cb} V_{cq}^*} \right)= \arg \left(- \frac{\lambda_t^{(q)}}{\lambda_c^{(q)}} \right)\,,
$}.
We denote by $T_q$ and $P_q$ the hadronic matrix elements accompanying the $\lambda_u^{(q)}$ and $\lambda_c^{(q)}$ CKM factors respectively, and we introduce the difference $\Delta_q=T_q-P_q$. 
Even though the notation $T$ and $P$ is reminiscent of the decomposition in tree and penguin contribution, we insist on the fact that both quantities involves penguin diagrams (and annihilation topologies) but no tree contributions  for $\bar{B}_q\to K^{*0} \bar{K}^{*0}$, as illustrated in Fig.~\ref{fig:kstarkstar}. 

The CP-conjugate amplitude is given by 
\begin{equation}
A_{\bar{f}}=(\lambda_u^{(q)})^* T_q + (\lambda_c^{(q)})^* P_q    =(\lambda_u^{(q)})^* \Delta_q - (\lambda_t^{(q)})^* P_q\,.
\end{equation}
$A_{\bar{f}}$ is related to $A=A(B_q\to  K^{*0}\bar{K}^{*0})=\eta_f A_{\bar{f}}$
where $\eta_f$ is the CP-parity of the final state, given for $j=0,||,\perp$ respectively as $1,1,-1$.

Two different theoretical tools can be used to determine the values of the hadronic matrix elements $T$ and $P$. On the one hand, $U$-spin symmetry can help to provide relationships among hadronic matrix elements that should hold up to corrections proportional to the difference between the masses of the $d$ and $s$ quarks. One thus expects that $T$ and $P$ hadronic matrix elements each differ by 30\% (at most) when one compares their values for $\bar{B}_d \to K^{*0} \bar{K}^{*0}$ and
$\bar{B}_s \to K^{*0} \bar{K}^{*0}$. 

On the other hand, factorisation approaches provide a way of computing these hadronic matrix elements. One starts with the Weak Effective Theory separating long and short distances at the scale $\mu=m_b$, described in App.~\ref{app:WET} in detail. It remains to determine the hadronic matrix elements corresponding to the operators $Q_i$ sandwiched between the initial $\bar{B}_q$ meson and the final $K^{*0}\bar{K}^{*0}$ pair. 
A naive factorisation approach would express them as proportional to the product of the form factor $A_0^{B_q\to \bar{K}^{*0}}(0)$ multiplied by the decay constant $f_{K^{*0}}$.
QCDF allows one to compute the hadronic matrix elements based on the heavy-quark limit $m_b\to\infty$ where the naive factorisation approach holds for some classes of decays, providing corrections to this picture. Indeed, the hadronic matrix elements $T_q$ and $P_q$
can be expressed as an expansion of $\alpha_s$ involving form factors and light-cone distribution amplitudes as hadronic inputs, up to  $1/m_b$-suppressed terms that contain long-distance contributions, corresponding to infrared divergences in the factorisation framework. 
In the case of vector modes, a clear hierarchy among polarisations occurs in the limit $m_b\to\infty$, so that only the longitudinal polarisation can actually be computed accurately within QCDF and will be our focus for the rest of the article (meaning that the longitudinal polarisation will be assumed implicitly for $VV$ final states in the following), whereas the transverse polarisations are $1/m_b$-suppressed and plagued at leading order with  long-distance (infrared-divergent) contributions. The expressions for the hadronic matrix elements within QCDF are discussed in App.~\ref{app:QCDF}.

As discussed extensively in Refs.~\cite{Descotes-Genon:2006spp,Descotes-Genon:2007iri,Descotes-Genon:2011rgs,Alguero:2020xca}, 
in the case of penguin-mediated decays, QCDF yields a further prediction concerning $\Delta_q$, which is the difference of two hadronic matrix elements differing only through the mass of the quark involved in the loop ($m_u$ versus $m_c$). $\Delta_q$ is protected from infrared divergences and is expected to be significantly smaller than both $T$ and $P$ (of similar sizes), as shown numerically in App.~\ref{app:PTDelta}.

All these considerations incite us to build a quantity involving the amplitudes for longitudinal polarisation only, of the form:
\begin{equation}\label{eq:Lgeneraldiscussion}
L_{K^*\bar{K}^*}=\frac{|A_0^s|^2+ |\bar A_0^s|^2}{|A_0^d|^2+ |\bar A_0^d|^2}
=\kappa \left|\frac{P_s}{P_d}\right|^2 
 \left[\frac{1+\left|\alpha^s\right|^2\left|\frac{\Delta_s}{P_s}\right|^2
 + 2 {\rm Re} \left( \frac{ \Delta_s}{P_s}\right) {\rm Re}(\alpha^s) 
 }{1+\left|\alpha^d\right|^2\left|\frac{\Delta_d}{P_d}\right|^2
  + 2 {\rm Re} \left( \frac{ \Delta_d}{P_d}\right) {\rm Re}(\alpha^d)} \right]\,,
\end{equation}
where $A^q_0$ corresponds to the amplitude for a $B_q$ meson decaying into a longitudinally polarised $K^{*0}\bar{K}^{*0}$ pair, and the CKM factors read 
\begin{eqnarray}
\kappa&=&\left|\frac{\lambda^s_u+\lambda^s_c}{\lambda^s_u+\lambda^s_c} \right|^2=22.91^{+0.48}_{-0.47}, \nonumber\\
\alpha^d&=&\frac{\lambda^d_u}{\lambda^d_u+\lambda^d_c}=-0.0135^{+0.0123}_{-0.0124} +0.4176^{+0.0123}_{-0.0124}i, \nonumber\\
\alpha^s&=&\frac{\lambda^s_u}{\lambda^s_u+\lambda^s_c}=0.0086^{+0.0004}_{-0.0004}-0.0182^{+0.0006}_{-0.0006}i.
\end{eqnarray}
As can be seen numerically, $\alpha^d$ is Cabibbo allowed, $\alpha^s$ is Cabibbo-suppressed $O(\lambda^2)$, whereas $\Delta_q/P_q$ is expected to be small in the case of penguin-mediated decays~\cite{Descotes-Genon:2006spp,Descotes-Genon:2007iri,Descotes-Genon:2011rgs,Alguero:2020xca} (see App.~\ref{app:PTDelta} for numerical estimates within QCD). Therefore $L_{K^*\bar{K}^*}$ is directly related to the ratio $|P_s/P_d|$, which can be predicted with good accuracy within QCDF and which is protected by $U$-spin symmetry from uncontrolled $1/m_b$-suppressed long-distance contributions.

\subsection{$\bar{B}_{d,s} \to K^{0} \bar{K}^{0}$}\label{sec:theory-PP}

The discussion is similar in the pseudoscalar case, though simpler, as there are no polarisations involved. Both $U$-spin and QCDF apply in a similar way to the determination of the $P$ and $T$ hadronic matrix elements. The computation within QCDF yields very similar expressions as in the $VV$ case up to the substitution $K^*\to K$ and $\bar{K}^*\to \bar{K}$, and the modification of the normalisation
\begin{equation}
A_{\bar{K} K}=\frac{G_F}{\sqrt{2}}m^2_{B_q}f_{K}F_0^{B_q\to K}(0)
 \end{equation}
 Let us stress that even though the formal structure detailed in App.~\ref{app:QCDF} still applies, the coefficients $\alpha$ and $\beta$ involved in the computation of the hadronic matrix elements $T_q$ and $P_q$ have different numerical values in the SM,
and their sensitivity to NP is different (as described in more detail in Sec.~\ref{sec:modelindep} and App.~\ref{app:alphacoeffs-PPVV}).

It proves thus interesting to create a quantity $L_{K\bar{K}}$ similar to Eq.~(\ref{eq:Lgeneraldiscussion}), without any $0$ since no polarisation is involved in the definition of the amplitudes. Following the same arguments, one would expect $L_{K\bar{K}}$ to be driven by the $U$-spin protected ratio of penguin amplitudes $|P_s/P_d|$ for $\bar{B}_{d,s} \to K^{0} \bar{K}^{0}$ (which is not necessary equal numerically to the one for $\bar{B}_{d,s} \to K^{*0} \bar{K}^{*0}$).

Two more comments are in order regarding the connection with the experimental measurements. First, the latter are performed in terms of $K_S$ rather than $K_0$ and $\bar{K}_0$, which can be seen as the sum of these two states divided by $\sqrt{2}$ (if we neglect the very small amount of CP-violation involved here). Secondly, the current measurements correspond in part to LHCb results integrated over time and the final state is a CP-eigenstate, which means that correction coming from the interference between $B_q$ mixing and its decay into $K^0\bar{K}^0$ must be applied~\cite{Descotes-Genon:2011rgs}. The effect is negligible for $B_d$, but it can have an impact of order $O(\Delta \Gamma_s/\Gamma_s)$ in the case where the $b\bar{b}$ pair is produced incoherently (as it is the case at LHCb). The effect is also present for $K^{*0}\bar{K}^{*0}$ as it occurs
for $B_s$ mesons stemming from
a $b\bar{b}$ incoherent production and
decaying into CP-eigenstates.
The exact size of the effect depends on the CP-asymmetry $A_{\Delta \Gamma}$ for the decay of interest~\cite{Dunietz:2000cr,Descotes-Genon:2011rgs,DeBruyn:2012wj,DeBruyn:2012wk,Descotes-Genon:2015hea,Descotes-Genon:2020tnz,Descotes-Genon:2022gcp}. We will not try to compute this asymmetry within QCDF in order to remain conservative.
In both cases, whenever necessary (Eqs.~(\ref{eq:expdataLKstKst}) and (\ref{eq:expdataLKK}) and Tabs.~\ref{tab:BrPP} and \ref{tab:BrVV}), we have added a 7\% relative uncertainty to the experimental measurement, corresponding to the correction of this $B_s$-mixing effect so that we can compare theory and experiment.

\subsection{$\bar{B}_{d,s} \to K^{0} \bar{K}^{*0}$ and $\bar{B}_{d,s} \to \bar{K}^{0} {K^{*0}}$}\label{sec:theory-PV-VP}

Finally, we turn to the pseudoscalar-vector case. The polarisation of the vector is necessarily longitudinal due to helicity conservation, so that the pair of final-state mesons is emitted in a $P$-wave. Once again, one can analyse the $P$ and $T$ hadronic matrix elements using $U$-spin and QCDF.
However, since the final state is not symmetric, we see that there are two different possibilities, depending on whether the spectator quark ends up in a (pseudoscalar) $K$ meson or a (vector) $K^*$ meson. Following Ref.~\cite{Beneke:2003zv}, we pay special attention to the ordering of the mesons in the final state: the decays will be written as $B_q\to M_1M_2$ and $\bar{B}_q\to M_1M_2$ with $M_1$ denoting the final meson carrying away the spectator quark.
Depending on whether this meson is pseudoscalar or vector, there is a different normalisation of the QCDF results with different form factors, which can be worked out already in naive factorisation~\cite{Beneke:2003zv}:
\begin{eqnarray}
A_{K^* K}&=&
-2\sqrt{2}G_F m_{K^*} \epsilon^*_{K^*} \cdot p_{B_q} A_0^{B_q\to K^*}(0)\\
A_{K K^*}&=&
-2\sqrt{2}G_F m_{K^*} \epsilon^*_{K^*} \cdot p_{B_q} F_+^{B_q\to K}(0)
\end{eqnarray}
The first case corresponds to $\bar{B}_d\to K^* \bar{K}$ and $\bar{B}_s\to \bar{K}^* K$, whereas the second case corresponds $\bar{B}_d\to K\bar{K}^*$ and $\bar{B}_s\to \bar{K}K^*$ (as well as their CP-conjugate).

The same arguments go for the protection of the difference between $T_q$ and $P_q$ from infrared divergences, and its smallness. It may thus prove interesting for each case to build a quantity similar to $L_{K^*\bar{K}^*}$, driven by the ratio of the corresponding matrix elements $P_s/P_d$ and protected by $U$-spin symmetry.

\section{Definition, SM predictions and experimental determinations of optimised observables} \label{sec:observables}

\subsection{Vector-vector case}\label{sec:LKstKst}

In Sec.~\ref{sec:theory-PV-VP}, we recalled the theoretical definition of the observable $L_{K^*\bar{K}^*}$. As shown in Ref.~\cite{Alguero:2020xca} this observable can be 
defined as a ratio of longitudinal amplitudes of two U-spin related decays:
\begin{equation}\label{eq:LKstarKstar}
L_{K^*\bar{K}^*}=\rho(m_{K^{*0}},m_{K^{*0}})\frac{{\cal B}({\bar{B}_s \to K^{*0} {\bar K^{*0}}})}{{\cal B}({\bar{B}_d \to K^{*0} {\bar K^{*0}})}}\frac{ f_L^{B_s}}{ f_L^{B_d}}=\frac{|A_0^s|^2+ |\bar A_0^s|^2}{|A_0^d|^2+ |\bar A_0^d|^2}\,,
\end{equation}
where $\rho(m_1,m_2)$ stands for the ratio of phase-space factors defined by
\begin{equation}
    \rho(m_1,m_2)=\frac{\tau_{Bd}}{
    \tau_{Bs}}\frac{m_{B_s}^3}{m_{B_d}^3}\frac{\sqrt{(m_{B_d}^2-(m_1+m_2)^2)(m_{B_d}^2-(m_1-m_2)^2)}}{\sqrt{(m_{B_s}^2-(m_1+m_2)^2)(m_{B_s}^2-(m_1-m_2)^2)}}
\end{equation}
The values for the masses, lifetimes and decay widths are all taken from PDG~\cite{Workman:2022ynf}, as well as the 
branching ratios which should be understood as  CP averages including also the CP-conjugate decay. PDG~\cite{Workman:2022ynf} and HFLAV~\cite{HFLAV:2022pwe} have different averaging procedures in the case of measurements from several collaborations. This leads to slightly different, though consistent, combinations with a larger (and thus more conservative) uncertainty in the PDG case.

The expression of this observable in terms of CKM factors and amplitudes is shown in Eq.~(\ref{eq:Lgeneraldiscussion}). 
The SM prediction for this observable within QCDF
using the inputs in App.~\ref{app:QCDF} and the expression of $L_{K^*\bar{K}^*}$
given in Eq.~(\ref{eq:Lgeneraldiscussion}) 
is~\footnote{The value quoted is slightly different from Ref.~\cite{Alguero:2020xca} due to the updated values of the inputs, see Tab.~\ref{tab:inputs}.}
 \begin{equation} 
L_{K^*\bar{K}^*}^{\rm SM}=
19.53^{+9.14}_{-6.64}
\end{equation}
  while the experimental measurement is taken as
 \begin{equation} \label{eq:expdataLKstKst}
L_{K^*\bar{K}^*}^{\rm exp}=
4.43\pm 0.92 
 \end{equation}
where we added a 7\% relative uncertainty due to $B_s$-mixing (see Sec.~\ref{sec:theory-PP}).

The uncertainty on the theoretical expression is obtained by assuming that all the sources of uncertainties are from a statistical origin, and modelled by random variables with normal distributions. There are obvious limits of such a treatment for theoretical estimates of hadronic quantities (see in particular Ref.~\cite{Charles:2016qtt} for a discussion of alternative models) but we keep it for simplicity which also allows us to provide p.d.f.s for both theoretical and experimental values of the observables.

\begin{figure} 
\begin{center}
\includegraphics[width=7cm]{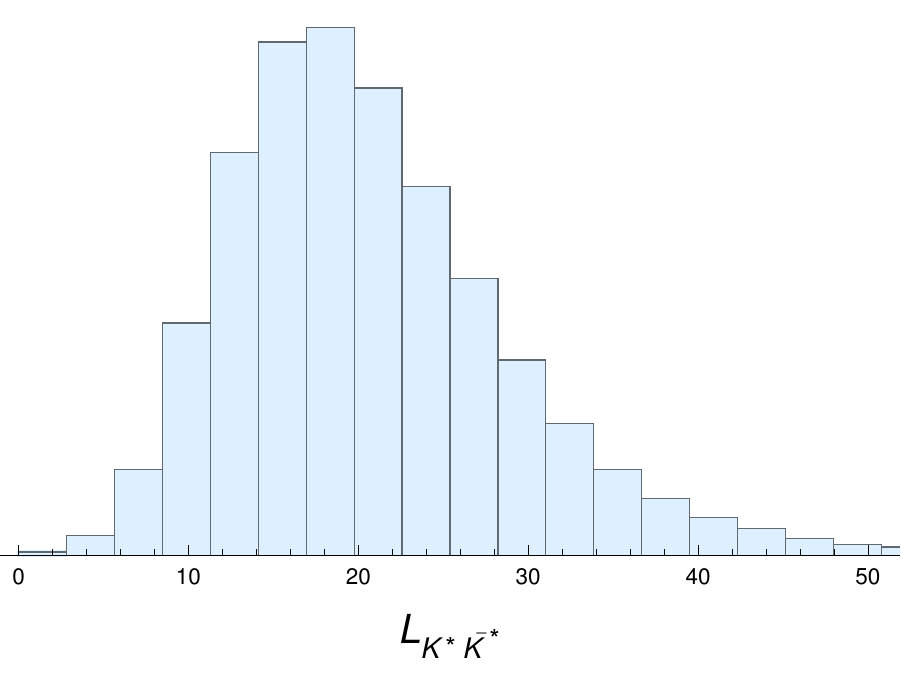}
\qquad \includegraphics[width=7cm]{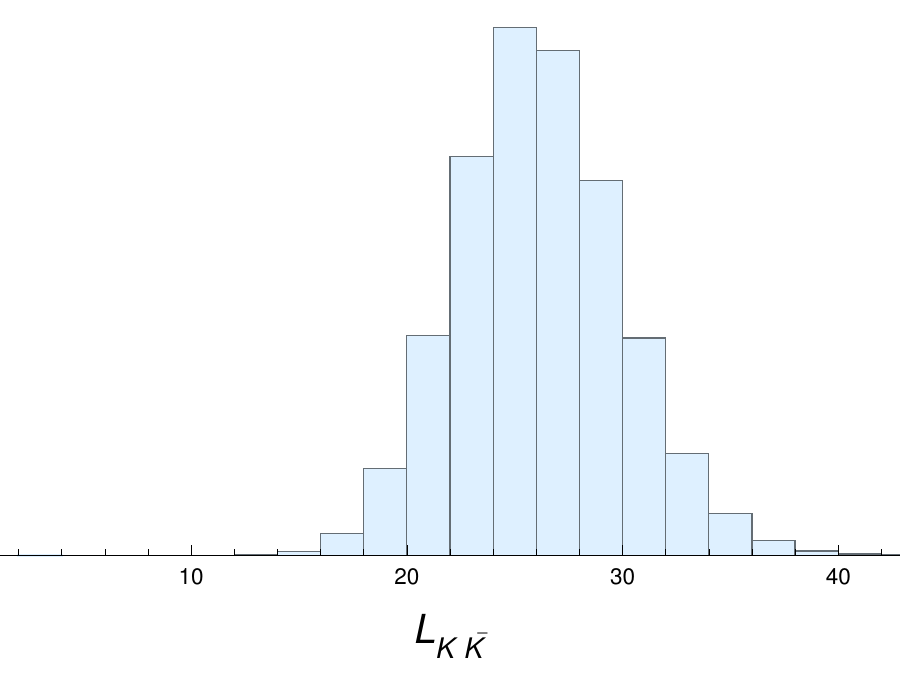}
\end{center}
\caption{$L_{K^*\bar{K}^*}$ (left) and $L_{K\bar{K}}$ (right) SM distributions.}
\label{fig:LKstKstLKKdistrib}
\end{figure}

Regarding the theoretical prediction, 
the large uncertainty on the form factors $A_0^{B_{d,s}\to {K^{*0}}}$ (and the lack of knowledge on their correlation, although they are related by $U$-spin) yields rather significant non-Gaussianities as they enter through the square of their ratios in the prediction for $L_{K^*\bar{K}^*}$ (see App.~\ref{app:nonGaussian} for more detail). As already discussed in Ref.~\cite{Alguero:2020xca}, a theoretical determination of both form factors would allow us to determine the correlation between these form factors, which is expected to be significant based on $U$-spin symmetry, and thus to reduce the uncertainty on the theoretical prediction of $L_{K^*\bar{K}^*}$.
The SM distribution for this observable is presented in Fig.~\ref{fig:LKstKstLKKdistrib} and the pull is 2.6 $\sigma$.

\subsection{Pseudoscalar-pseudoscalar case}\label{sec:LKK}

As discussed in Sec.~\ref{sec:theory-PP} the corresponding observable for a decay to two pseudoscalars, in particular for $P=K^0$ (and its CP conjugate) is a combination of branching ratios and the respective phase space factors, obviously without any polarisation involved:
\begin{equation}\label{eq:LKtKt}
L_{K\bar{K}}=\rho(m_{K^0},m_{K^0})\frac{{\cal B}({\bar{B}_s \to K^{0} {\bar K^{0}}})}{{\cal B}({\bar{B}_d \to K^{0} {\bar K^{0}}})} =\frac{|A^s|^2+ |\bar A^s|^2}{|A^d|^2+ |\bar A^d|^2}\,,
\end{equation}

The SM prediction for this observable is obtained within QCDF \cite{Beneke:2003zv} using the same expression as Eq.~(\ref{eq:Lgeneraldiscussion}) once $P_{s,d}$ (and $\Delta_{d,s}$) are substituted to the matrix elements for the corresponding decays into pseudoscalar mesons. Using the updated values collected in Tab.~\ref{tab:inputs} we find:
 \begin{equation} 
L_{K\bar{K}}^{\rm SM}=26.00^{+3.88}_{-3.59}
\end{equation}
The SM distribution for this observable is presented in Fig.~\ref{fig:LKstKstLKKdistrib}. As can be seen from our results, the distribution can be approximated as Gaussian.

 On the experimental side, following PDG~\cite{Workman:2022ynf}, we have:
$
    {\cal B}({\bar{B}_d\to K^0\bar{K}^0}) = (1.21 \pm 0.16)\times 10^{-6}
$
where PDG includes Belle~\cite{Belle:2012dmz} and Babar~\cite{BaBar:2006enb} results in calculating their global average. In the $B_s$ case, we have
$
    {\cal B}({\bar{B}_s\to K^0\bar{K}^0}) = (1.76 \pm 0.33)\times 10^{-5},
$ 
where the PDG average includes the results from the LHCb~\cite{LHCb:2020wrt} and Belle collaborations~\cite{Belle:2015gho}, with an additional 7\% relative uncertainty due to $B_s$-mixing (see Sec.~\ref{sec:theory-PP})~\footnote{As discussed in Sec.~\ref{sec:theory-PP}, the modulation due to the time dependence induced by $B_s$-mixng  arises only for $B_s$-mesons produced incoherently and decaying into CP-eigenstates. 
In all rigour, this modulation affects only the LHCb result, but not the Belle result. However, the former dominates the average, and we checked explicitly that adding a 7\% relative uncertainty to the PDG average is a very good approximation.}.
Using Eq.~(\ref{eq:LKtKt}) we obtain the following experimental value:
\begin{equation}\label{eq:expdataLKK}
    L_{K\bar{K}}^{\rm exp} = 14.58\pm 3.37\,.
\end{equation}
The SM distribution for the theoretical prediction is given in Fig.~\ref{fig:LKstKstLKKdistrib}.
In this case we observe a deficit similar to the decays into vector mesons but less acute, with a pull of 2.4 $\sigma$.
 
 \subsection{Vector-pseudoscalar case}
 
It is natural to consider also the decays into a pseudoscalar and a vector mesons to complete the analysis. As seen in Sec.\ref{sec:theory}, one has in principle to distinguish the modes depending on which meson (pseudoscalar or vector) caries away the spectator quark. Although this separation is possible in principle at LHCb, it requires tagging to determine the flavours involved, which reduces the number of accessible events drastically: indeed, tagging has a cost of at least 1/20 in terms of statistics (see Table 1 in Ref.~\cite{Fazzini:2018dyq}). The effect is even more important for $\bar{B}_d\to K^{(*)0} \bar{K}^{(*)0}$ decays which are rarer as they are mediated by a CKM-suppressed $b\to d$ transition. 
 We will thus organise the discussion in two steps. We start by considering the observables that will require tagging for the $B_s$ and $B_d$ modes, i.e, will be accessible during LHCb Run 3. We will then propose related observables relaxing the tagging requirement, first for the $B_d$- and then also for the $B_s$-mesons so that these observables may already be accessible with the current LHCb data. 

The observables with an optimised NP sensitivity can be separated in two cases depending on the meson that collects the spectator quark:
\begin{itemize}
    \item $M_1=K^{*0}$ 
     case. We will denote this observable ${\hat L}_{K^*}$:
    \begin{equation}\label{eq:LKst-def}
{\hat L}_{{K}^{*}}=\rho(m_{K^0},m_{K^{*0}})\frac{{\cal B}({{\bar B}_s\to {{ K^{*0}}\bar K}^{0})}
}{{\cal B}({{\bar B}_d \to {\bar K}^{*0} { K^{0}})}} =\frac{|A^s|^2+ |\bar A^s|^2}{|A^d|^2+ |\bar A^d|^2}\,,
\end{equation} 
  The SM prediction for this observable is obtained from Eq.~\ref{eq:Lgeneraldiscussion} simply changing $P_{d,s}$ and $\Delta_q$ by the corresponding ones of these modes:
     \begin{equation}
    {\hat L}^{\rm SM}_{{K}^{*}} =21.30^{+7.19}_{-6.30}  
    \end{equation} 
   \item $M_1=K^0$ case: We will denote this observable ${\hat L}_K$:
    \begin{equation}\label{eq:LK-def}
{\hat L}_{K}=\rho(m_{K^0},m_{K^{*0}})\frac{{\cal B}({{\bar B}_s \to 
{ K^{0}}{\bar K}^{*0} )}}{{\cal B}({{\bar B}_d \to {\bar K}^{0} { K^{*0}})}} =\frac{|A^s|^2+ |\bar A^s|^2}{|A^d|^2+ |\bar A^d|^2}\,,
\end{equation}   
      The SM prediction for this observable is:
    \begin{equation}
    {\hat L}^{\rm SM}_{K}= 25.01^{+4.21}_{-4.07}\end{equation}
   \end{itemize}

\begin{figure} \begin{center}
\includegraphics[width=7cm]{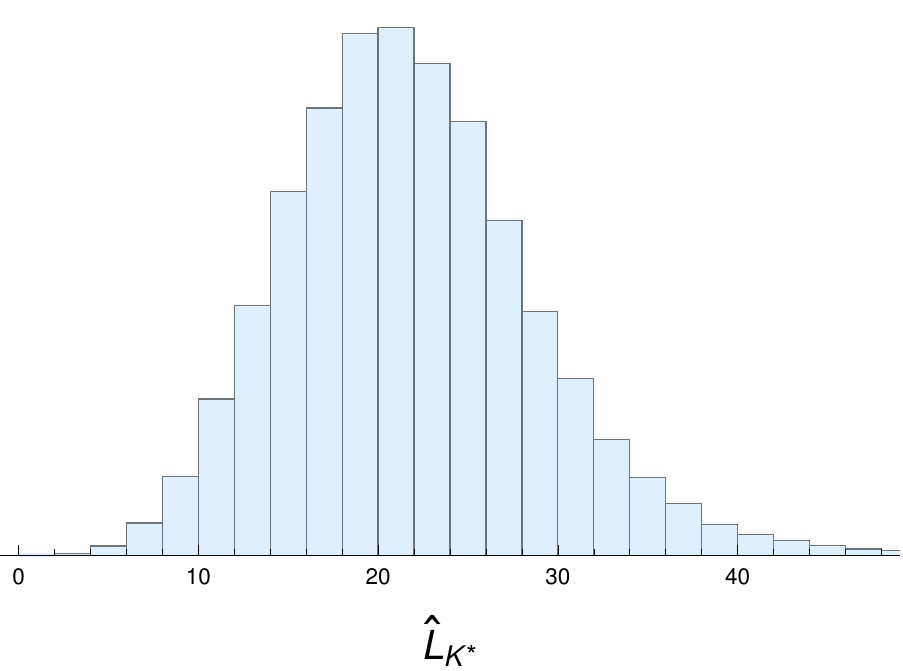} \includegraphics[width=7cm]{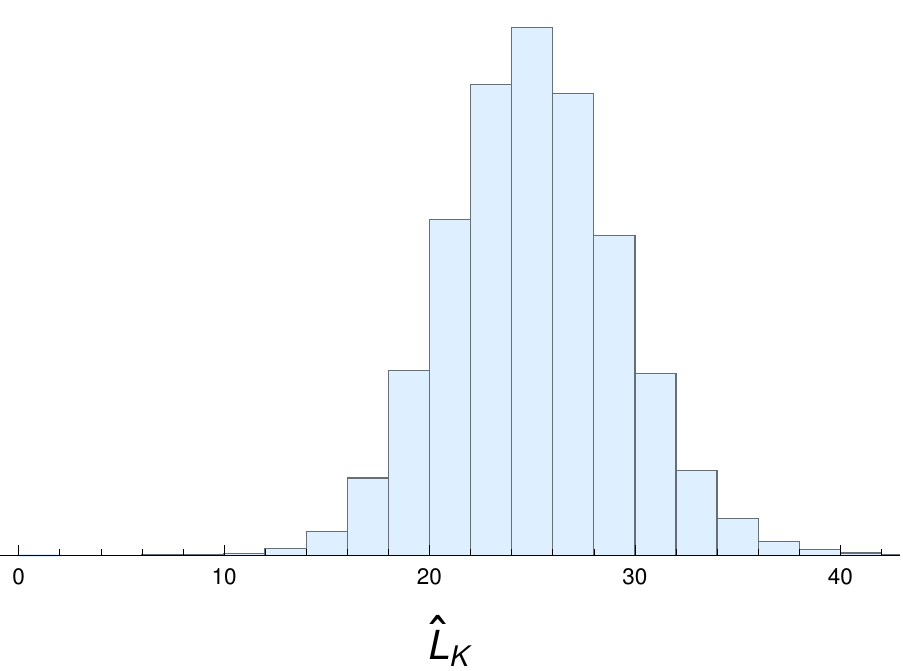} 
\end{center}
\caption{$\hat{L}_{K^*}$ (left) and $\hat{L}_{K}$  (right) SM distributions.}
\label{fig:hatLKstKLKstKdistrib}
\end{figure}

\begin{figure} \begin{center}
\includegraphics[width=7cm]{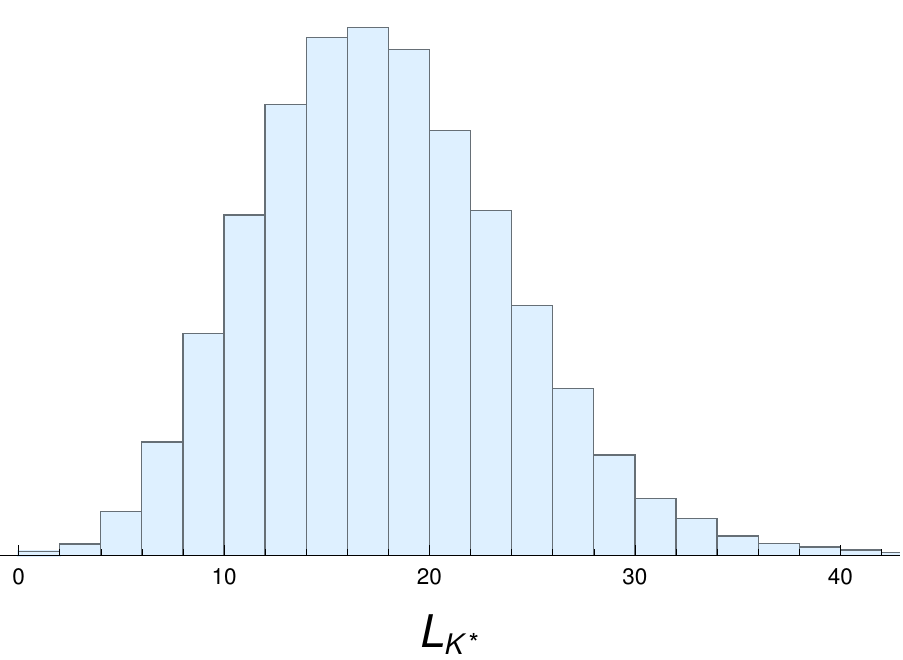} \includegraphics[width=7cm]{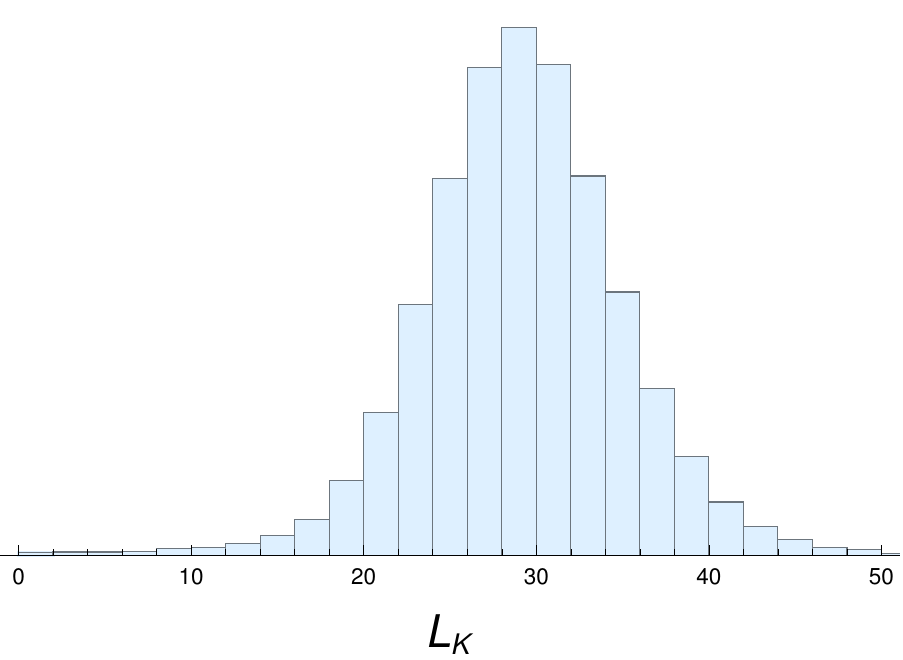} 
\includegraphics[width=7cm]{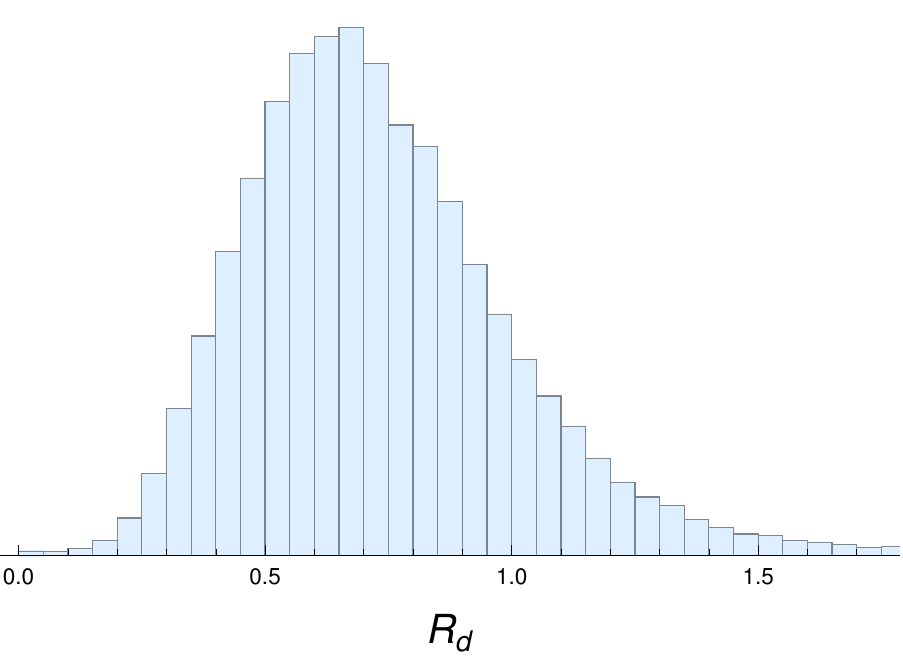} 
\includegraphics[width=7cm]{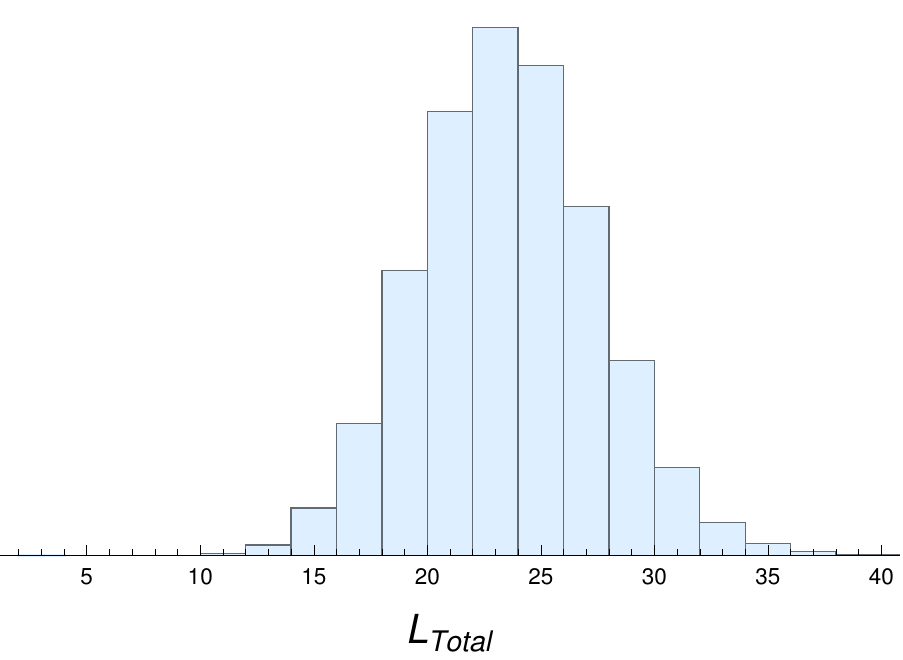} 
\end{center}
\caption{From left to right, and top to bottom: ${L}_{K^*}$, ${L}_{K}$, $R_d$ and $L_{\rm total}$ SM distributions.}
\label{fig:LKstKLKstKdistrib}
\end{figure}

The corresponding distributions are shown in Fig.~\ref{fig:hatLKstKLKstKdistrib}.
We see that $\hat{L}_{K^*}$ is more affected by non-Gaussianities than $\hat{L}_K$. The former (resp. latter) observable involves modes where the spectator quark is carried away by a vector (resp. pseudoscalar) meson, which means that the prediction in factorisation approaches involves the form factors $A_0^{B_q\to K^{*0}}$ (resp. $F_+^{B_q\to K}$), which have large (resp. small) uncertainties. Therefore it is not surprising that the distribution of $\hat{L}_{K^*}$ suffers from similar non-Gaussianities as $L_{K^*\bar{K}^*}$, whereas $\hat{L}_{K}$ is less affected.

As discussed at the beginning of this section, requiring tagging for $B_d$ modes is particularly challenging: it yields a very small number of events, as tagging is applied to decays mediated by a CKM-suppressed $b\to d$ transition.
We may thus relax this requirement by defining related observables easier to access using the currently available data from Run 1 and Run 2, although we may lose sensitivity to some NP scenarios on the way:
\begin{equation}\label{eq:LhatKstar} 
    L_{K^{*}}= 2\,\rho(m_{K^0},m_{K^{*0}})\frac{{\cal B}({{\bar B}_s\to  { K^{*0}}
    {\bar K}^{0}    )}}{{\cal B}({{\bar B}_d \to {\bar K}^{*0} { K^{0}})}
    +{\cal B}({{\bar B}_d \to {\bar K}^{0} { K^{*0}})}
    } 
    =\frac{2 R_d}{1+ R_d}\hat{L}_{K^{*}}\,,\end{equation}
and 
\begin{equation} \label{eq:LhatK} 
    L_K= 2\, \rho(m_{K^0},m_{K^{*0}})\frac{{\cal B}({{\bar B}_s\to  { K^{0}}
    {\bar K}^{*0}    )}}{{\cal B}({{\bar B}_d \to {\bar K}^{*0} { K^{0}})}
    +{\cal B}({{\bar B}_d \to {\bar K}^{0} { K^{*0}})}
    }     =\frac{2}{1+ R_d}\hat{L}_K\,,
\end{equation}
reexpressing them in terms of the optimal observables, using $R_d$:
\begin{equation}
    R^d=\frac{{\cal B}({{\bar B}_d \to {\bar K}^{*0} {{ K}^{0}})}}{{\cal B}({{\bar B}_d \to {\bar K}^{0} { K^{*0}})}}\,.
 \end{equation}
      The SM prediction for the three observables is:
      \begin{equation}
  L^{\rm SM}_{{K}^{*}} = 17.44^{+6.59}_{-5.82},   \quad    L^{\rm SM}_K=29.16^{+5.49}_{-5.25},  \quad    R^{d \, \rm SM} = 0.70^{+0.30}_{-0.22}\,.
      \end{equation}
so that SM values for $L_{K^*}$ and $L_{K}$ are close to ${\hat L}_{K^*}$ and ${\hat L}_{K}$, respectively. The corresponding distributions are shown in Fig.~\ref{fig:LKstKLKstKdistrib}.  

Finally, we can also consider another observable that can be accessed in the short term, but with an even more limited sensitivity to NP. It combines both $B_s$ (together with their CP conjugates) and both $B_d$ decays:
\begin{eqnarray}
    {L}_{\rm total} &=& \rho(m_{K^0},m_{K^{*0}}) \left(\frac{{\cal B}({\bar{B}_s \to K^{*0} {\bar K^{0}})}+ {\cal B}({\bar{B}_s \to K^{0} {\bar K^{*0}}})}{{\cal B}({\bar{B}_d \to {\bar K}^{*0} { K^{0}})}+ {\cal B}({\bar{B}_d \to {\bar K}^{0} { K^{*0}})}}\right)\nonumber
    \\&=&
    \frac{L_{K^*}+L_K}{2}=
    \frac{\hat{L}_{K}+ \hat{L}_{K^*} R^d}{1+R^d}
\end{eqnarray}
whose SM prediction is:
    \begin{equation}
       {L}^{\rm SM}_{\rm total} = 23.48^{+3.95}_{-3.82}. 
    \end{equation}
 
 The experimental situation for the vector-pseudoscalar decays is rather different from the previous modes. For the moment, there is neither experimental information on  $\hat{L}_{K^*}$, $\hat{L}_{K}$ nor on $L_{K^*}$, $L_{K}$. For
 $L_{\rm total}$ there exists a measurement in Ref.~\cite{LHCb:2019vww} for the combined CP-averaged $B_s$ modes and an upper bound for the $B_d$ modes in Ref.~\cite{LHCb:2015oyu}, but one cannot infer an experimental value/bound for $L_{\rm total}$ easily. Interestingly if one combines the central value for $B_s$ branching ratios and the upper bound for $B_d$ branching ratios, the result that we obtain
 for $L_{\rm total}$ is around 21. Although we cannot assign a clear statistical meaning to this ``lower bound'', it is interesting to notice that this value is quite close to the SM prediction.
We consider thus the measurement of these observables of prime interest: they will provide distinctive patterns of deviations in different NP scenarios, as will be discussed in the following sections.

\begin{figure}[t]
\begin{center}
\includegraphics[width=0.40\linewidth]{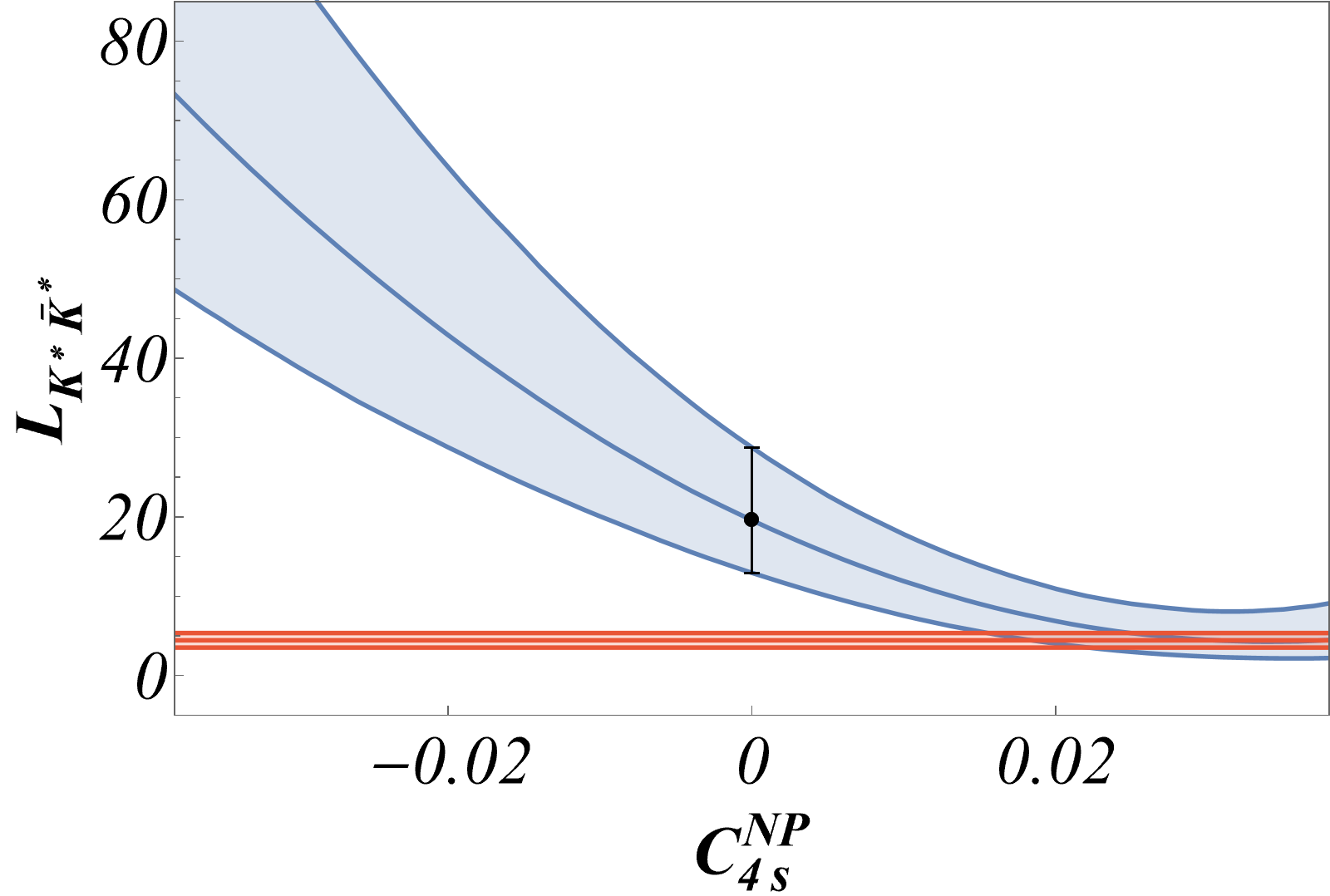}
\includegraphics[width=0.40\linewidth]{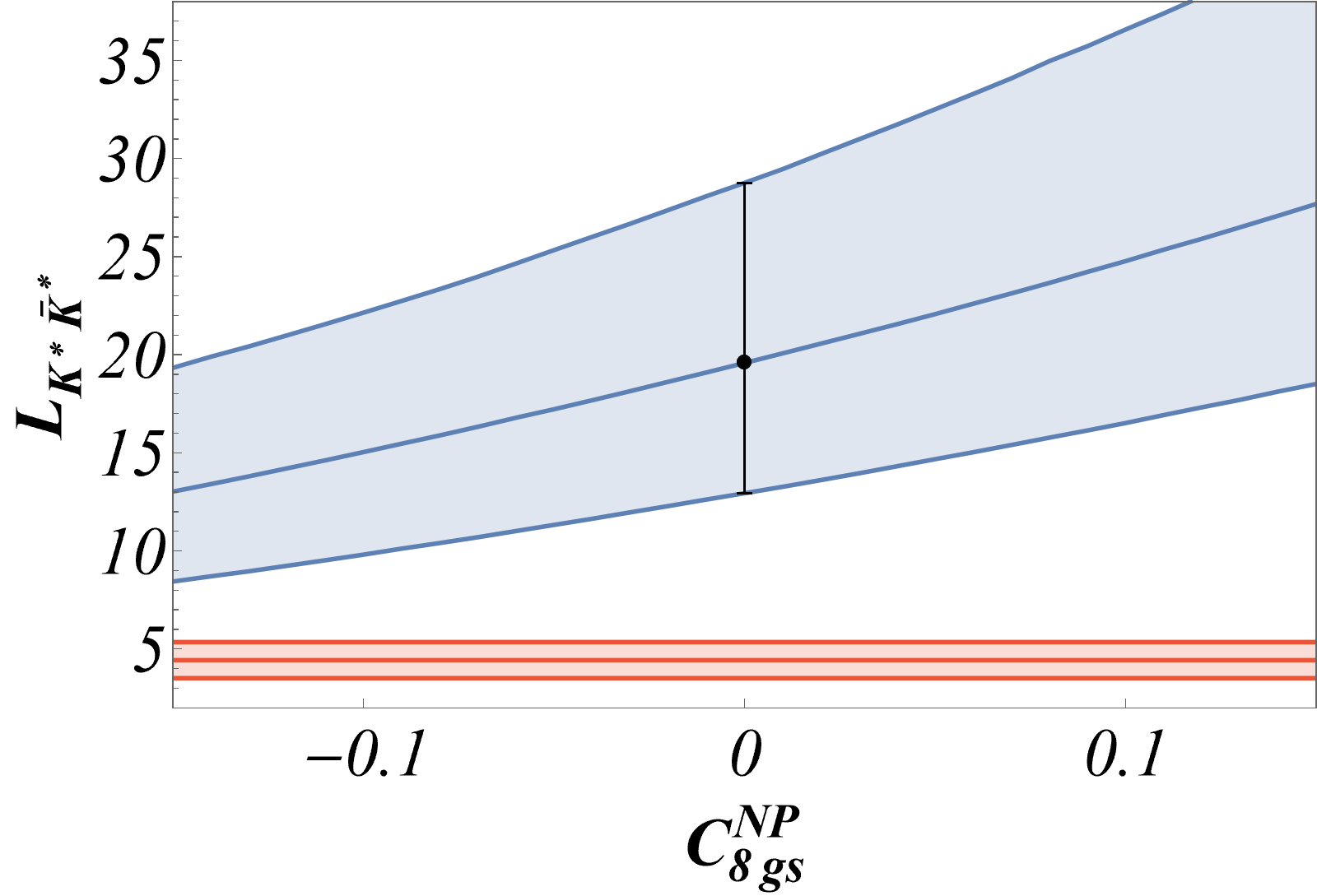}
\end{center} 
\caption{
Sensitivity of $L_{K^*\bar{K^*}}$ to individual NP contributions for the most relevant Wilson coefficients: ${\cal C}^{\rm NP}_{4s}$,  ${\cal C}^{\rm NP}_{8gs}$. 
 For each coefficient, the range of variation considered for the NP contribution corresponds to 100\% of its SM value. The uncertainty of the theoretical prediction (blue band) is computed for each value of the NP contribution. The red band corresponds to the experimental 1$\sigma$ range.
} \label{fig:figLKstarKstarrel}
\end{figure}
 
\begin{figure}[t]
\begin{center}
\includegraphics[width=0.40\linewidth]{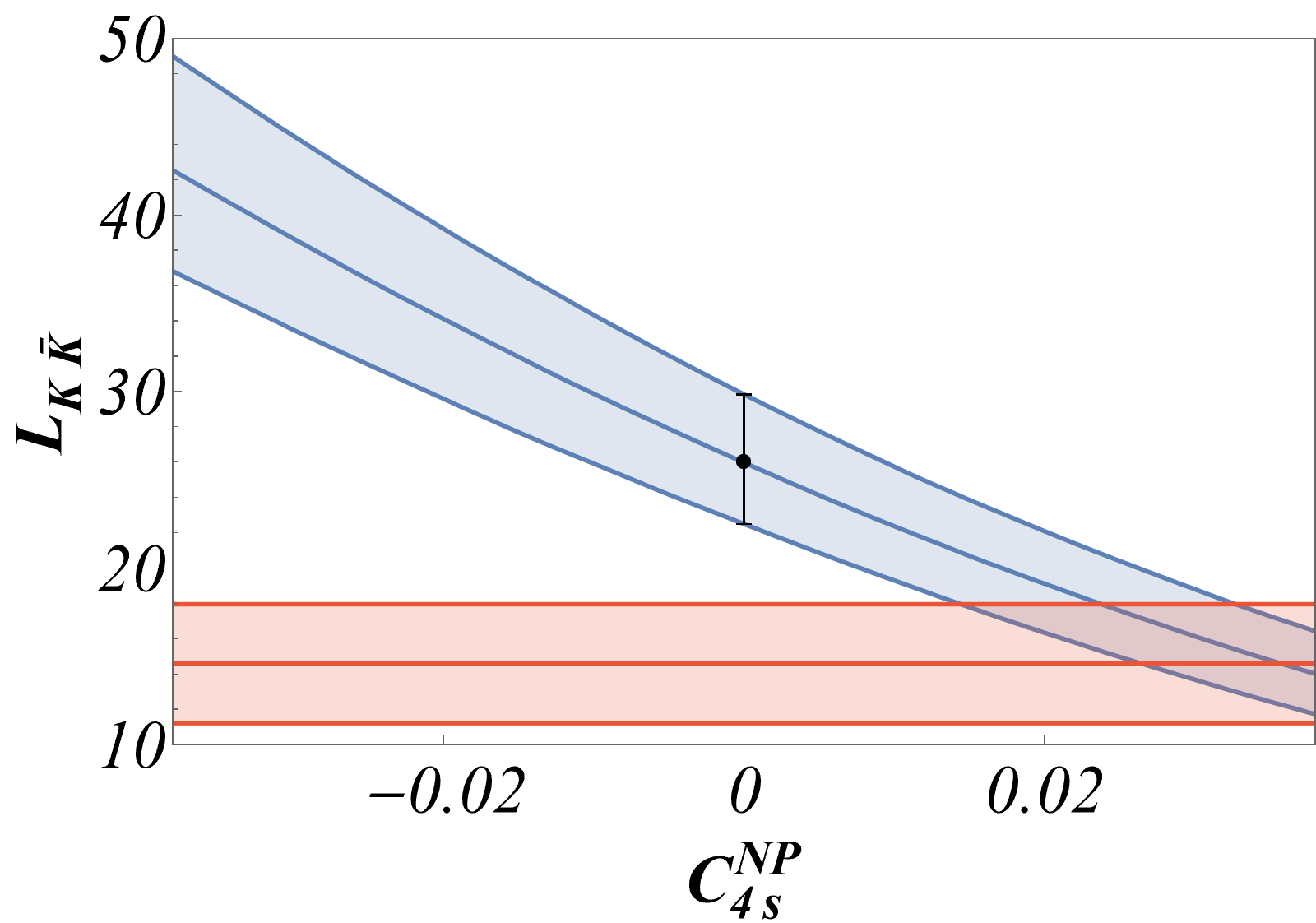}
\includegraphics[width=0.40\linewidth]{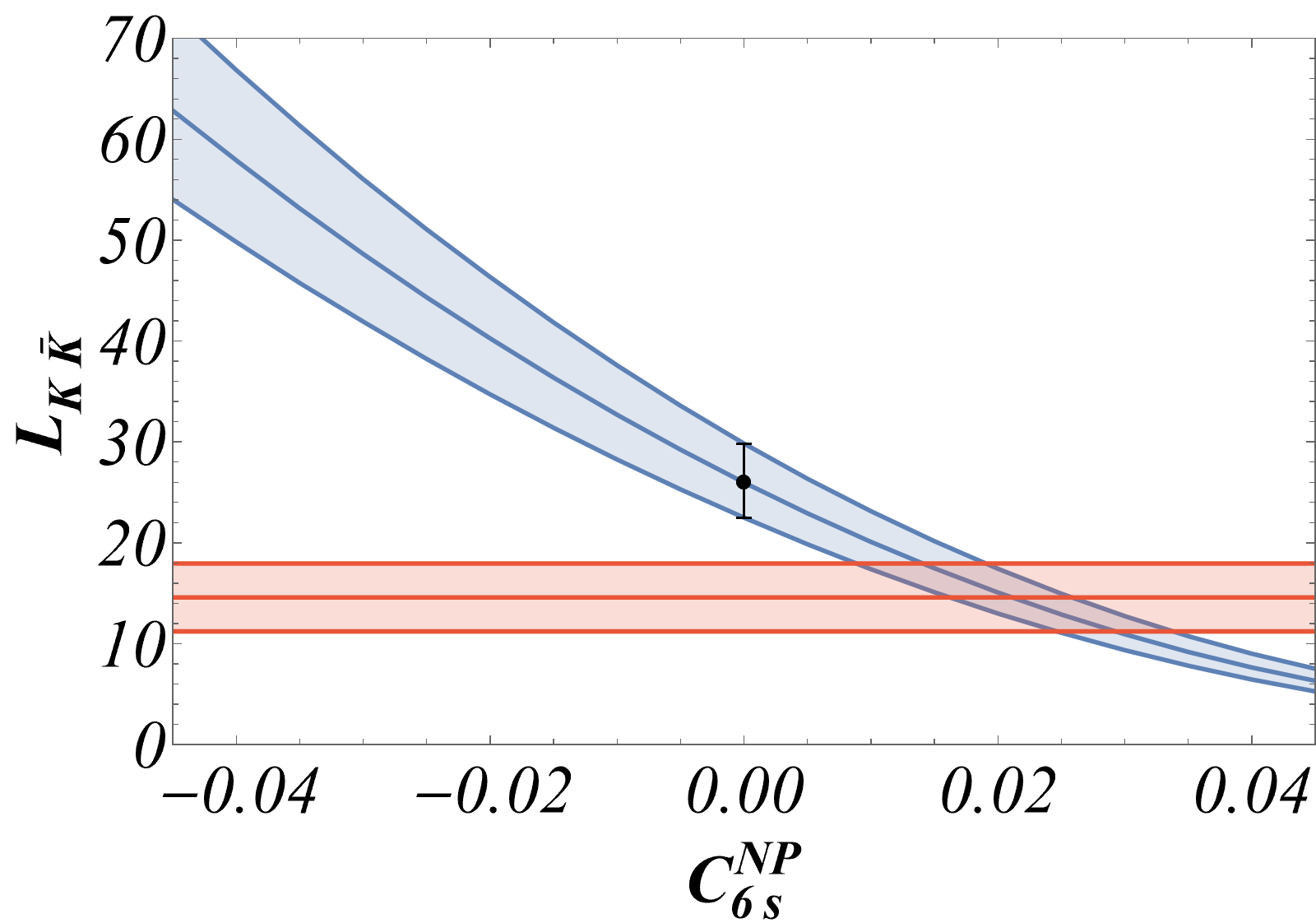}
\includegraphics[width=0.40\linewidth]{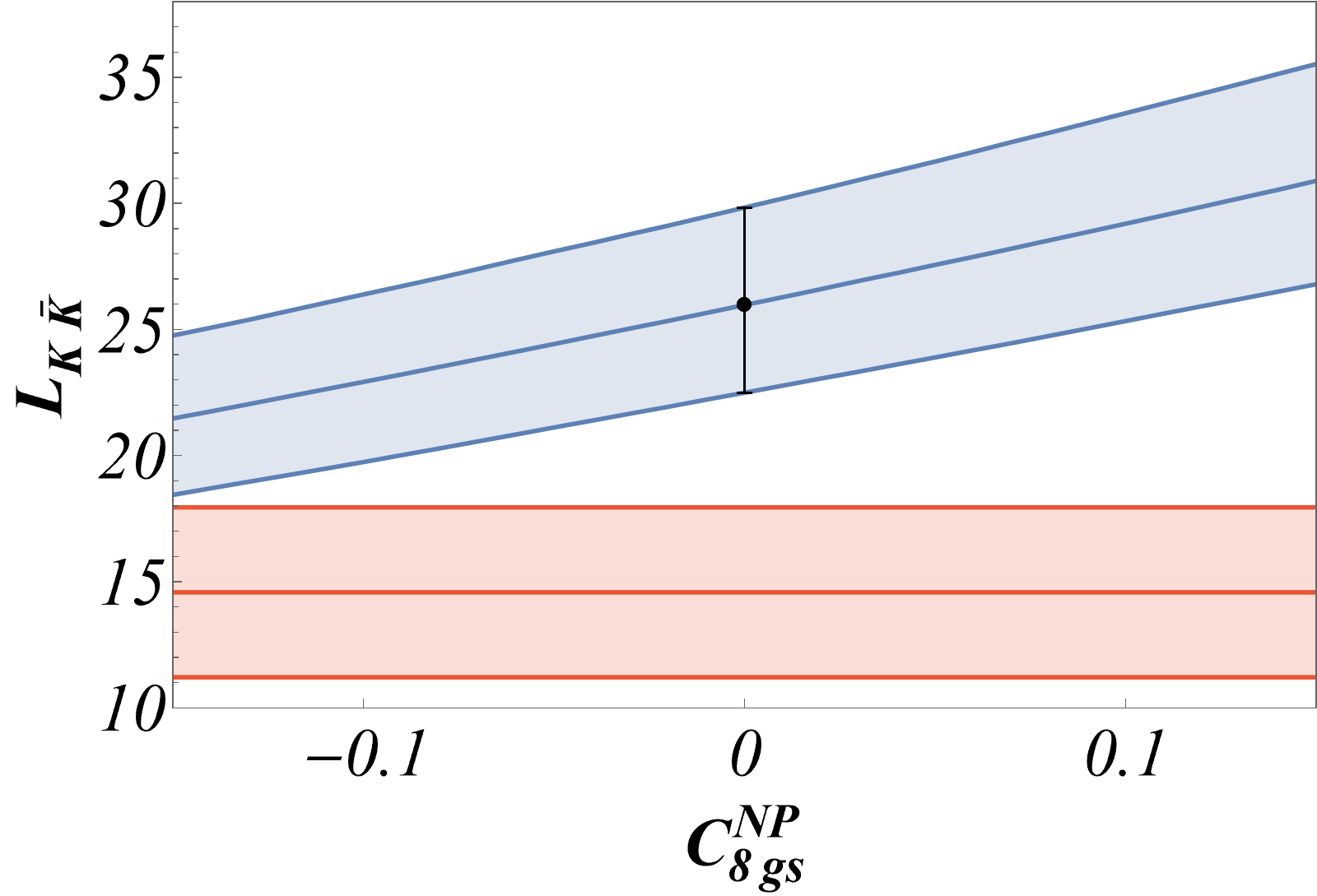}
\end{center} 
\caption{
Sensitivity of $L_{K\bar{K}}$ to individual NP contributions in the most relevant Wilson coefficients: ${\cal C}^{\rm NP}_{4s}$, ${\cal C}^{\rm NP}_{6s}$,  ${\cal C}^{\rm NP}_{8gs}$.  See Fig.~\ref{fig:figLKstarKstarrel} for more detail.
} \label{fig:figLKKrel}
\end{figure}

\begin{figure}
\centering
\includegraphics[width=0.40\textwidth]{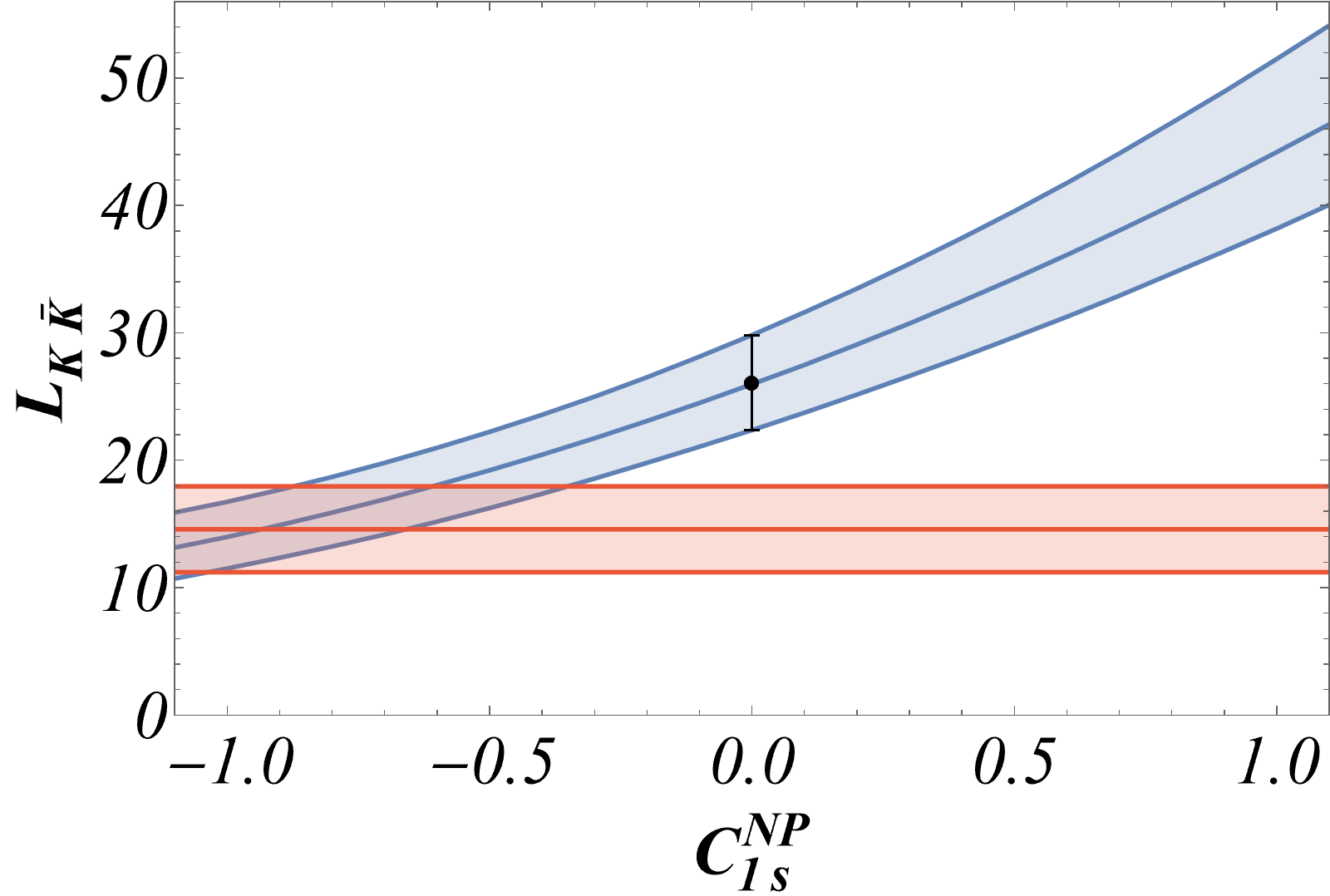}\qquad
\includegraphics[width=0.40\textwidth]{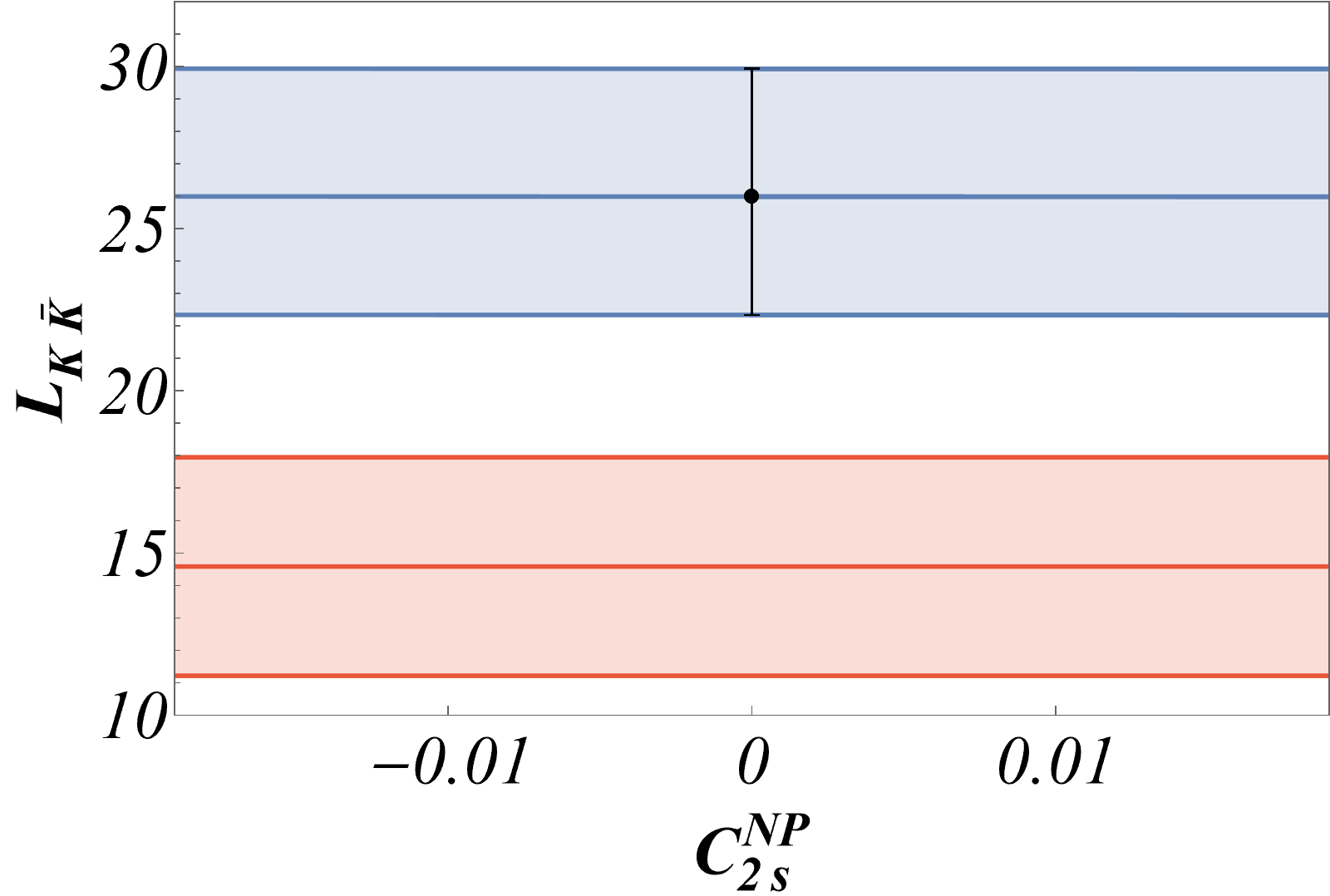}\qquad
\includegraphics[width=0.40\textwidth]{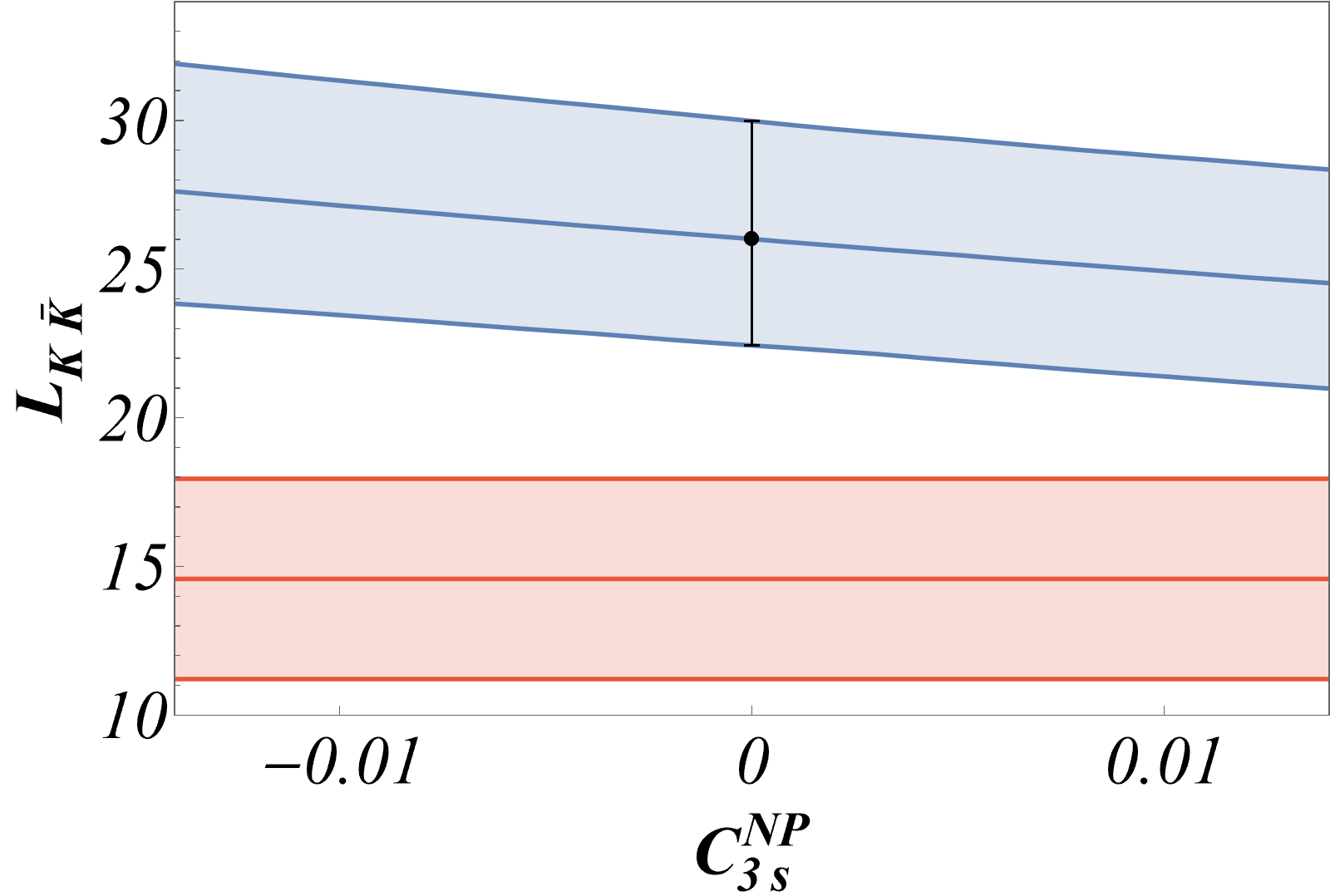}\qquad
\includegraphics[width=0.40\textwidth]{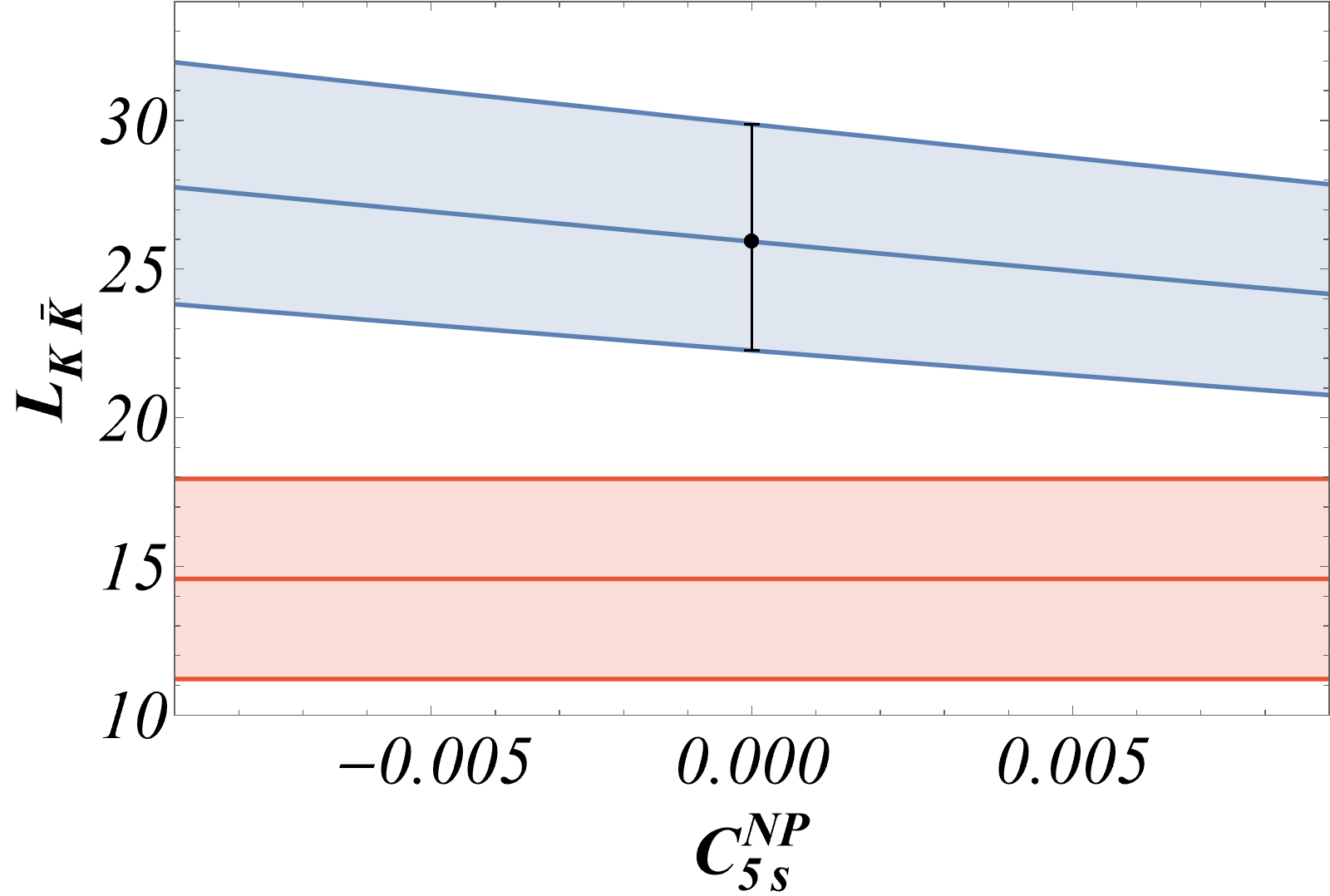}\qquad\qquad
\includegraphics[width=0.40\textwidth]{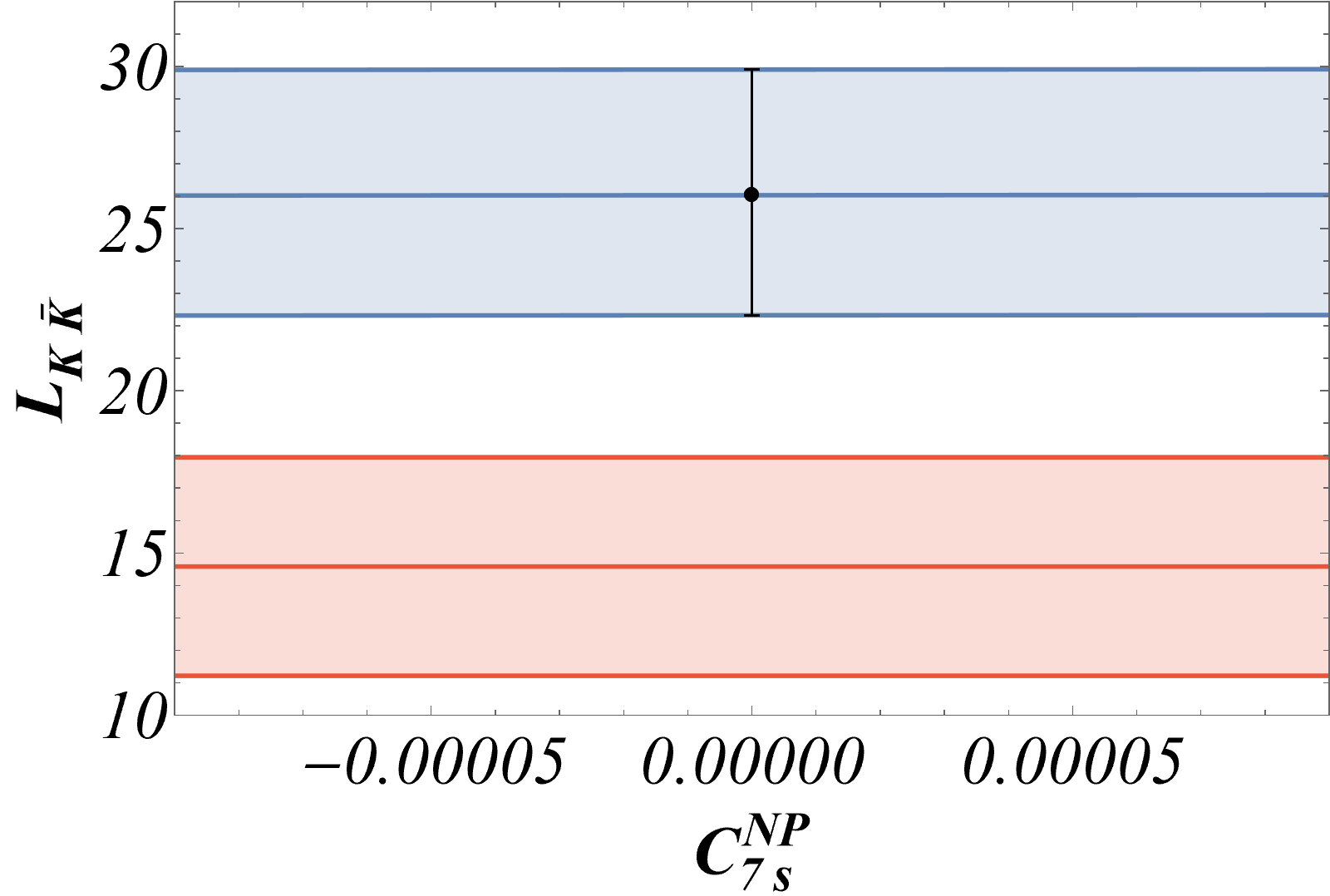}\qquad
\includegraphics[width=0.40\textwidth]{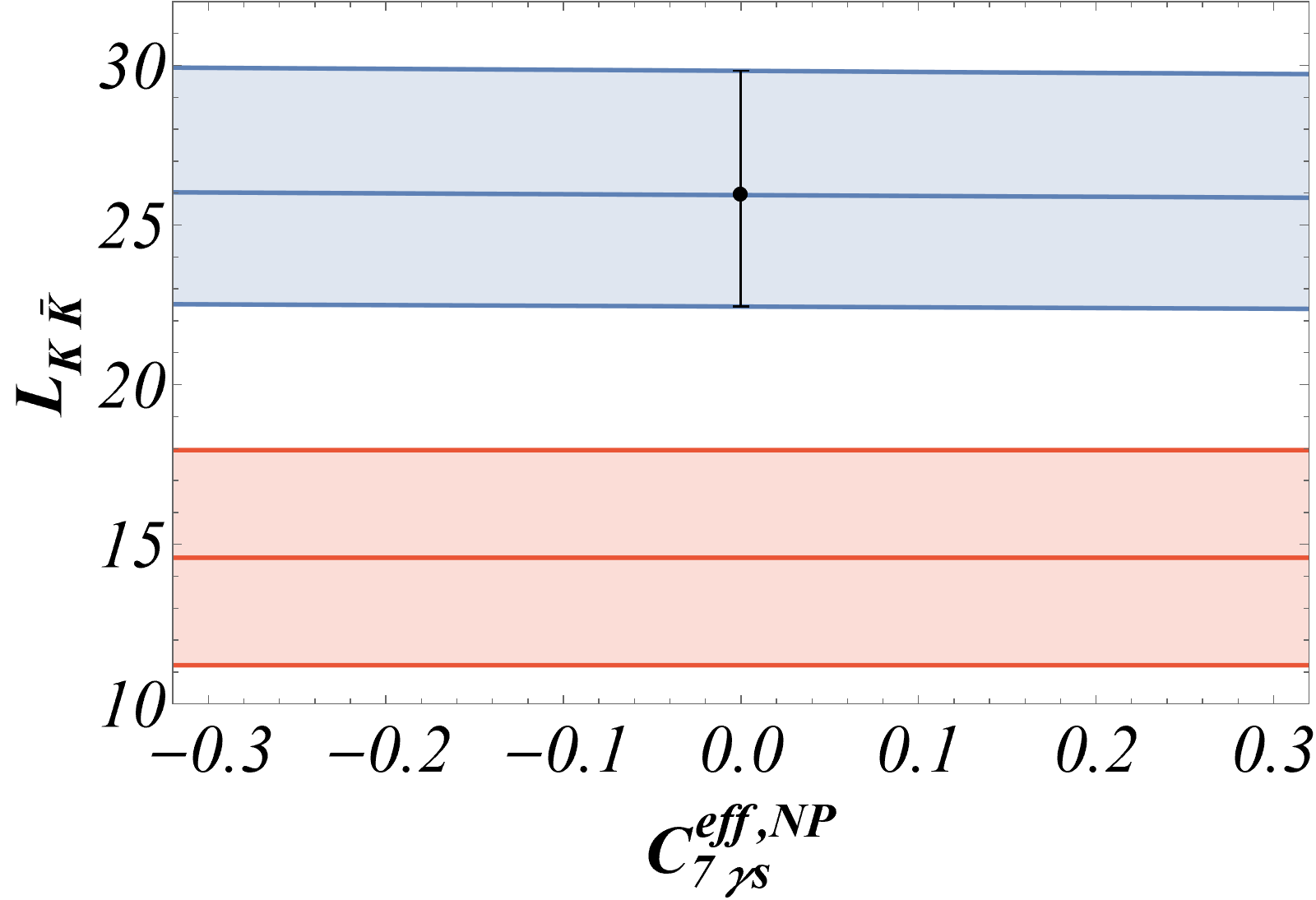}\qquad
\caption{Sensitivity of $L_{K\bar{K}}$ to NP individual contributions for other Wilson coefficients ${\cal C}^{\rm NP}_{is}$. 
See Fig.~\ref{fig:figLKstarKstarrel} for more detail. }
\label{fig:figLKK}
\end{figure} 

\section{Optimised observables within an effective field theory approach}\label{sec:modelindep}

After our discussion of the SM predictions for these observables and their deviations from experiment (when available), we can now consider their sensitivity to NP contributions within the Weak Effective Theory described in App.~\ref{app:WET}. For simplicity, in this section, we will consider only shifts to Wilson coefficients associated to SM operators, and we will express deviations in terms of the Wilson coefficients ${\cal C}_{is}^{\rm NP}$ for $b\to s$ transition (this initial assumption will be generalised in Sec.~\ref{sec:indivBR}). We will emphasise the differences among the observables in terms of size and signs (and its origin) of their Wilson coefficient dependence. 

\subsection{$\bar{B}_q\to VV$ and $\bar{B}_q\to PP$} \label{sec:BVVBPP}

In Ref.~\cite{Alguero:2020xca}, we tested the sensitivity of $L_{K^*\bar{K}^*}$ to NP shifts for a single Wilson coefficient. We found that
three coefficients (tree ${\cal C}_{1s}$, QCD penguin ${\cal C}_{4s}$ and chromomagnetic ${\cal C}_{8gs}$) could explain the corresponding tension between experiment and the SM central value through an NP contribution of at most 100\% of the SM value, as can be seen in Fig.~\ref{fig:figLKstarKstarrel} (similar to Ref.~\cite{Alguero:2020xca} but recomputed with the inputs given in Tab.~\ref{tab:inputs}). As discussed in Ref.~\cite{Alguero:2020xca} and following Ref.~\cite{Lenz:2019lvd}, the value of ${\cal C}_{1s}$ cannot deviate so much from the SM value, leaving ${\cal C}_{4s}$ and ${\cal C}_{8gs}$ as potential candidates for an explanation of the anomaly in $L_{K^*\bar{K}^*}$.

We may perform the same exercise for $L_{K \bar{K}}$, where four Wilson coefficients (tree ${\cal C}_{1s}$, QCD penguins ${\cal C}_{4s}$, ${\cal C}_{6s}$ and chromomagnetic ${\cal C}_{8gs}$)
can satisfy the same conditions, see  Figs.~\ref{fig:figLKKrel} and \ref{fig:figLKK}. In these plots we have not assumed the SM uncertainty at each NP point, as we did for the sensitivity plots in Ref.~\cite{Alguero:2020xca} but rather recomputed the uncertainty point by point to get a more precise result~\footnote{For completeness we recomputed these uncertainties also for $L_{K^*{\bar K}^*}$ in Fig.~\ref{fig:figLKstarKstarrel}, recovering  the same conclusion as in Ref.~\cite{Alguero:2020xca}.}. 
The large values needed for the tree-level coefficient ${\cal C}_{1s}$ are excluded here
for the same reasons as in the previous case~\cite{Lenz:2019lvd}, so that the only remaining possibilities consist of the Wilson coefficients of two QCD penguin operators ($Q_{4s}$ and $Q_{6s}$) and the chromomagnetic operator ($Q_{8gs}$).

We can illustrate the dependence of these two observables on the relevant  Wilson coefficients taking the central values of the inputs~\footnote{In these expressions, we quote the SM value obtained by taking the central values of the input parameters. This is slightly different from the central value of the predictions in the previous section where we used the median of the distribution to define the central value, see App.~\ref{app:nonGaussian}.}:
\begin{align} \label{eq:LKstLKst-Cis}
\!\!L_{K^* \bar{K}^*} = &\; 19.25 - 936.23\; {\cal C}_{4s}^{\rm NP} + 14383.60 \;({\cal C}_{4s}^{\rm NP})^2 + 55.44\; {\cal C}_{6s}^{\rm NP} + 73.70\; ({\cal C}_{6s}^{\rm NP})^2 
\nonumber \\
& + 50.53\; {\cal C}_{8gs}^{\rm NP}
  + 39.38\; ({\cal C}_{8gs}^{\rm NP})^2 - 711.45\; {\cal C}_{4s}^{\rm NP}\; {\cal C}_{6s}^{\rm NP}
  - 1502.07\; {\cal C}_{4s}^{\rm NP}\; {\cal C}_{8gs}^{\rm NP} \nonumber \\ & + 
 43.76\; {\cal C}_{6s}^{\rm NP}\; {\cal C}_{8gs}^{\rm NP} 
 \end{align}
 \begin{align} \label{eq:LKLK-Cis}
\!\!L_{K \bar{K}} = &\; 25.90 - 380.76\; {\cal C}_{4s}^{\rm NP} + 1646.11\; ({\cal C}_{4s}^{\rm NP})^2 - 631.58\; {\cal C}_{6s}^{\rm NP} + 4313.58\; ({\cal C}_{6s}^{\rm NP})^2
\nonumber \\
& + 31.92\; {\cal C}_{8gs}^{\rm NP} 
  + 10.38\; ({\cal C}_{8gs}^{\rm NP})^2 + 
 5318.62\; {\cal C}_{4s}^{\rm NP}\; {\cal C}_{6s}^{\rm NP} - 257.90\; {\cal C}_{4s}^{\rm NP}\; {\cal C}_{8gs}^{\rm NP} 
 \nonumber \\
& - 
 421.08\; {\cal C}_{6s}^{\rm NP}\; {\cal C}_{8gs}^{\rm NP}
\end{align}
These expressions can be used as guidelines to understand the sensitivity to NP affecting only $b \to s$ transitions, which can be visualised by plotting these observables as function of NP contributions to the different Wilson coefficients of interest. 
These plots will span fairly wide ranges for these NP contributions in order to show the behaviour of the observables. In general, we will restrict ourselves to NP contributions of (at most) the same size as the SM value for the QCD penguins and (at most) three times larger for the chromomagnetic operator.

We observe three main features:
\begin{itemize}
\item In the region of interest, quadratic terms
are subleading but non-negligible for $L_{K^*\bar{K}^*}$ and rather suppressed for $L_{K\bar{K}}$. The interference terms among QCD penguins are rather small for $L_{K^*\bar{K}^*}$
and $L_{K\bar{K}}$ but not totally negligible in the latter case.
The expressions are therefore mostly dominated by the linear terms 
for NP values inside the region of interest.

\item $L_{K^*\bar{K}^*}$ is dominated by  ${\cal C}_{4s}^{\rm NP}$ and ${\cal C}_{8gs}^{\rm NP}$ and it is rather insensitive  to ${\cal C}_{6s}^{\rm NP}$ (in agreement with Ref.~\cite{Alguero:2020xca}), whereas $L_{K\bar{K}}$ is also affected by ${\cal C}_{6s}^{\rm NP}$. This difference between $L_{K^*\bar{K}^*}$ and $L_{K\bar{K}}$ comes from the absence/presence of the contribution ${\cal C}_6+{\cal C}_5/N_c$ in the expression of hadronic matrix elements: indeed, the QCD penguin operators $Q_{5s}$ and $Q_{6s}$ can be fierzed into scalar/pseudoscalar operators that vanish when applied to a final state containing vector mesons (see App.~\ref{app:alphacoeffs-PPVV} for a more detailed explanation).

\item The linear term of ${\cal C}_{4s}^{\rm NP}$ in $L_{K^*\bar{K}^*}$ is approximately three times larger than that of $L_{K\bar{K}}$, which explains the larger impact of varying ${\cal C}_{4s}^{\rm NP}$ for $L_{K^*\bar{K}^*}$. Similarly for ${\cal C}_{8gs}^{\rm NP}$ that exhibits a factor near twice larger in $L_{K^*\bar{K}^*}$.
 The explicit reasons for this enhancement factor are discussed in App.~\ref{app:alphacoeffs-PPVV}. 
\end{itemize}

\subsection{$\bar{B}_q\to VP$ and $\bar{B}_q\to PV$}

We have considered several observables for 
the pseudoscalar-vector case, starting from the ones requiring tagging for both $B_d$- and $B_s$-meson decays:
\begin{align} \label{eq:hatLKst-Cis}
\hat{L}_{K^*} = &\; 21.00 + 1040.25\; {\cal C}_{4s}^{\rm NP} + 12886.60\; ({\cal C}_{4s}^{\rm NP})^2 - 1504.72\; {\cal C}_{6s}^{\rm NP} + 27037.90\; ({\cal C}_{6s}^{\rm NP})^2
\nonumber \\
& - 26.72\; {\cal C}_{8gs}^{\rm NP}   
   + 8.52\; ({\cal C}_{8gs}^{\rm NP})^2\; - 37304.70\; {\cal C}_{4s}^{\rm NP}\; {\cal C}_{6s}^{\rm NP}
 - 662.39\; {\cal C}_{4s}^{\rm NP}\; {\cal C}_{8gs}^{\rm NP}
 \nonumber \\
 & + 
 959.60\; {\cal C}_{6s}^{\rm NP}\; {\cal C}_{8gs}^{\rm NP}\,,
\end{align}
\begin{align}   \label{eq:hatLK-Cis}
\hat{L}_{K} = &\;25.04 - 1201.22\; {\cal C}_{4s}^{\rm NP} + 15994.20\; ({\cal C}_{4s}^{\rm NP})^2 + 149.47\; {\cal C}_{6s}^{\rm NP} + 240.53\; ({\cal C}_{6s}^{\rm NP})^2 \nonumber \\
&
+ 66.04\; {\cal C}_{8gs}^{\rm NP} 
  + 46.59\; ({\cal C}_{8gs}^{\rm NP})^2 - 3252.68\; {\cal C}_{4s}^{\rm NP}\; {\cal C}_{6s}^{\rm NP} - 1723.21\; {\cal C}_{4s}^{\rm NP}\; {\cal C}_{8gs}^{\rm NP} \nonumber \\ & + 182.57\; {\cal C}_{6s}^{\rm NP}\; {\cal C}_{8gs}^{\rm NP}\,.
\end{align}
While  ${\hat L}_K$ resembles in behaviour and dependencies to $L_{K^*\bar{K}^*}$ with a slightly increased sensitivity to ${\cal C}_{6s}$, we observe significant differences in  $\hat{L}_{K^*}$. For instance,
 ${\hat L}_{K^*}$ exhibits a linear term of ${\cal C}_{4s}$ of same size but opposite sign compared to $L_{K^*\bar{K}^*}$ as well as linear term of opposite sign for ${\cal C}_{8gs}$. $\hat{L}_{K^*}$ exhibits also a large sensitivity to ${\cal C}_{6s}$ like $L_{K\bar{K}}$ (and contrary to $L_{K^*\bar{K}^*}$).
  The former property, induced by the different sign of the chirally enhanced factor in $\alpha_4^c$ (see App.~\ref{app:alphacoeffs-PVVP}), provides a very powerful probe of various NP scenarios at LHCb through the measurements of pseudoscalar-vector modes.

We can move now to the observables requiring only $B_s$-meson tagging:  
\begin{align} \label{eq:LKstar}
{L}_{K^*} = &\;18.38 + 910.71\; {\cal C}_{4s}^{\rm NP} + 11281.80\; ({\cal C}_{4s}^{\rm NP})^2 - 1317.34\; {\cal C}_{6s}^{\rm NP} + 23670.80\; ({\cal C}_{6s}^{\rm NP})^2 
\nonumber \\
& - 23.40\; {\cal C}_{8gs}^{\rm NP} 
  + 7.45\; ({\cal C}_{8gs}^{\rm NP})^2 - 
 32659.10\; {\cal C}_{4s}^{\rm NP} {\cal C}_{6s}^{\rm NP} - 579.90\; {\cal C}_{4s}^{\rm NP}\; {\cal C}_{8gs}^{\rm NP}
 \nonumber \\
 & + 
 840.10\; {\cal C}_{6s}^{\rm NP}\; {\cal C}_{8gs}^{\rm NP}
\end{align}
\begin{align} \label{eq:LK}
{L}_{K} = &\;28.16 - 1350.81\; {\cal C}_{4s}^{\rm NP} + 17986.00\; ({\cal C}_{4s}^{\rm NP})^2 + 168.09\; {\cal C}_{6s}^{\rm NP} + 270.48\; ({\cal C}_{6s}^{\rm NP})^2 
\nonumber \\
& + 74.27\; {\cal C}_{8gs}^{\rm NP} 
  + 52.40\; ({\cal C}_{8gs}^{\rm NP})^2 - 
 3657.73\; {\cal C}_{4s}^{\rm NP}\; {\cal C}_{6s}^{\rm NP} - 1937.80\; {\cal C}_{4s}^{\rm NP}\; {\cal C}_{8gs}^{\rm NP}
 \nonumber \\
 & + 
 205.30\; {\cal C}_{6s}^{\rm NP}\; {\cal C}_{8gs}^{\rm NP}
\end{align}
showing similar sensitivities to the Wilson coefficients of interest in the $b \to s$ case. As explained  in Sec.~\ref{sec:bdbsNP} and App.~\ref{app:benchmark}, the measurement of $\hat{L}_{K}$ and $\hat{L}_{K^*}$ will provide a much stronger signal compared to $L_K$ and $L_{K^*}$ if there is NP in both $b \to s$ and  $b\to d$ transitions, as $R_d$ entering the latter observables is then modified, which 
tends to reduce the relative sensitivity of $L_K$ and $L_{K^*}$ to NP in such scenarios. Even if $L_K$ and $L_{K^*}$ are easier to obtain currently at LHCb, it is thus essential to measure $\hat{L}_{K}$ and $\hat{L}_{K^*}$ accurately.

Finally $L_{\rm total}$ exhibits a reduced sensitivity to all relevant Wilson coefficients, as it combines observables with opposite sensitivities to NP shifts:
\begin{align} \label{eq:Ltotal}
{L}_{\rm total} = &\;23.27 - 220.05\; {\cal C}_{4s}^{\rm NP} + 14633.90\; ({\cal C}_{4s}^{\rm NP})^2 - 574.63\; {\cal C}_{6s}^{\rm NP} + 11970.70\; ({\cal C}_{6s}^{\rm NP})^2 
\nonumber \\
& + 25.44\; {\cal C}_{8gs}^{\rm NP}
  + 29.93\; ({\cal C}_{8gs}^{\rm NP})^2 - 
 18158.40\; {\cal C}_{4s}^{\rm NP}\; {\cal C}_{6s}^{\rm NP} - 1258.85\; {\cal C}_{4s}^{\rm NP}\; {\cal C}_{8gs}^{\rm NP} \nonumber \\ &+ 
 522.70\; {\cal C}_{6s}^{\rm NP}\; {\cal C}_{8gs}^{\rm NP}
\end{align}
Indeed  one finds some sensitivity to ${\cal C}_{4s}^{\rm NP}$ only for very large NP contributions, where the strong cancellation among the quadratic and linear terms is reduced. Therefore, unless exceptionally large NP values are allowed for ${\cal C}_{4s}^{\rm NP}$, this observable will be quite SM-like in the relevant region if the NP contribution to only one Wilson coefficient is switched on (with NP values up to the same size as the SM contribution).
Notice that even if there seems to be a large quadratic term proportional to ${\cal C}_{6s}^{\rm NP}$, it mostly cancels with the corresponding linear term in the relevant region (the same cancellation occurs for ${\cal C}_{8gs}^{\rm NP}$). However, as discussed in the next section, the large interference between ${\cal C}_{4s}^{\rm NP}$ and ${\cal C}_{6s}^{\rm NP}$ can provide an interesting and distinctive signal.

Finally, since we consider only NP shifts in $b\to s$ Wilson coefficients in this section,  we may set $R_d$ to its SM value (we will reconsider these assumptions in Sec.~\ref{sec:indivBR}).

 \begin{figure}[ht]
\centering
\includegraphics[width=0.75\textwidth,height=0.5\textwidth]{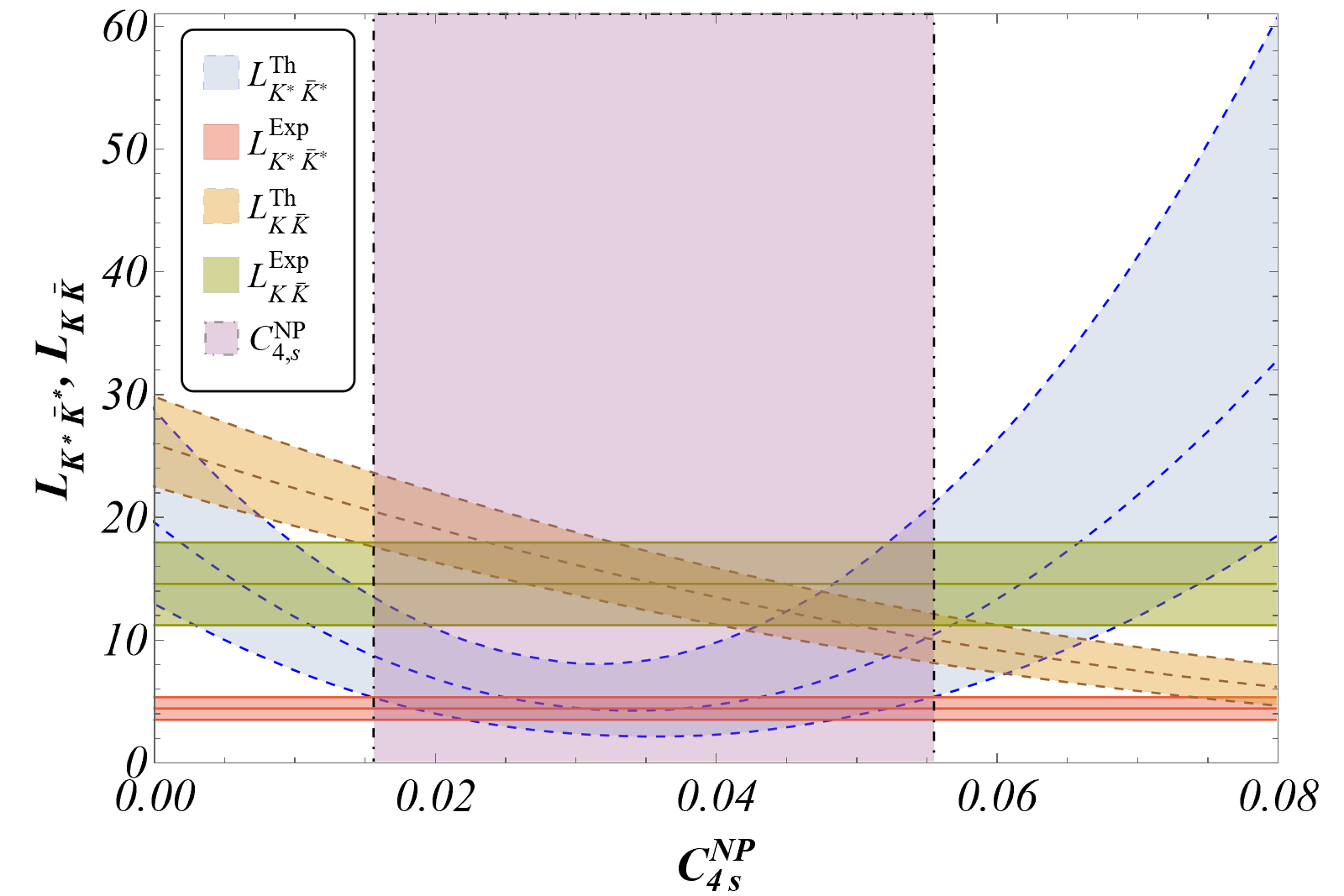}
\caption{ Variation of $L_{K^*\bar{K}^*}$ and $L_{K\bar{K}}$ w.r.t ${\cal C}_{4s}^{\rm NP}$. The range of ${\cal C}_{4s}^{\rm NP}$ where both observables are compatible theoretically and experimentally within $1\sigma$ is: ${\cal C}_{4s}^{\rm NP}\in [0.016, 0.055]$ (corresponding to values of 0.4 to 1.5 times the SM value). Notice that this allowed (magenta) region is determined assuming NP in $b\to s$ transitions only and taking $L_{K\bar{K}}$ and $L_{K^*\bar{K}^*}$ as inputs. A more general case allowing also for NP in $b \to d$ transitions
and including constraints from individual branching ratios will be discussed in Sec.~\ref{sec:indivBR}}.
\label{fig:C4_range}
\end{figure}

\begin{figure}[ht]
\centering
\includegraphics[width=0.75\textwidth,height=0.5\textwidth]{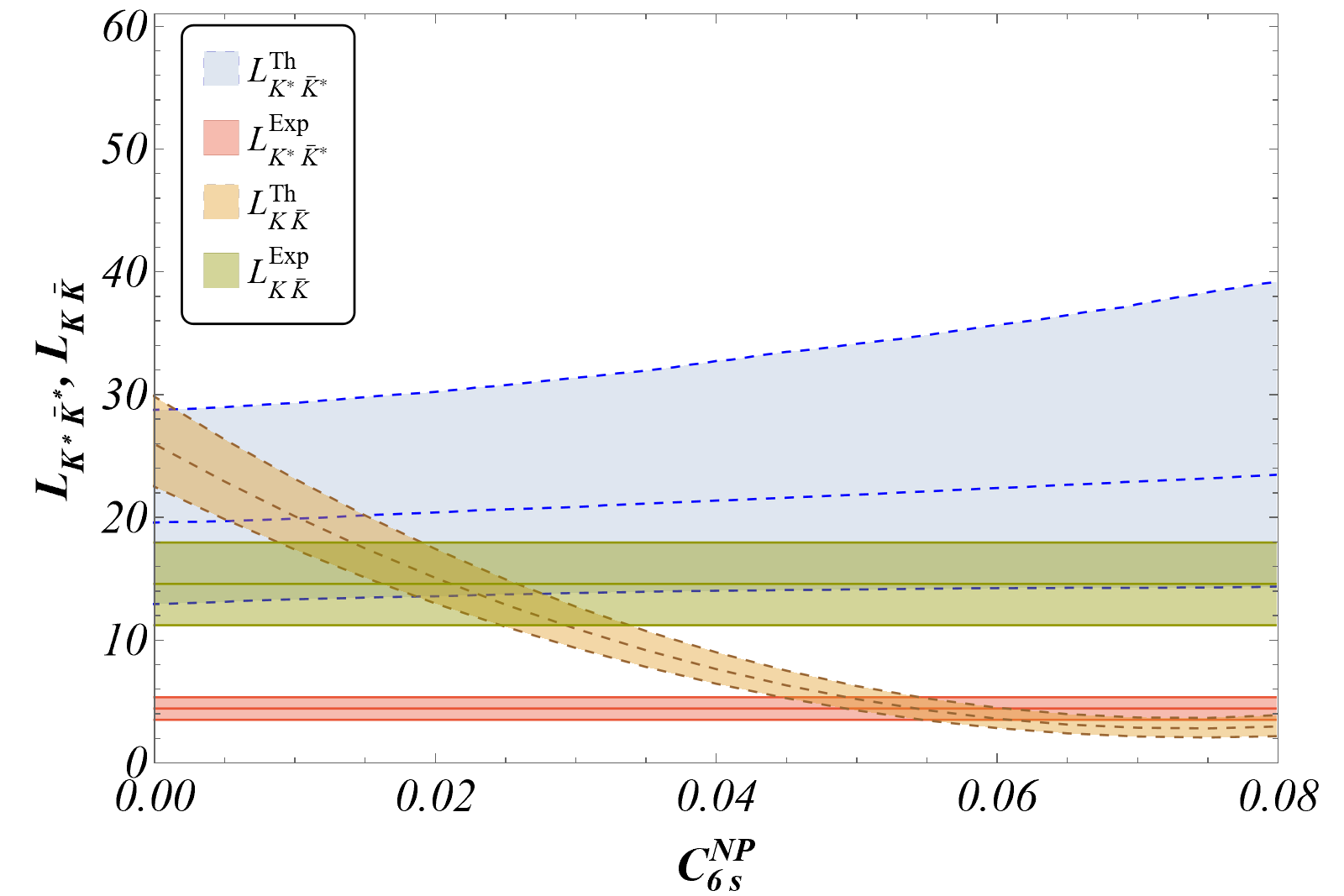}
\caption{ Variation of $L_{K^*\bar{K}^*}$ and $L_{K\bar{K}}$ w.r.t ${\cal C}_{6s}^{\rm NP}$. See Fig.~\ref{fig:C4_range} for further comments on the assumptions and observables considered. No value of ${\cal C}_{6s}^{\rm NP}$ provides values compatible at 1$\sigma$ for both observables if ${\cal C}_{4s}^{\rm NP}=0$.}
\label{fig:C6_range}
\end{figure}

\begin{figure}[ht]
\centering
\includegraphics[width=0.75\textwidth,height=0.6\textwidth]{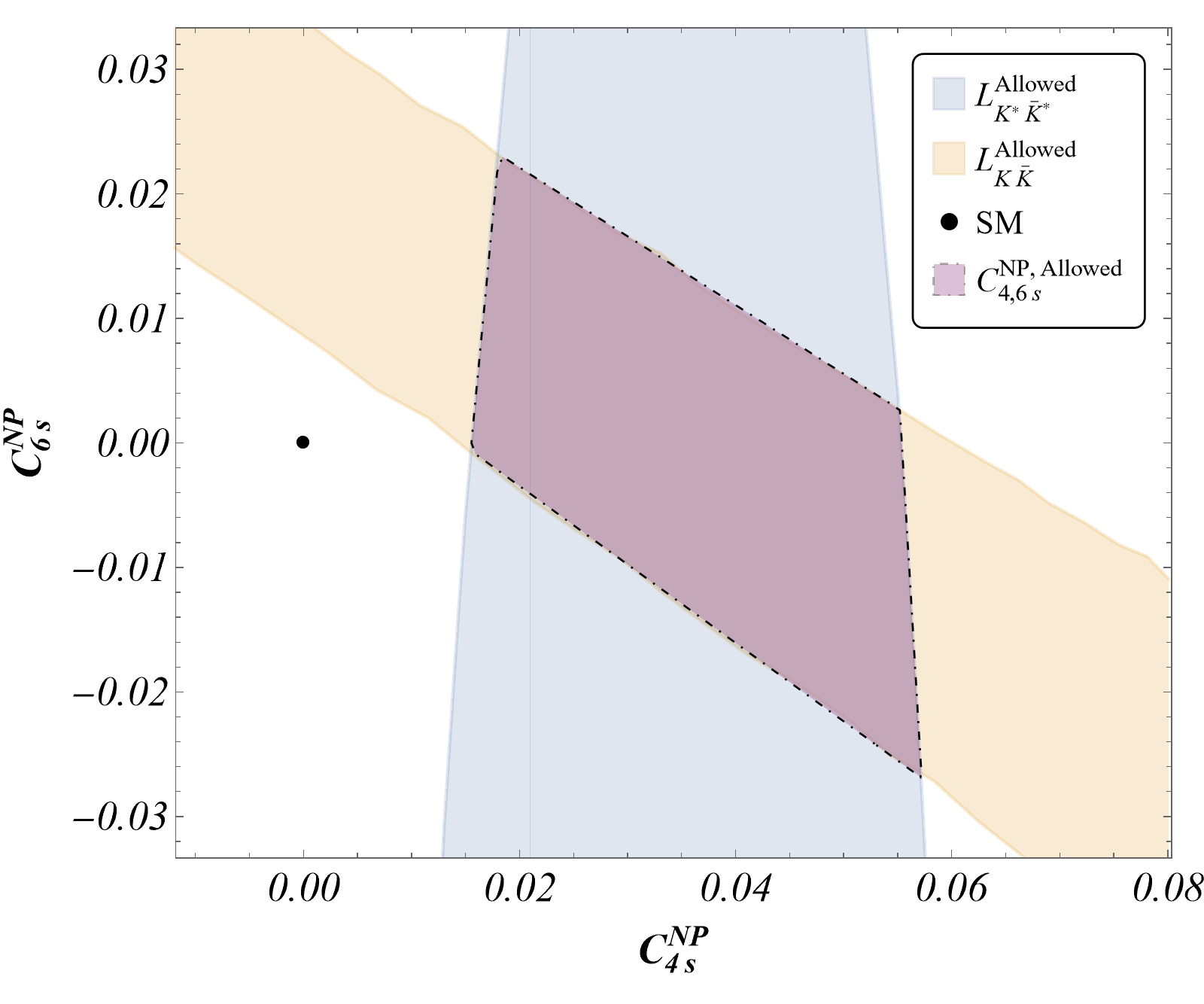}
\caption{ Allowed region for ${\cal C}_{4s}^{\rm NP}$ and ${\cal C}_{6s}^{\rm NP}$ where 
 both $L_{K^*\bar{K}^*}$ and $L_{K\bar{K}}$ are compatible theoretically and experimentally within $1\sigma$.
The SM point is represented by a black dot at ${\cal C}_{4,6s}^{\rm NP} = 0$. See Fig.~\ref{fig:C4_range} for further comments on the assumptions and observables considered. }
\label{fig:C46s_pred2}
\end{figure}

\begin{figure}[ht]
\centering
\includegraphics[width=0.75\textwidth,height=0.5\textwidth]{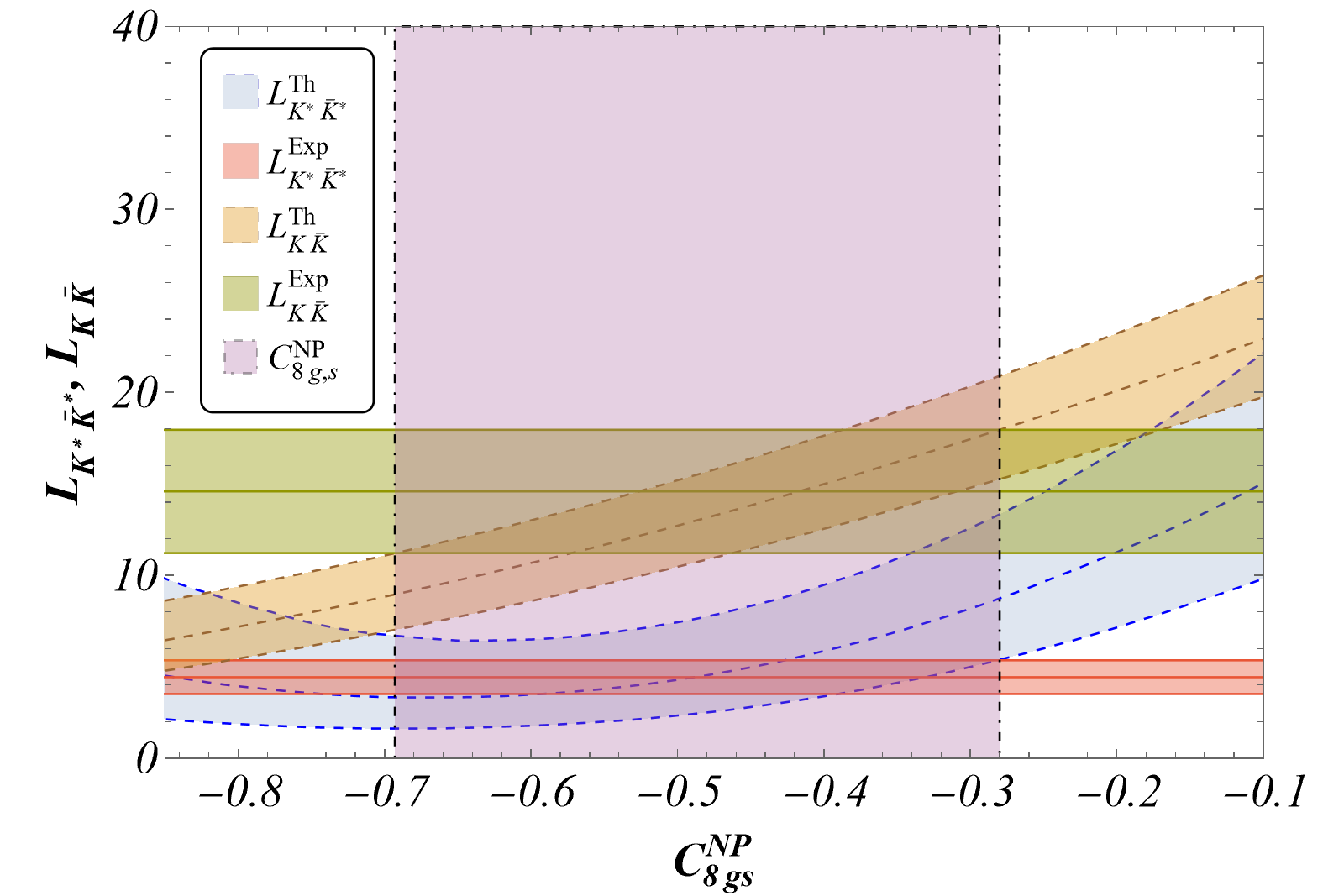}
\caption{Variation of $L_{K^*\bar{K}^*}$ and $L_{K\bar{K}}$ w.r.t ${\cal C}_{8gs}^{\rm NP}$. The range of ${\cal C}_{8gs}^{\rm NP}$ where both observables are compatible theoretically and experimentally within $1\sigma$ is 
 ${\cal C}_{8gs}^{\rm NP}\in [-0.69, -0.28]$
(corresponding to values of 1.8 to 4.5 times the SM value).
Like in Fig.~\ref{fig:C4_range}, notice that this allowed (magenta) region is determined assuming NP in $b\to s$ transitions only and taking $L_{K\bar{K}}$ and $L_{K^*\bar{K}^*}$ as inputs. A more general case allowing also for NP in $b \to d$ transitions
and including constraints from individual branching ratios will be discussed in Sec.~\ref{sec:indivBR}}.
\label{fig:C8_range}
\end{figure}

\section{Combined analysis of the modes}\label{sec:combined}

\subsection{Sensitivity of the optimised observables to NP in $b \to s$ transitions}\label{sec:sensitivity}

%Allowed regions for NP assuming NP only in $b \to s$}

In this section we focus on determining allowed regions considering QCD penguin operators ($Q_{4s}$ or $Q_{6s}$) or the chromomagnetic operator ($Q_{8gs}$), using the two measured $L$ observables ($L_{K^* \bar{K}^*}$ and $L_{K\bar{K}}$) as inputs and assuming that NP enters only $b\to s$ transitions. 

 In Fig.~\ref{fig:C4_range} we explore the region allowed for ${\cal C}_{4s}^{\rm NP}$ taking as inputs the experimental $1\sigma$ estimates for $L_{K^*\bar{K}^*}$ and $L_{K\bar{K}}$. The region accounting for both measured observables within $1\sigma$ is shaded and corresponds to ${\cal C}_{4s}^{\rm NP}\in [0.016, 0.055]$.
Similarly, Fig.~\ref{fig:C6_range} shows the marginal sensitivity to ${\cal C}_{6s}^{\rm NP}$ of $L_{K^*\bar{K^*}}$, so that it appears impossible to find a common region accounting for both observables due to NP contributions only to this Wilson coefficient. 

Fig.~\ref{fig:C46s_pred2} shows the region for ${\cal C}_{4s}^{\rm NP}$ and ${\cal C}_{6s}^{\rm NP}$ allowed by $L_{K^*\bar{K}^*}$ and $L_{K\bar{K}}$. The SM is represented by a black dot at ${\cal C}_{4,6s}^{\rm NP} = 0$. In order to generate the figure, values for ${\cal C}_{4s}^{\rm NP}$ and ${\cal C}_{6s}^{\rm NP}$ were generated within the ranges $[-0.03,0.08]$ and $[-0.05,0.05]$ respectively at an interval of $0.001$. For each pair of values, the $1\sigma$ range for the theoretical prediction of the observables $L_{K^*\bar{K}^*}$ and $L_{K\bar{K}}$ was determined. The contour encompasses the values for which these ranges overlap with the $1\sigma$ experimental ranges. The region of overlap between the contours thus obtained for $L_{K^*\bar{K}^*}$ and $L_{K\bar{K}}$ represent the values of ${\cal C}_{4s}^{\rm NP}$ and ${\cal C}_{6s}^{\rm NP}$ that accommodate the experimental values of both observables taking into account the uncertainties from the theoretical SM parameters (at 1$\sigma$).

The NP contributions considered are smaller than the SM values of these two Wilson coefficients, and compatible with the current bounds (see Ref.~\cite{Descotes-Genon:2020tnz} for further discussion). Interestingly, Fig.~\ref{fig:C46s_pred2} shows that current data is compatible with NP contributions to the Wilson coefficients of $Q_{4s}$ and $Q_{6s}$ of same sign and size, i.e,  ${\cal C}_{4s}^{\rm NP}={\cal C}_{6s}^{\rm NP}\sim 0.02$. This would correspond to an NP scenario where the axial contribution of the $q_i\bar{q_j}$ cancels, i.e. NP would occur through a 4-quark vector operator with a structure  similar to the semileptonic vector operator $O_9$ (changing leptons by quarks) involved in $b\to s\mu\mu$ transitions.

Fig.~\ref{fig:C8_range} shows the allowed region for NP contribution to the Wilson coefficient 
${\cal C}_{8gs}$ based on $L_{{K^*}\bar{K}^{*}}$ and $L_{{K}\bar{K}}$. In this case there is also an allowed region ${\cal C}_{8gs}^{\rm NP}\in [-0.69, -0.28]$, corresponding to NP contributions larger than the SM value, but still compatible (in part of the allowed region) with the current bounds (see Ref.~\cite{Descotes-Genon:2020tnz} for further discussion).

\begin{figure}[ht]
\centering
\includegraphics[width=0.75\textwidth,height=0.5\textwidth]{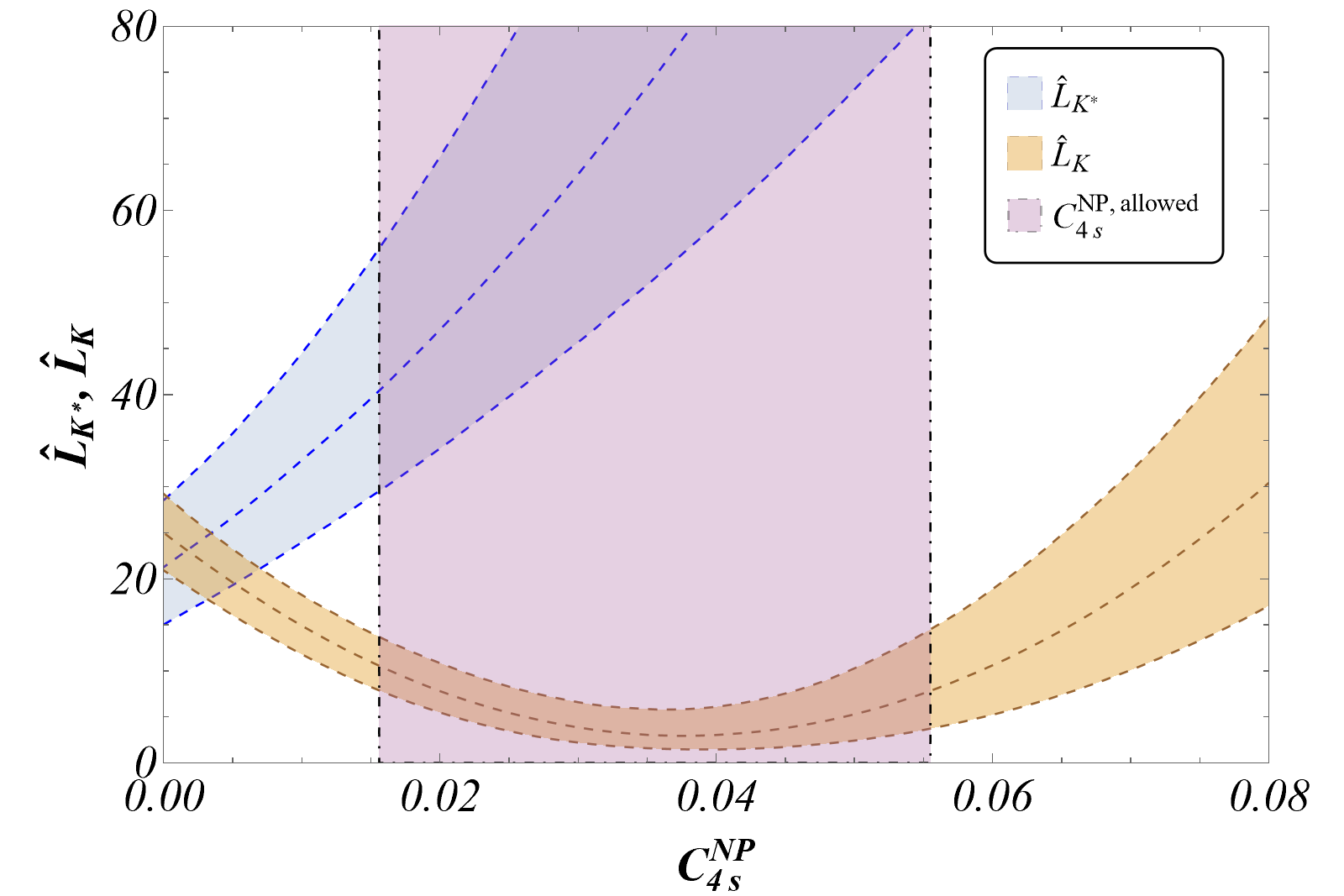}
\caption{Scenario 1:
Variation of $\hat{L}_{K^*}$ and $\hat{L}_K$ w.r.t ${\cal C}_{4s}^{\rm NP}$ and their behaviour in the range for ${\cal C}_{4s}^{\rm NP}\in [0.016, 0.055]$ from Fig.~\ref{fig:C4_range}. }
\label{fig:C4_pred}
\end{figure}
\begin{figure}[ht]
\centering
\includegraphics[width=0.75\textwidth,height=0.5\textwidth]{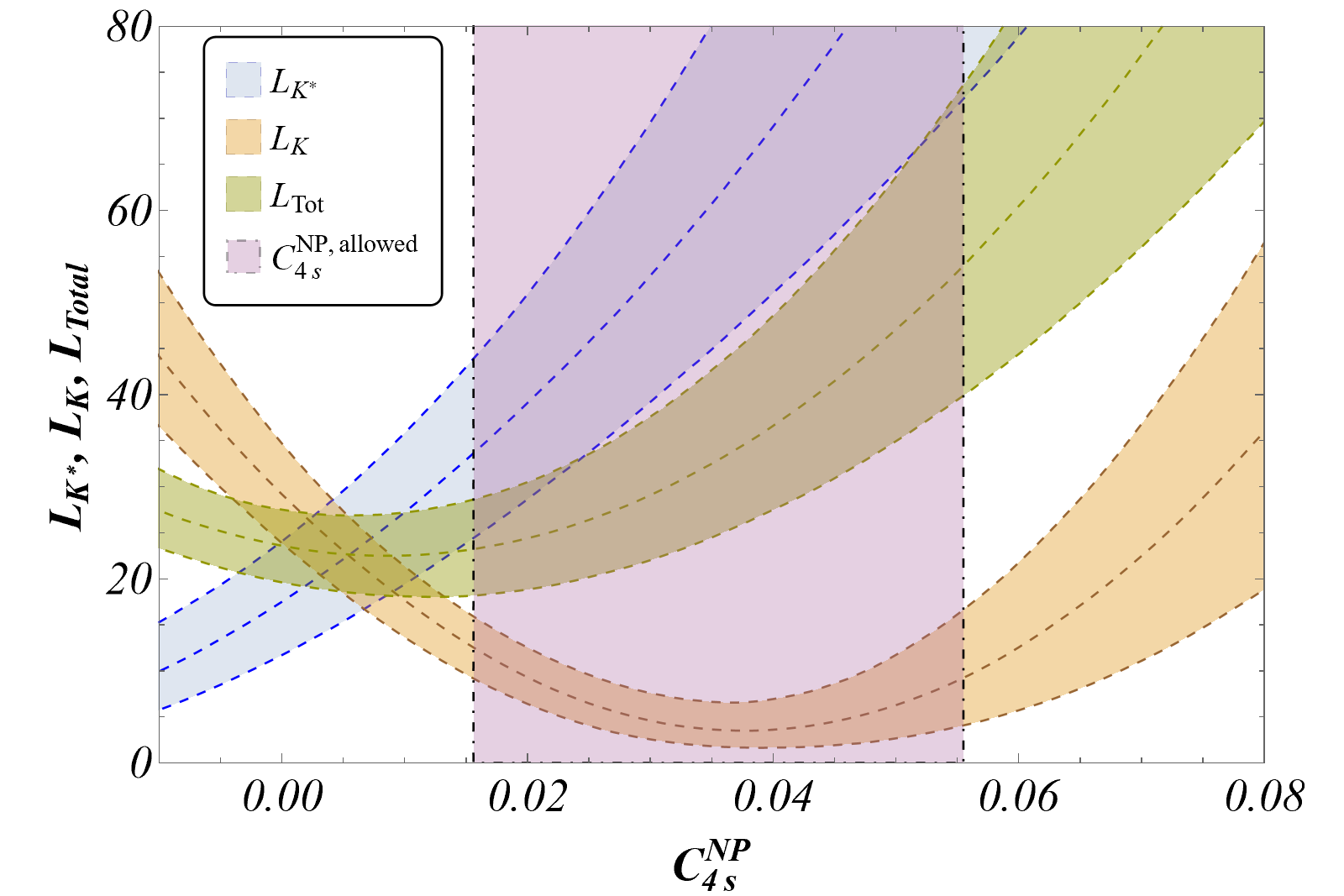}
\caption{Scenario 1:
Variation of ${L}_{K^*}$, ${L}_K$  and $L_{\rm total}$ w.r.t ${\cal C}_{4s}^{\rm NP}$ and their behaviour in the range for ${\cal C}_{4s}^{\rm NP}\in [0.016, 0.055]$ from Fig.~\ref{fig:C4_range}. }
\label{fig:C4_pred_Ltot}
\end{figure}
\begin{figure}[ht]
\centering
\includegraphics[width=0.75\textwidth,height=0.5\textwidth]{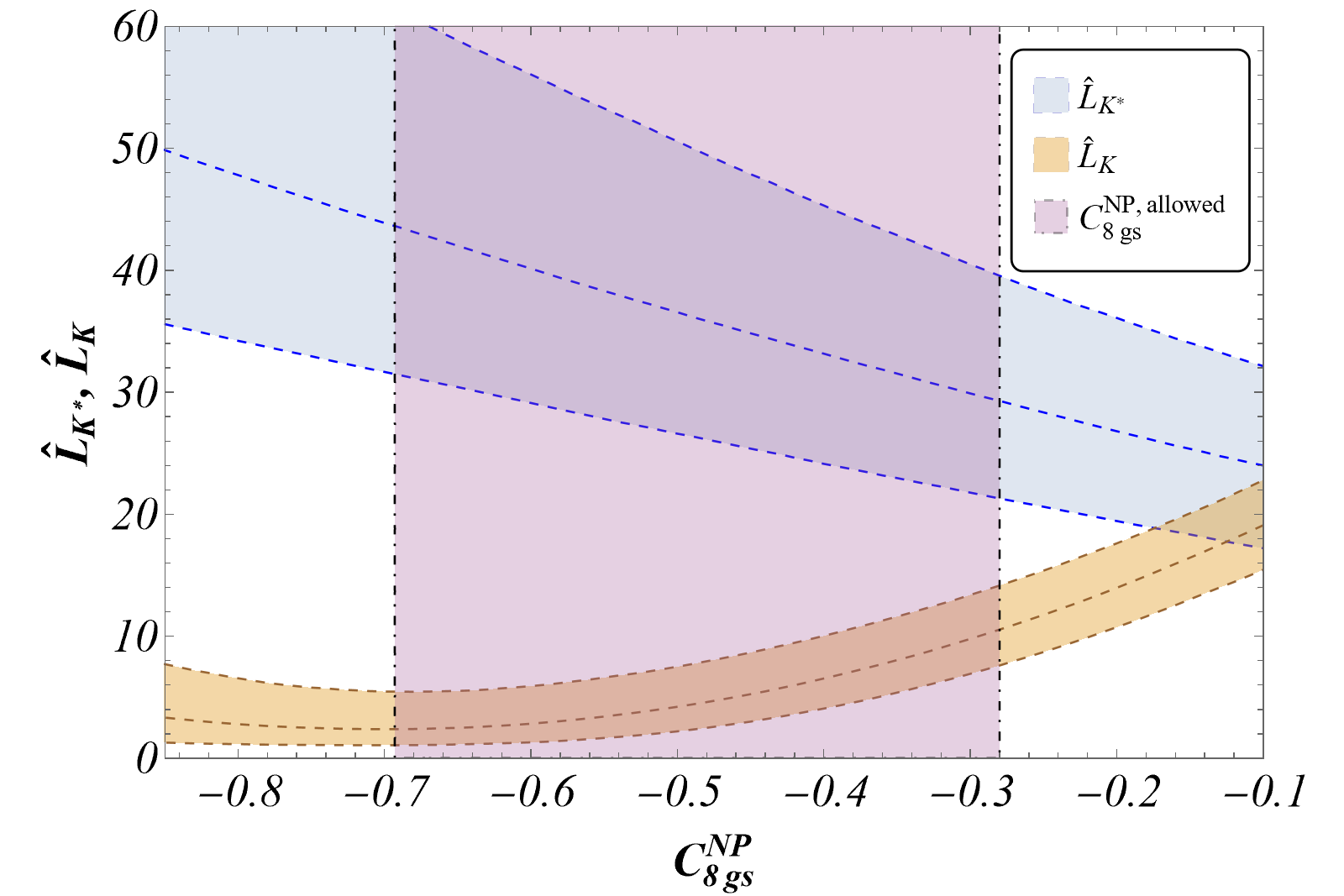}
\caption{Scenario 2: Variation of $\hat{L}_{K^*}$ and $\hat{L}_K$ w.r.t ${\cal C}_{8gs}^{\rm NP}$ and their behaviour in the range for ${\cal C}_{8gs}^{\rm NP}\in [-0.69, -0.28]$ from Fig.~\ref{fig:C8_range}. }
\label{fig:C8_pred}
\end{figure}
\begin{figure}[ht]
\centering
\includegraphics[width=0.75\textwidth,height=0.5\textwidth]{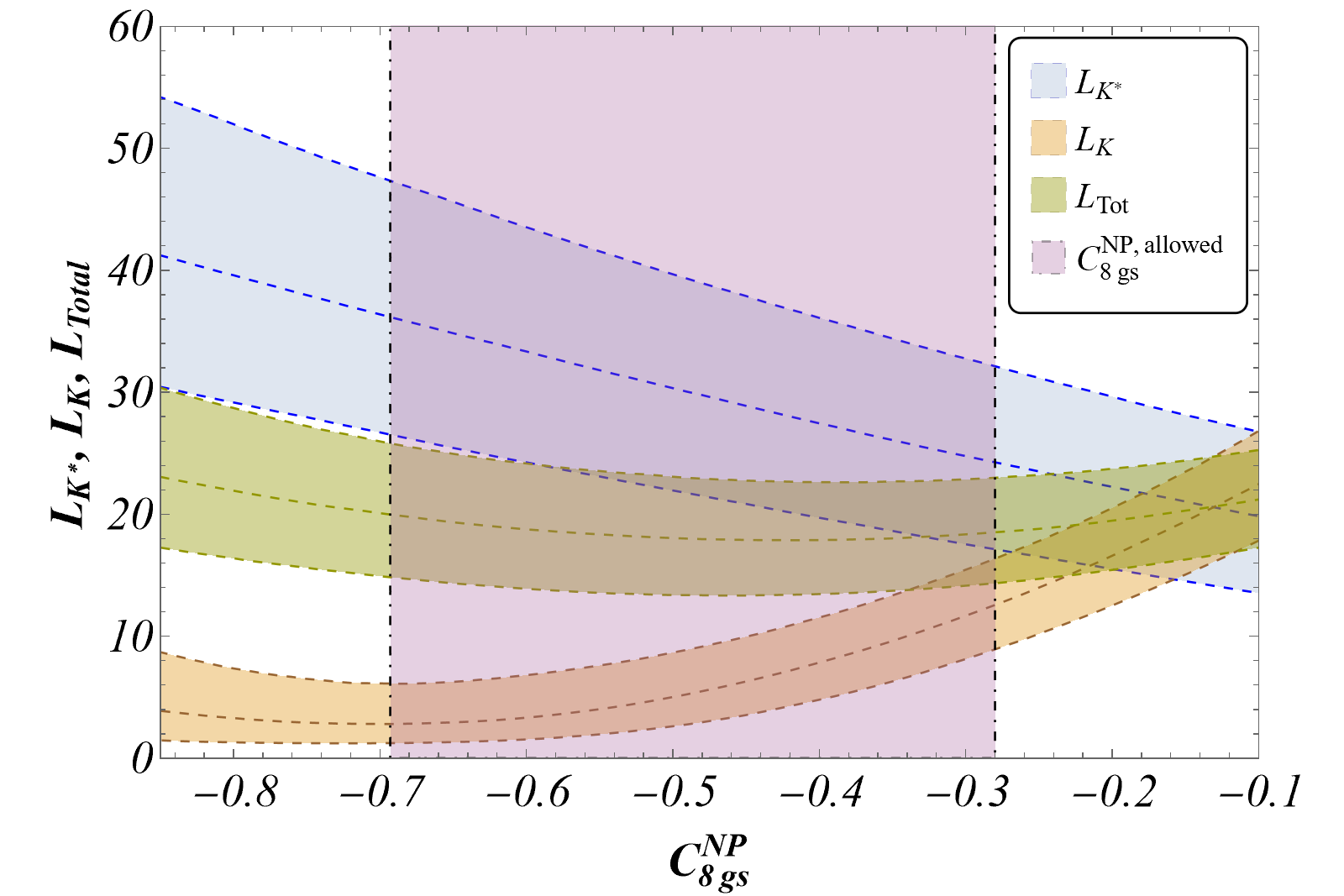}
\caption{Scenario 2: Variation of ${L}_{K^*}$, ${L}_K$  and $L_{\rm total}$ w.r.t ${\cal C}_{8gs}^{\rm NP}$ and their behaviour in the range for ${\cal C}_{8gs}^{\rm NP}\in [-0.69, -0.28]$ from Fig.~\ref{fig:C8_range}.}
\label{fig:C8_pred_Ltot}
\end{figure}
\begin{figure}[ht]
\centering
\includegraphics[width=0.75\textwidth,height=0.5\textwidth]{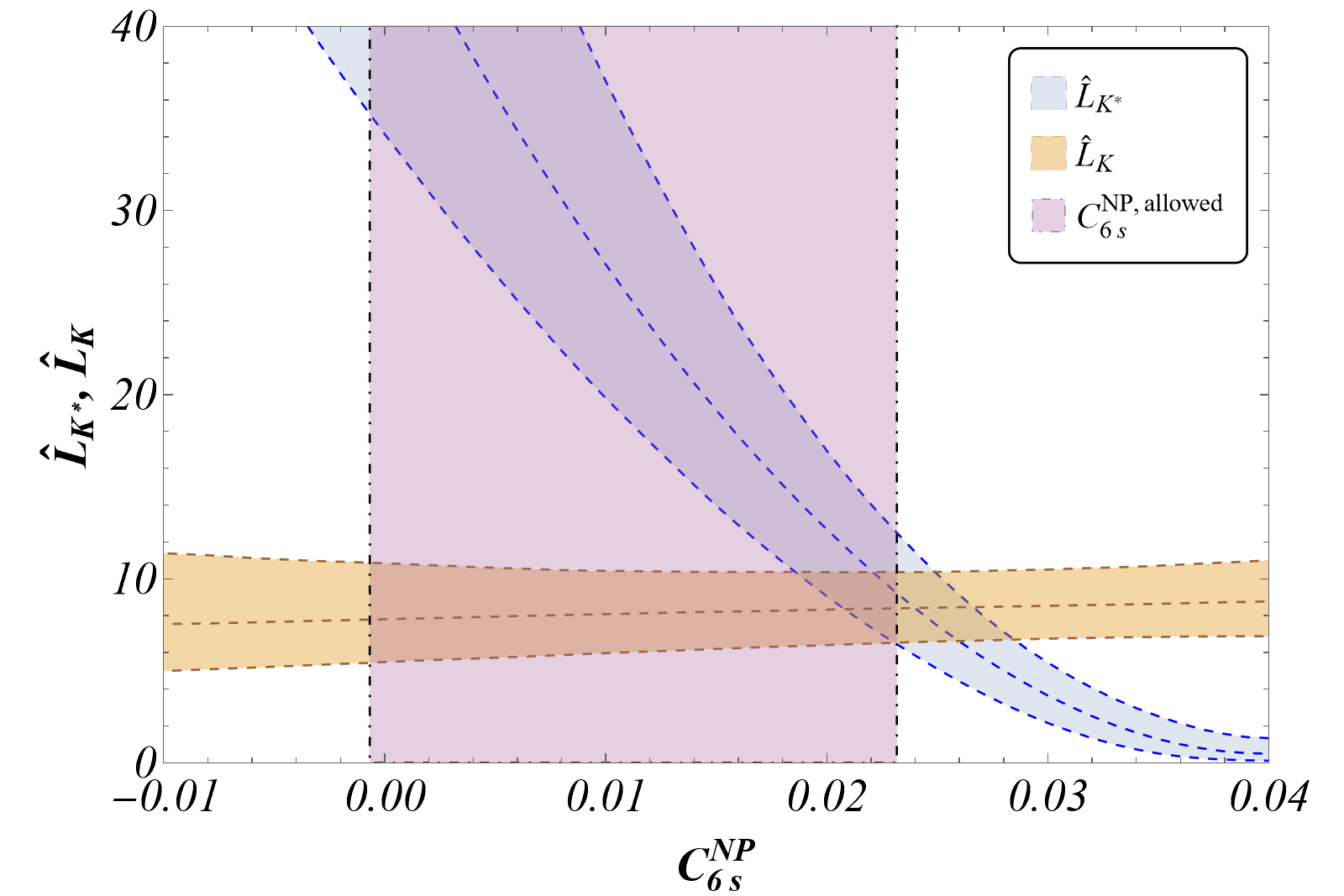}
\caption{Scenario 3: Variation of $\hat{L}_{K^*}$ and $\hat{L}_K$ w.r.t ${\cal C}_{6s}^{\rm NP}$, assuming ${\cal C}_{4s}^{\rm NP} = 0.02$ (the case of ${\cal C}_{4s}^{\rm NP} = 0$ 
 is not displayed because it does not yield any common solution to both observables). The range for ${\cal C}_{6s}^{\rm NP}$ is taken in accordance to the y-axis of Fig.~\ref{fig:C46s_pred2} such that a part of this figure corresponds to the allowed region from Fig.~\ref{fig:C46s_pred2}.} 
\label{fig:C6_pred}
\end{figure}
\begin{figure}[ht]
\centering
\includegraphics[width=0.75\textwidth,height=0.5\textwidth]{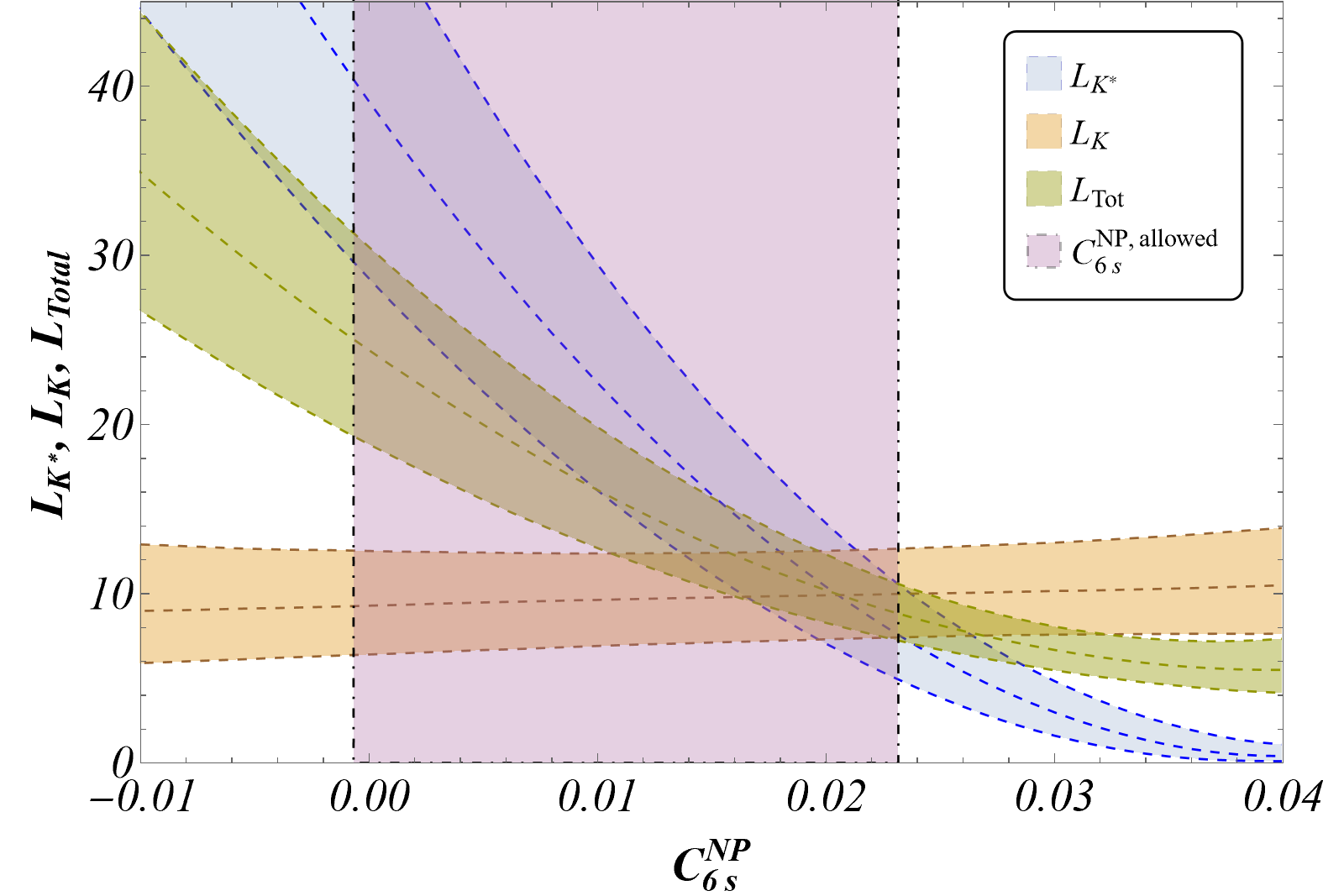}
\caption{Scenario 3: Variation of ${L}_{K^*}$, ${L}_K$  and $L_{\rm total}$ w.r.t ${\cal C}_{6s}^{\rm NP}$, assuming
 ${\cal C}_{4s}^{\rm NP} = 0.02$. The ranges are identical to Fig.~\ref{fig:C6_pred}. }
\label{fig:C6_pred_Ltot}
\end{figure}

\begin{figure}
\includegraphics[width=1.0\textwidth,height=0.8\textwidth]{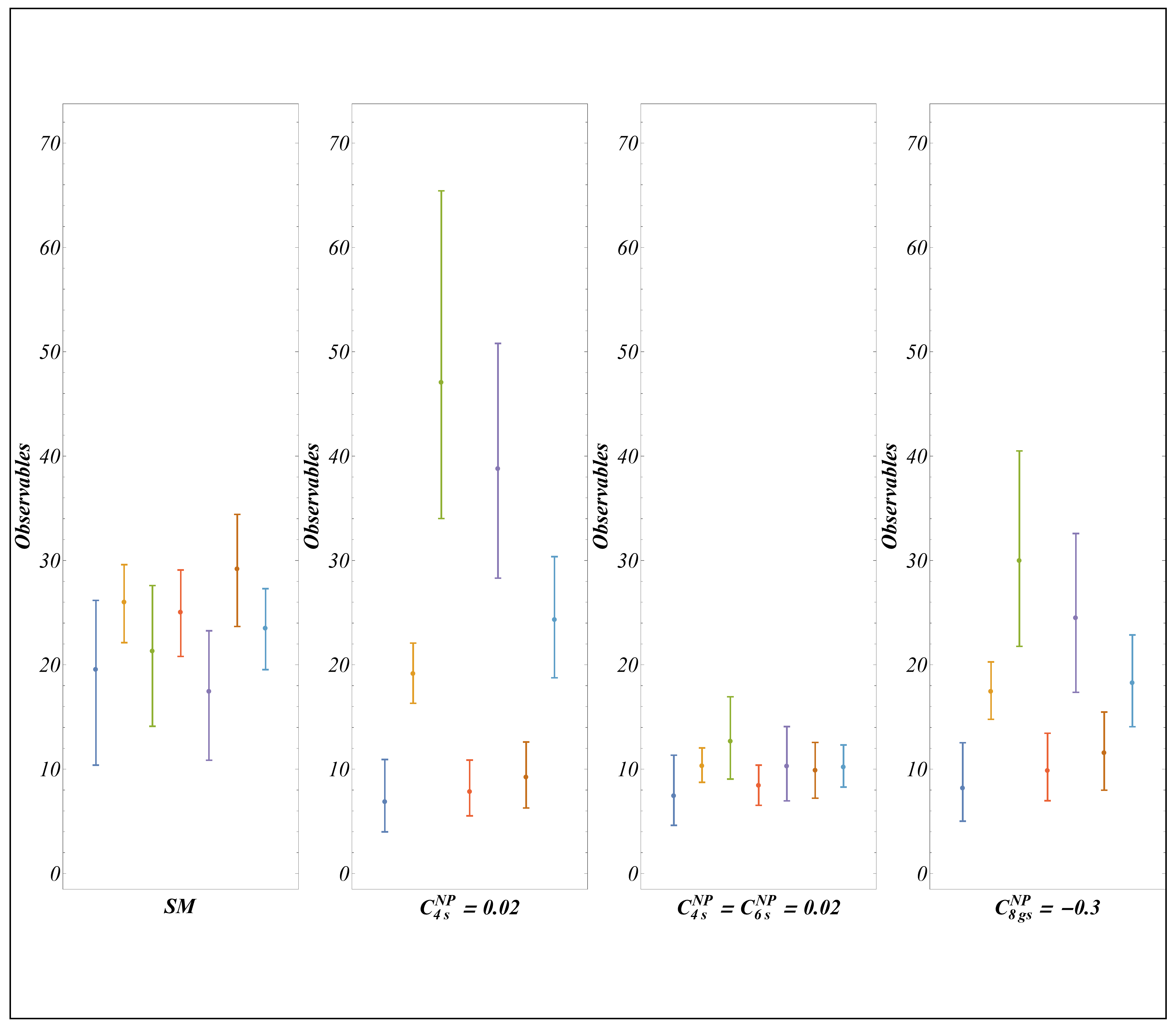}\qquad\\
\centering
\includegraphics[width=0.5\textwidth,height=0.15\textwidth]{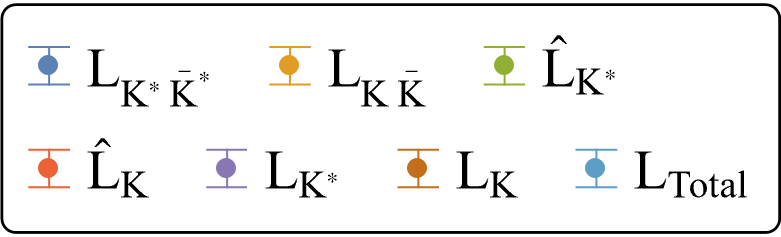}
\caption{Predictions within the SM and different scenarios at specific NP points illustrating the patterns to be expected in each case, assuming NP enters only $b\to s$ transitions}.
\label{fig:dev-patternsI}
\end{figure}

\begin{figure}
\includegraphics[width=1.0\textwidth,height=0.8\textwidth]{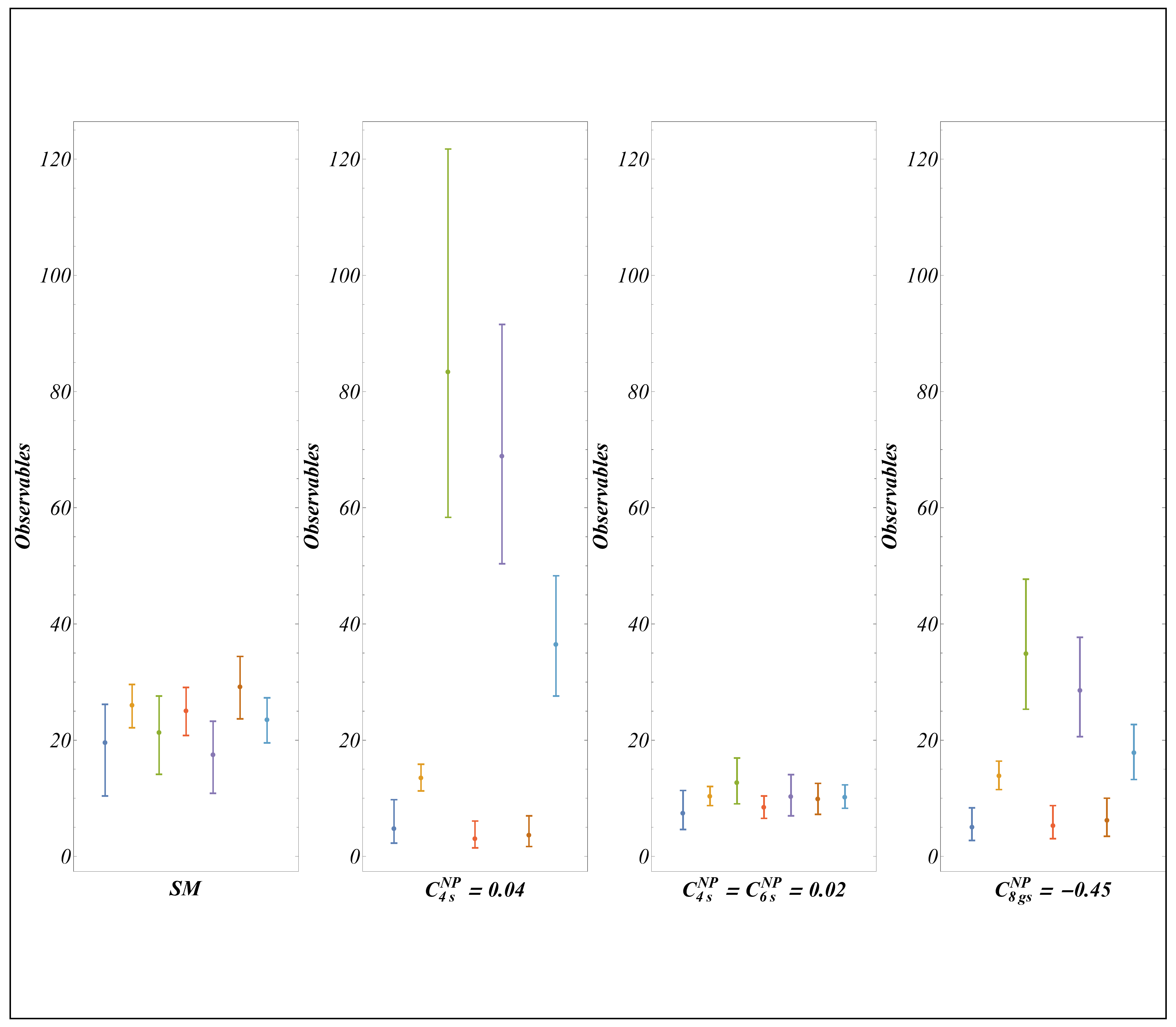}\qquad\\
\centering
\caption{Predictions within the SM and different scenarios at specific NP points illustrating the patterns to be expected in each case. 
}
\label{fig:dev-patternsII}
\end{figure}

\subsection{Possible patterns of deviations}\label{sec:pattern}

One may notice that the SM predictions of all observables fall in a rather narrow range. We will explore in this section how they split and define a pattern of deviations in presence of NP.
In the following,``NP scenario" means a particular hypothesis of NP contributions to specific short-distance Wilson coefficients, whereas ``pattern of deviations of observables" corresponds to the specific and correlated deviations exhibited by a set of related observables in the presence of NP. The (experimental) observation of a given pattern may inform us on the underlying NP scenario (theoretically defined).
 
Assuming that an NP contribution to only one Wilson coefficient is responsible for the tension in $L_{K^*\bar{K^*}}$ and $L_{K\bar{K}}$, we may explore how $\hat{L}_K$, $\hat{L}_{K^*}$, ${L}_K$, ${L}_{K^*}$ and $L_{\rm total}$ are affected according to three different scenarios of NP:

\begin{itemize}
\item Scenario 1 (${\cal C}_{4s}^{\rm NP}$):
In Fig.~\ref{fig:C4_pred}, we observe that the range of values explaining both $\hat{L}_{K\bar{K^*}}$ and $\hat{L}_{K\bar{K}}$ yields a very noticeable splitting between $\hat{L}_K$ and $\hat{L}_{K^*}$, whereas their values are very close in the SM. For instance, for the value ${\cal C}_{4s}^{\rm NP}\sim 0.036$ reproducing the central value of both $\hat{L}_{K\bar{K^*}}$ and $\hat{L}_{K\bar{K}}$ approximately, there is a factor 30 of discrepancy between
$\hat{L}_K$ and $\hat{L}_{K^*}$. Fig.~\ref{fig:C4_pred_Ltot} shows the behaviour of the observables ${L}_K$, ${L}_{K^*}$ and $L_{\rm total}$ under this scenario.

\item Scenario 2 (${\cal C}_{8gs}^{\rm NP}$): 
In Fig.~\ref{fig:C8_pred}, the splitting between $\hat{L}_K$ and $\hat{L}_{K^*}$ 
can be observed, but it is less significant than in the previous scenario. 
A clear-cut separation among the observables requires huge NP contributions (up to the limit of 300\% of the SM size). For instance, for the value ${\cal C}_{8gs}^{\rm NP} \sim -0.45$ reproducing approximately the central value of both $L_{K\bar{K^*}}$ and $L_{K\bar{K}}$, there is at most  a factor 7 of discrepancy between the observables $\hat{L}_K$ and $\hat{L}_{K^*}$. Fig.~\ref{fig:C8_pred_Ltot} shows the behaviour of the observables ${L}_K$, ${L}_{K^*}$ and $L_{\rm total}$ under this scenario.

\item Scenario 3 (${\cal C}_{6s}^{\rm NP}$ with a non-vanishing contribution ${\cal C}_{4s}^{\rm NP}=0.02$): 
In Fig.\ref{fig:C6_pred}, we can see the rather different outcome of this scenario. $\hat{L}_K$ is rather insensitive to ${\cal C}_{6s}^{\rm NP}$ and remains SM-like, while $\hat{L}_{K^*}$ exhibits an important suppression at the edge of the allowed region ${\cal C}_{6s}^{\rm NP}\sim 0.02$. 
Fig.~\ref{fig:C6_pred_Ltot} shows the similar behaviour of the observables ${L}_K$, ${L}_{K^*}$ and $L_{\rm total}$ under this scenario.
\end{itemize}

These NP scenarios yield different patterns of deviations assuming that ${\cal C}_{4s}^{\rm NP}$ and ${\cal C}_{6s}^{\rm NP}$ may reach a size similar to the SM values and ${\cal C}_{8g,s}^{\rm NP}$ can get up to three times its SM value. 
Measuring the observables and noticing their deviations provides direct information on the possible NP mechanisms responsible for the pattern observed. Moreover, these patterns provide useful guidelines 
to identify observables where clear signals of NP may be looked for experimentally.
Figs.~\ref{fig:dev-patternsI} and \ref{fig:dev-patternsII} provide an illustration of the patterns of deviations, showing the predicted values of 
the observables within the SM (on the far left) and according to different NP benchmark points of  NP scenarios. 
  We consider either modest NP contributions (Fig.~\ref{fig:dev-patternsI}) or larger values (Fig.~\ref{fig:dev-patternsII}).

\begin{itemize}

\item SM Pattern: All observables have their central values in a fairly narrow range between 17 to 30.

\item Pattern 1:
An NP contribution to ${\cal C}_{4s}$
of half its SM value induces a very specific pattern:
a huge deviation in both $\hat{L}_{K^*}$ (up to twice the SM prediction)  and $\hat{L}_K$ (half the SM prediction), enhancing the difference between these two observables. A similar pattern occurs for $L_{K^*}$ and $L_K$, albeit slightly less pronounced. On the other hand,
 $L_{\rm total}$ remains basically unchanged and both $L_{K^*\bar{K}^*}$ and $L_{K\bar{K}}$ get suppressed w.r.t. to their SM prediction (down to half the SM prediction for $L_{K^*\bar{K}^*}$, and a more moderate suppression for $L_{K\bar{K}}$). For ${\cal C}_{4s}^{\rm NP}$ of same size as the SM value, the pattern of deviation is even more enhanced, with a splitting between $\hat{L}_K$ and $\hat{L}_{K^*}$  up to an order of magnitude of difference. The variations must be significant as the current data requires non-vanishing
  minimal values of ${\cal C}_{4s}^{\rm NP}$ 
 (close to 0.02 if one assumes NP affecting only $b\to s$ transitions) and cannot be accommodated by smaller values of these coefficients.
 \bigskip

\item Pattern 2: The point ${\cal C}_{4s}^{\rm NP}={\cal C}_{6s}^{\rm NP}=0.02$ yields a pattern where the (central values of the) $L$ observables get all suppressed down to values around 10. This NP point corresponds to a cancellation of the axial contribution between the two operators, i.e. a 4-quark vector operator with a structure  similar to the semileptonic vector operator $O_9$ involved in $b\to s\mu\mu$ transitions.

\item Pattern 3: The pattern in the case of a large or very large NP contribution to ${\cal C}_{8gs}$ is very similar to the case of an NP contribution to ${\cal C}_{4s}$, though with milder differences among the $L$ observables.
In other words, a large dispersion would point to NP in ${\cal C}_{4s}$, whereas a moderate dispersion would rather correspond to ${\cal C}_{8gs}$. Indeed, reproducing deviations of the same size as in pattern 1 is difficult here, as it would require extreme values of ${\cal C}_{8gs}^{\rm NP}$  (of order 300\% of its SM value). Moreover, ${L}_K$ and $L_{\rm total}$ are compatible at 1 $\sigma$ or less for ${\cal C}_{8gs}^{\rm NP}$ (for values of ${\cal C}_{8gs}^{\rm NP}\sim -0.3$),
and separated at more than 2 $\sigma$ for ${\cal C}_{4s}^{\rm NP}$ (for values of ${\cal C}_{4s}^{\rm NP}\sim 0.02$).
\end{itemize}

\begin{table}[t]
\begin{center}
\tabcolsep=1.3cm\begin{tabular}{|c|c|}
\hline\multicolumn{2}{|c|}{
$\mathcal{B}(\bar{B}_d\rightarrow K^0 \bar{K}^0 )$ $[10^{-6}]$}  \\ \hline
SM (QCDF)& Experiment \\
\hline
$1.09^{+0.29}_{-0.20}$&$1.21\pm 0.16$ \cite{Workman:2022ynf,Belle:2012dmz,BaBar:2006enb}\\
\hline
\end{tabular}
\vskip 1pt
\tabcolsep=1.35cm\begin{tabular}{|c|c|}
\hline\multicolumn{2}{|c|}{
$\mathcal{B}(\bar{B}_s\rightarrow K^0 \bar{K}^0  )$ $[10^{-5}]$}  \\ \hline
SM (QCDF)& Experiment \\
\hline
$2.80^{+0.89}_{-0.62}$&$1.76\pm 0.33$~\cite{Workman:2022ynf, LHCb:2020wrt,Belle:2015gho}\\
\hline
\end{tabular}
\end{center}
\caption{Branching ratios for pseudoscalar-pseudoscalar final states. A 7\% relative uncertainty is added in quadrature for the $B_s$ decay due to $B_s$-mixing (see Secs.~\ref{sec:theory-PP} and \ref{sec:LKK}).}
\label{tab:BrPP}
\end{table}

\begin{table}[t]
\begin{center}

\tabcolsep=2.1cm\begin{tabular}{|c|c|}
\hline\multicolumn{2}{|c|}{Measured longitudinal polarisation fractions}  \\ 
\hline
$f_L(\bar{B}_d\rightarrow K^{0*}\bar{K}^{0*})$&
$f_L(\bar{B}_s\rightarrow  K^{0*}\bar{K}^{0*})$\\
\hline
$0.73\pm0.05$\cite{Alguero:2020xca,Aaij:2019loz,Aubert:2007xc}&$0.240\pm 0.040$ \cite{LHCb:2015exo}\\
\hline
\end{tabular}

\vskip 1pt

\tabcolsep=1.67cm\begin{tabular}{|c|c|}
\hline\multicolumn{2}{|c|}{Measured branching ratios (all polarisations included)}  \\ 
\hline
$\mathcal{B}(\bar{B}_d\rightarrow K^{*0} \bar{K}^{*0})_{\rm all \ pol}$&
$\mathcal{B}(\bar{B}_s\rightarrow  K^{*0}\bar{K}^{*0} )_{\rm all\ pol}$\\
\hline
$(0.83\pm0.24)\times 10^{-6}$\cite{Workman:2022ynf}&$(1.11\pm 0.28)\times 10^{-5}$ \cite{Workman:2022ynf}\\
\hline
\end{tabular}

\vskip 1pt

\tabcolsep=2.33cm\begin{tabular}{|c|c|}
\hline\multicolumn{2}{|c|}{Longitudinal
$\mathcal{B}(\bar{B}_d\rightarrow K^{*0} \bar{K}^{*0})$ $[10^{-7}]$ }  \\ 
\hline
SM (QCDF)& Experiment\\
\hline
$2.27^{+0.98}_{-0.74}$&$6.04^{+1.81}_{-1.78}$\\
\hline
\end{tabular}

\vskip 1pt

\tabcolsep=2.33cm\begin{tabular}{|c|c|}
\hline\multicolumn{2}{|c|}{Longitudinal
$\mathcal{B}(\bar{B}_s\rightarrow K^{*0}\bar{K}^{*0} )$ $[10^{-6}]$ }  \\ 
\hline
SM (QCDF)& Experiment\\
\hline
$4.36^{+2.23}_{-1.65}$&$2.62^{+0.85}_{-0.75}$\\
\hline
\end{tabular}
\caption{Branching ratios for Vector-Vector final states. A 7\% relative uncertainty is added in quadrature due to $B_s$-mixing for the $B_s$ decay (see Secs.~\ref{sec:theory-VV} and \ref{sec:LKstKst}).
}
\label{tab:BrVV}
\end{center}
\end{table}

\begin{figure}[t]
\centering
\includegraphics[width=0.8\textwidth]{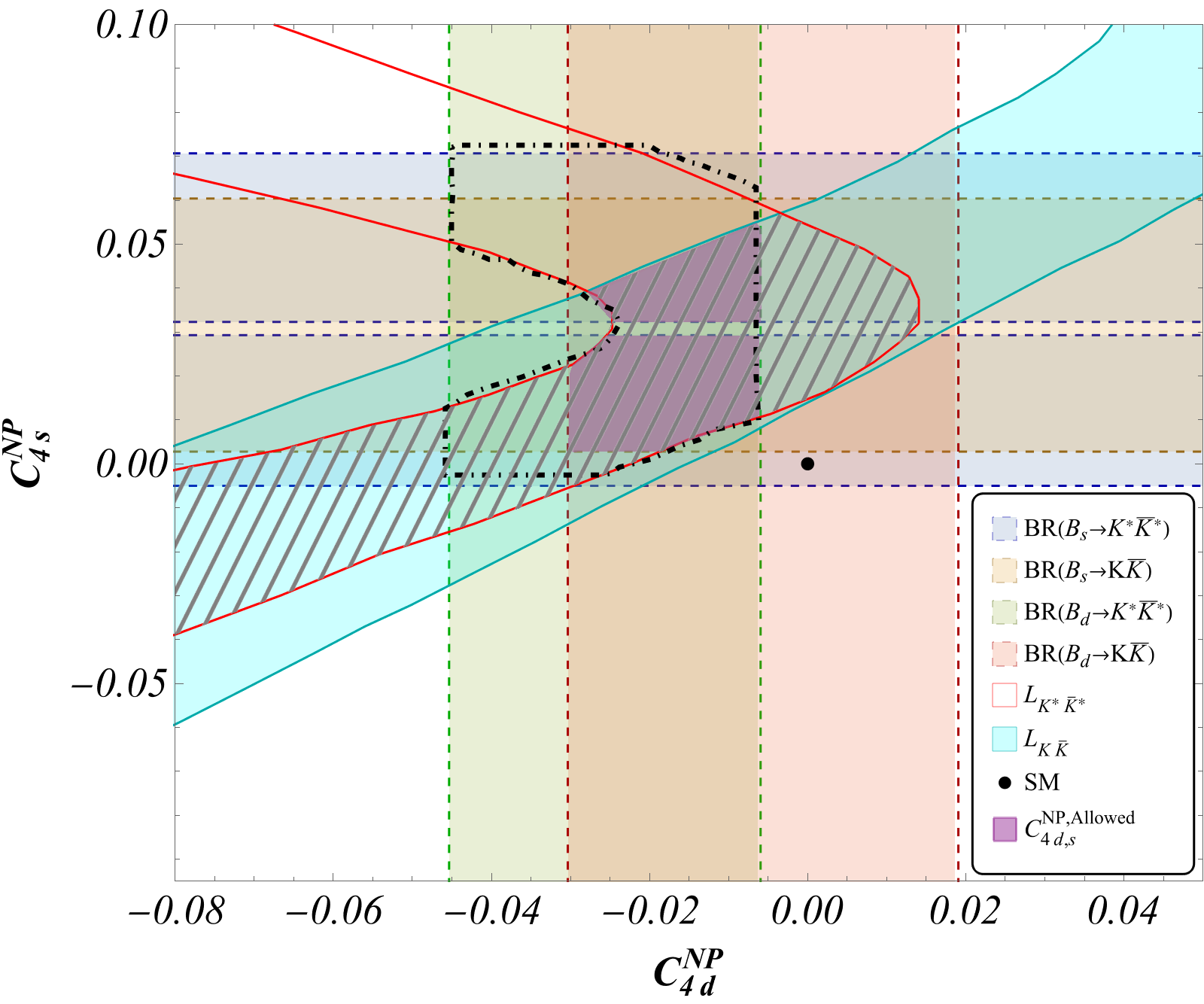}
\caption{
Allowed region for ${\cal C}_{4d}^{\rm NP}$-${\cal C}_{4s}^{\rm NP}$ accommodating the constraints from the measured $L$ observables and individual branching ratios, fixing ${\cal C}_{6d,6s}^{\rm NP}=0$ (magenta region) and letting ${\cal C}_{6d,6s}$ float freely (enlarged region delimited by a black dot-dashed line). We used the full expression of the $L$ observables. Notice that the enlarged region does not expand closer to the SM point. The hatched region represents the values allowed by the two measured $L$ observables only.}
\label{fig:BR1}
\end{figure}

\begin{figure}[t]
\centering
\includegraphics[width=0.8\textwidth]{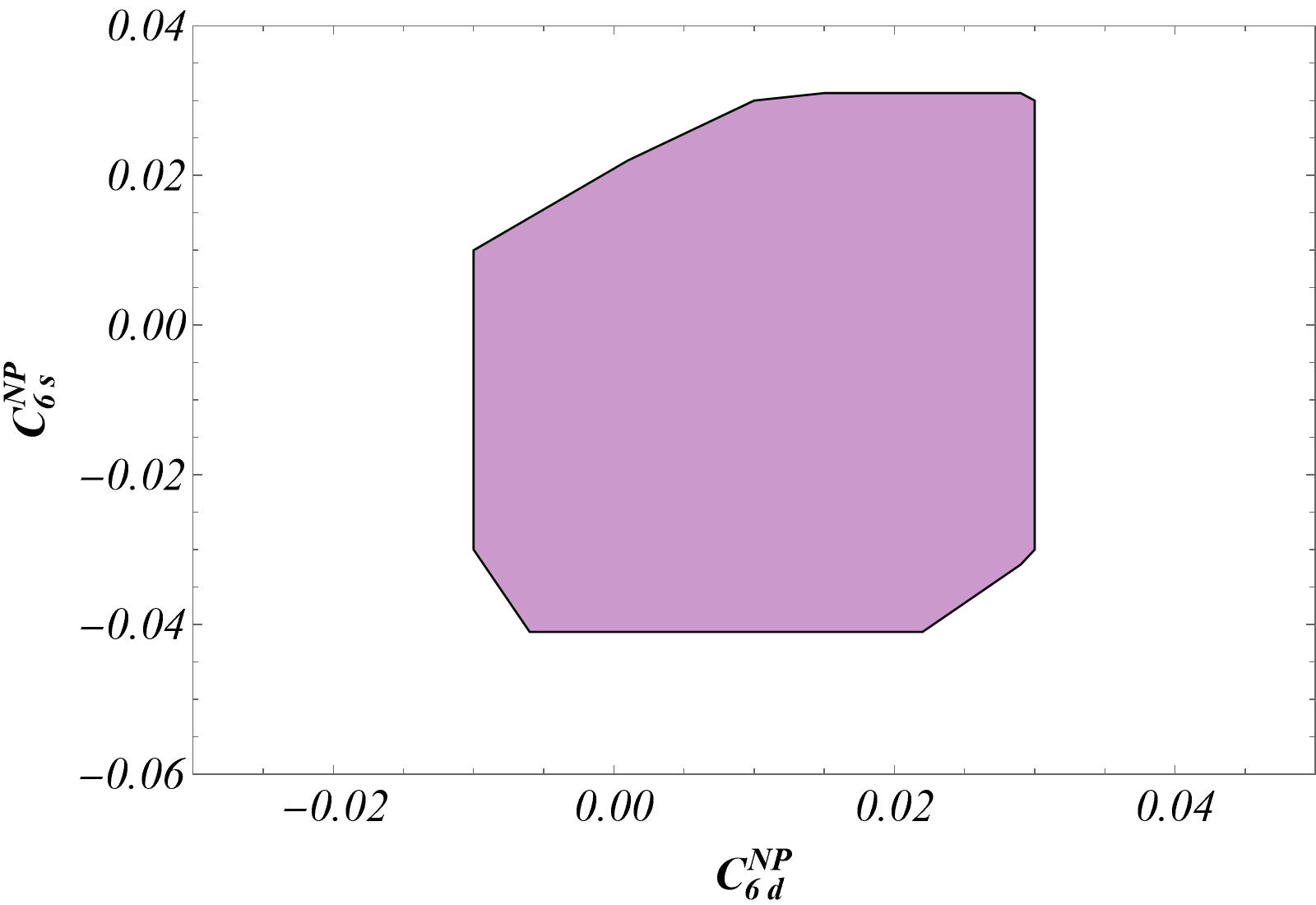}
\caption{
Allowed region for ${\cal C}_{6d}^{\rm NP}$-${\cal C}_{6s}^{\rm NP}$ 
accommodating the constraints from the measured $L$ observables  and individual branching ratios, letting ${\cal C}_{4d,4s}^{\rm NP}$ float freely. We used the full expression of the $L$ observables. Notice from Fig.~\ref{fig:BR1} that the point ${\cal C}_{4d,4s}^{\rm NP}=0$ is not included.}
\label{fig:BR2}
\end{figure}

\begin{figure}[t]
\centering
\includegraphics[width=0.8\textwidth]{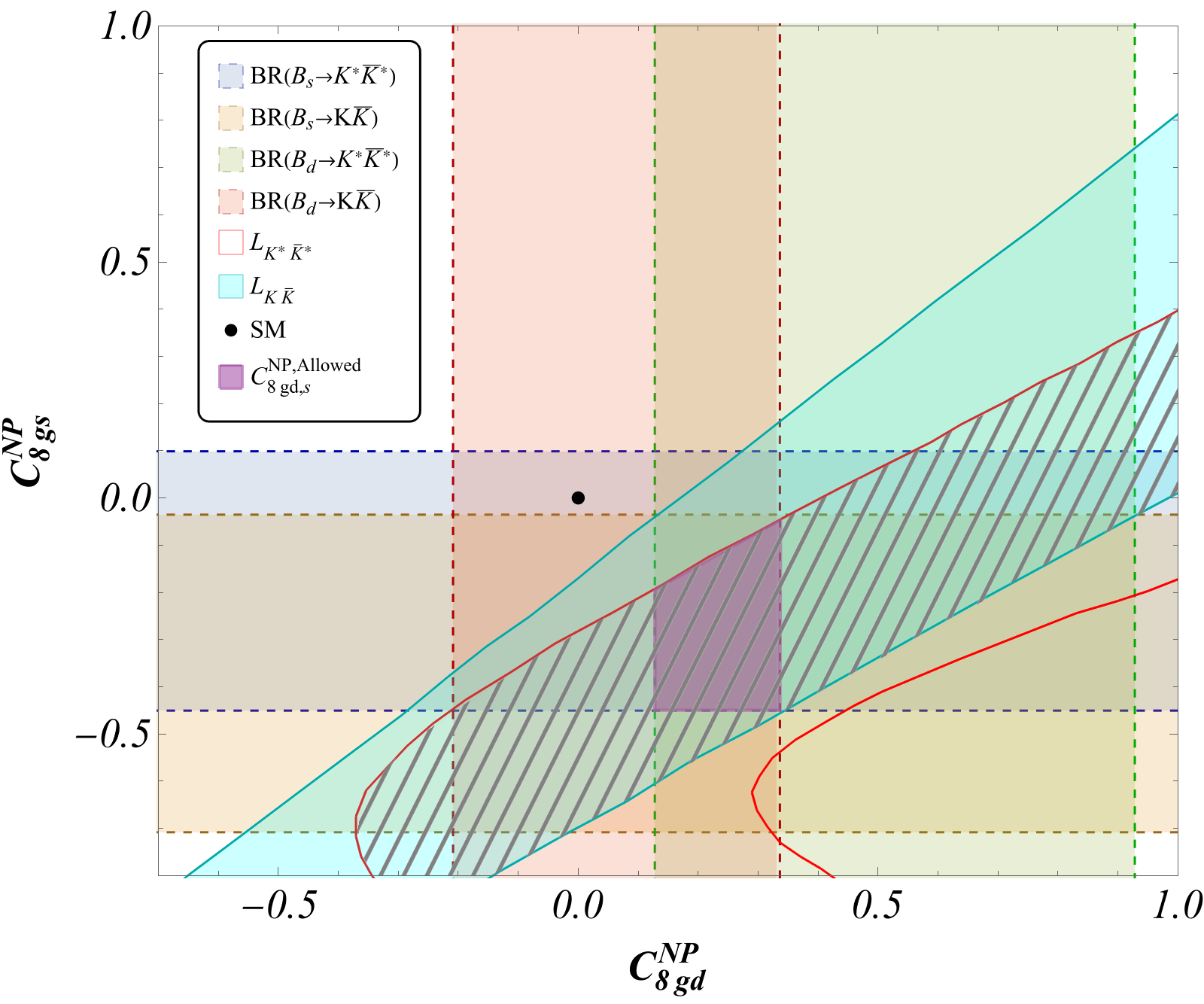}
\caption{
Allowed region for NP contributions to ${\cal C}_{8gd}^{\rm NP}$-${\cal C}_{8gs}^{\rm NP}$ accommodating the constraints from the measured $L$ observables and individual branching ratios (magenta region). We used the full expression of the $L$ observables. The hatched region represents the values allowed by the two measured $L$ observables only.}
\label{fig:BR3}
\end{figure}

\section{Additional constraints from individual branching ratios} \label{sec:indivBR}

Up to now, we have considered ratios of branching ratios as these quantities are more protected from large hadronic uncertainties than individual branching ratios. It is however interesting to consider also the individual branching ratios to see if additional trends in NP can be gathered from these observables, keeping in mind that they are less robust from the theoretical point of view.

\subsection{Hadronic uncertainties and SM predictions}
\label{sec:hadronic}

We have built the optimised $L$ observables from ratios of branching ratios that have a limited sensitivity to hadronic uncertainties, based on $SU(3)$-flavour symmetry and confirmed by QCDF. However, when we observe deviations in some of these observables ($L_{K\bar{K}}$ and $L_{K^*\bar{K}^*}$), one can in principle attribute the deviations to either (or both) branching ratios. It is thus important to discuss also the individual branching ratios in this context, even though their theoretical estimate is far more challenging than the $L$-observables. There is no protection from $SU(3)$-flavour symmetry, so that there are potentially large uncertainties from the form factors (involved in the normalisation of the naive factorisation prediction) as well as from $1/m_b$-suppressed long-distance corrections.

We can compare the experimental measurement and the theoretical prediction for individual branching ratios. We give the results for the $\bar{B}_{d,s} \to K^0\bar{K}^0$ case in Table~\ref{tab:BrPP}, whereas we focus on longitudinal branching ratios for vector-vector modes in Table~\ref{tab:BrVV} (as explained in Sec.~\ref{sec:theory-VV}, only longitudinal amplitudes can be computed reliably in QCDF). We see that there are deviations for $\bar{B}_s\to K^0\bar{K}^0$ and (the longitudinal part of) $\bar{B}_s\to K^{*0}\bar{K}^{*0}$ at the level of 1.6  $\sigma$ and 0.9 $\sigma$ respectively, but also for 
(the longitudinal part of) $\bar{B}_d\to K^{*0}\bar{K}^{*0}$ at the level of 1.8 $\sigma$. Assuming that all measurements are correct, we need NP in $b\to s$ but also $b\to d$ to explain the discrepancies observed.

A further comment is in order regarding the accuracy of our predictions within QCDF.
In Ref.~\cite{Amhis:2022hpm}, it is argued that rescattering effects could in principle explain some of the deviations discussed here. This is however an unsatisfying scenario for two different reasons. On  the one hand, the CKM factors associated with rescattering are quite different for $B_d$ and $B_s$ decays, as can be seen in Eq.~(\ref{eq:LKstarKstar}): rescattering is Cabibbo-suppressed for $B_s$ decays ($\alpha_s=O(\lambda^2)$) whereas it is Cabibbo-allowed for $B_d$ decays ($\alpha_d=O(\lambda^0)$), so that rescattering would be a natural explanation if the $B_d$ decay deviates from the SM prediction whereas the $B_s$ decay agrees with the SM expectations. However this is not the pattern of deviations that we observe for $K\bar{K}$ (large deviation for $B_s$) or for $K^*\bar{K}^*$ (with deviations  for both $B_d$ and $B_s$ decays).

On the other hand, the rescattering effects discussed in Ref.~\cite{Amhis:2022hpm} are huge, corresponding to $\Delta_q/P_q$ as large as 2. This is in disagreement with QCDF expectations, which states that such rescattering effects should be of order $\Lambda_{QCD}/m_b$ or $\alpha_s$~\cite{Beneke:2000ry}. As discussed extensively in Refs.~\cite{Descotes-Genon:2006spp,Descotes-Genon:2007iri,Descotes-Genon:2011rgs,Alguero:2020xca} and checked again in App.~\ref{app:PTDelta}, this suppression is even more pronounced for penguin-mediated decays where cancellations occur in $\Delta_q$ between the long-distance contributions from $T_q$ and $P_q$. In the absence of a specific enhancement mechanism that could yield such rescattering effects yet to be observed, we will consider that $\Delta_q/P_q$ has indeed the order of magnitude suggested by QCDF and that rescattering is not the source of the deviations observed.

\subsection{NP in $b \to s$ and in $b \to d$ transitions
}\label{sec:bdbsNP}

Given the previous results, we would like to discuss 
the allowed regions for the Wilson coefficients taking into account not only the optimised $L$ observables but also the individual branching ratios. This is only an exploratory (but very instructive) study and we leave a complete statistical analysis (global fit) of all the observables for future work, as it will require a thorough discussion of the role played by the sources of hadronic uncertainties (form factors and $1/m_b$ suppressed corrections) that are less suppressed for branching ratios than for the $L$ observables. 

 We focus also here and in App.~\ref{app:benchmark} on some benchmark points that can help to provide a first hint of possible scenarios of interest. Given the tension observed in Tabs.\ref{tab:BrPP} and \ref{tab:BrVV} we must allow 
 for NP both in $b\to s$ and $b\to d$ to be able to accommodate the individual branching ratios. This requires us to lift the initial restriction of NP entering only $b\to s$ decays.

We can start by showing the explicit dependence of individual branching ratios on the Wilson coefficients (we recall that we consider only longitudinal polarisations for $K^*\bar{K}^*$):
\begin{eqnarray}
{\cal B}(\bar{B}_s \to  K^{*0}\bar{K}^{*0}) \times 10^{6} &=&
4.36\times\Bigl[
1.00 + 0.85\, \mathcal{C}^{\rm NP}_{1s} + 0.26 \,(\mathcal{C}^{\rm NP}_{1s})^2 - 48.64 \,\mathcal{C}^{\rm NP}_{4s}  \nonumber\\
&& - 13.65 \,\mathcal{C}^{\rm NP}_{1s} \mathcal{C}^{\rm NP}_{4s} + 
  747.23 \,(\mathcal{C}^{\rm NP}_{4s})^2 + 2.88 \,\mathcal{C}^{\rm NP}_{6s}  \nonumber\\
&& + 1.96 \,\mathcal{C}^{\rm NP}_{1s} \mathcal{C}^{\rm NP}_{6s} - 36.96 \,\mathcal{C}^{\rm NP}_{4s} \mathcal{C}^{\rm NP}_{6s} + 
  3.83\,(\mathcal{C}^{\rm NP}_{6s})^2 \nonumber\\
&&  + 2.62\, \mathcal{C}^{\rm NP}_{8gs}  + 
  0.79\,  \mathcal{C}^{\rm NP}_{1s}  \mathcal{C}^{\rm NP}_{8gs} - 78.03\, \mathcal{C}^{\rm NP}_{4s}  \mathcal{C}^{\rm NP}_{8gs} \nonumber\\
&& + 2.27\, \mathcal{C}^{\rm NP}_{6s}  \mathcal{C}^{\rm NP}_{8gs} + 2.05\, (\mathcal{C}^{\rm NP}_{8gs})^2
\Bigl],
\nonumber \cr
\end{eqnarray}
\begin{eqnarray}
{\cal B}(\bar{B}_d \to  K^{*0} \bar{K}^{*0}) \times 10^{7} &=& 2.29\times\Bigl[1.00 + 0.86\, \mathcal{C}^{\rm NP}_{1 d} + 0.26 \,(\mathcal{C}^{\rm NP}_{1 d})^2 - 
 48.17 \,\mathcal{C}^{\rm NP}_{4 d} \nonumber\\
&& - 13.78\, \mathcal{C}^{\rm NP}_{1 d} \mathcal{C}^{\rm NP}_{4 d} + 
 738.05\, (\mathcal{C}^{\rm NP}_{4 d})^2 + 3.15 \,\mathcal{C}^{\rm NP}_{6 d} \nonumber\\
&& + 
 2.06\, \mathcal{C}^{\rm NP}_{1 d} \mathcal{C}^{\rm NP}_{6 d} - 44.36\, \mathcal{C}^{\rm NP}_{4 d} \mathcal{C}^{\rm NP}_{6 d} + 
 4.11 \,(\mathcal{C}^{\rm NP}_{6 d})^2 \nonumber\\
 &&
 + 2.63\, \mathcal{C}^{\rm NP}_{8 g d} + 
 0.81\, \mathcal{C}^{\rm NP}_{1 d} \mathcal{C}^{\rm NP}_{8 g d} - 79.00 \,\mathcal{C}^{\rm NP}_{4 d} \mathcal{C}^{\rm NP}_{8 g d}\nonumber\\
 && + 
 2.70\, \mathcal{C}^{\rm NP}_{6 d} \mathcal{C}^{\rm NP}_{8 g d} + 2.07 \,(\mathcal{C}^{\rm NP}_{8 g d})^2\Bigl],
\nonumber \cr
{\cal B}(\bar{B}_s \to K^0 \bar{K}^0) \times 10^{5}&=&2.86\times \Bigl[1.00 + 0.59 \,\mathcal{C}^{\rm NP}_{1s} + 
   0.13 \,(\mathcal{C}^{\rm NP}_{1s})^2 - 14.70 \,\mathcal{C}^{\rm NP}_{4s} \nonumber\\
   &&- 
   3.14 \,\mathcal{C}^{\rm NP}_{1s} \mathcal{C}^{\rm NP}_{4s} + 63.55\, (\mathcal{C}^{\rm NP}_{4s})^2 - 
   24.38 \,\mathcal{C}^{\rm NP}_{6s}\nonumber\\
   &&- 5.56 \,\mathcal{C}^{\rm NP}_{1s} \mathcal{C}^{\rm NP}_{6s} + 
   205.32 \,\mathcal{C}^{\rm NP}_{4s} \mathcal{C}^{\rm NP}_{6s} + 166.52 \,(\mathcal{C}^{\rm NP}_{6s})^2\nonumber\\
   &&+ 
   1.23 \,\mathcal{C}^{\rm NP}_{8gs} + 0.31 \,\mathcal{C}^{\rm NP}_{1s} \mathcal{C}^{\rm NP}_{8gs} - 
   9.96 \,\mathcal{C}^{\rm NP}_{4s} \mathcal{C}^{\rm NP}_{8gs}\nonumber\\
   &&- 16.25 \,\mathcal{C}^{\rm NP}_{6s} \mathcal{C}^{\rm NP}_{8gs} + 
   0.40 \,(\mathcal{C}^{\rm NP}_{8gs})^2\Bigl],
\nonumber \cr
{\cal B}(\bar{B}_d \to K^0 \bar{K}^0) \times 10^{6}&=&1.11\times\Bigl[1.00 + 0.62 \,\mathcal{C}^{\rm NP}_{1d} + 
   0.13\, (\mathcal{C}^{\rm NP}_{1d})^2 - 14.81\, \mathcal{C}^{\rm NP}_{4 d}\nonumber\\
   &&- 
   3.33\, \mathcal{C}^{\rm NP}_{1 d} \mathcal{C}^{\rm NP}_{4 d} + 64.61\, (\mathcal{C}^{\rm NP}_{4 d})^2 - 
   24.28\, \mathcal{C}^{\rm NP}_{6 d}\nonumber\\
  && - 5.81 \,\mathcal{C}^{\rm NP}_{1 d} \mathcal{C}^{\rm NP}_{6 d} + 
   206.48\, \mathcal{C}^{\rm NP}_{4 d} \mathcal{C}^{\rm NP}_{6 d} + 165.61 \,(\mathcal{C}^{\rm NP}_{6 d})^2 \nonumber\\
   &&
   + 
   1.34 \,\mathcal{C}^{\rm NP}_{8 g d} + 0.35 \,\mathcal{C}^{\rm NP}_{1 d} \mathcal{C}^{\rm NP}_{8 g d} - 
   10.91\, \mathcal{C}^{\rm NP}_{4 d} \mathcal{C}^{\rm NP}_{8 g d}\nonumber\\
   &&- 17.62\, \mathcal{C}^{\rm NP}_{6 d} \mathcal{C}^{\rm NP}_{8 g d} + 
   0.47\,(\mathcal{C}^{\rm NP}_{8 g d})^2\Bigl]
\end{eqnarray}

\begin{figure}[t]
\includegraphics[width=0.47\textwidth,height=0.34\textwidth]{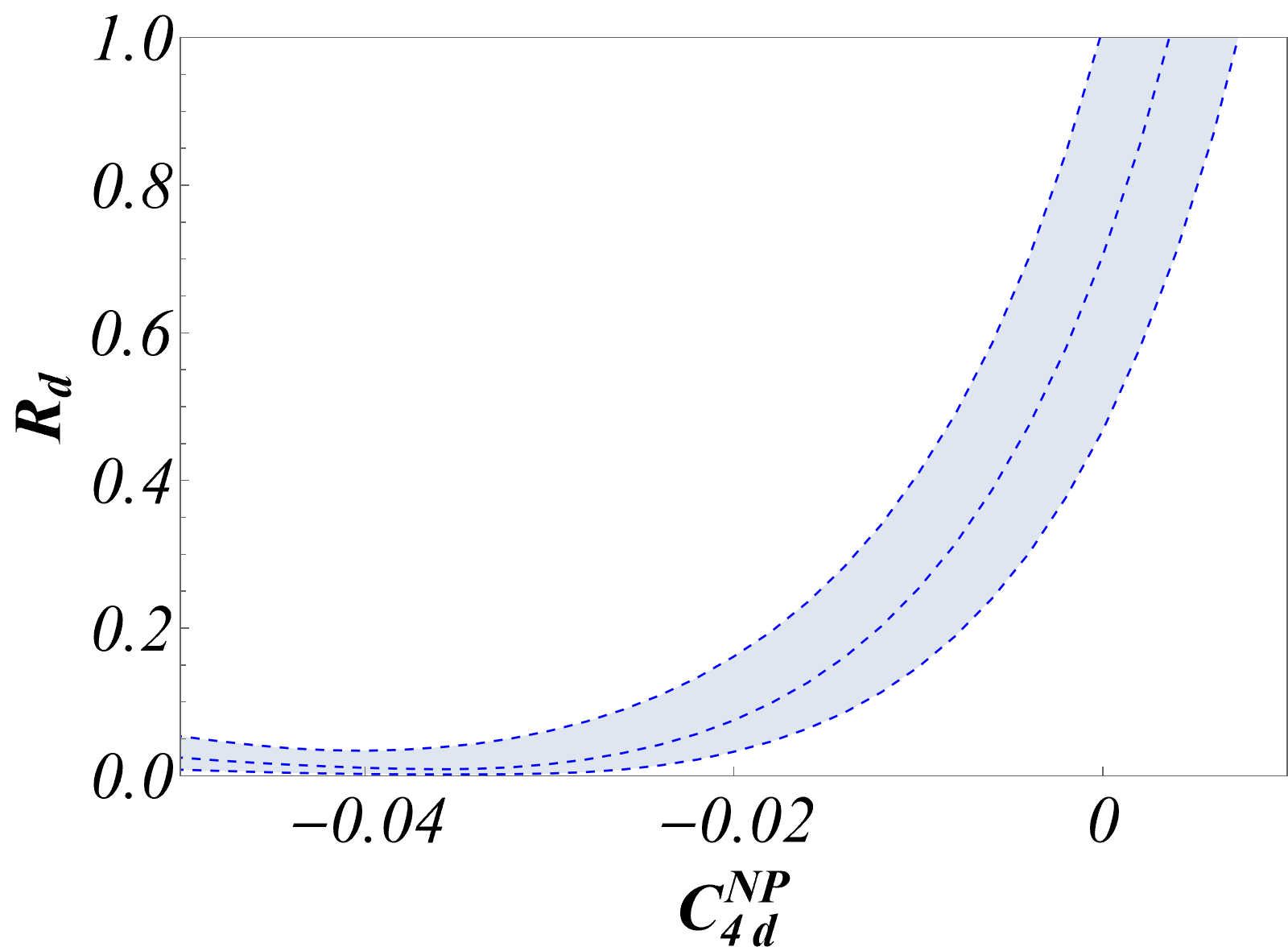}
\includegraphics[width=0.47\textwidth,height=0.34\textwidth]{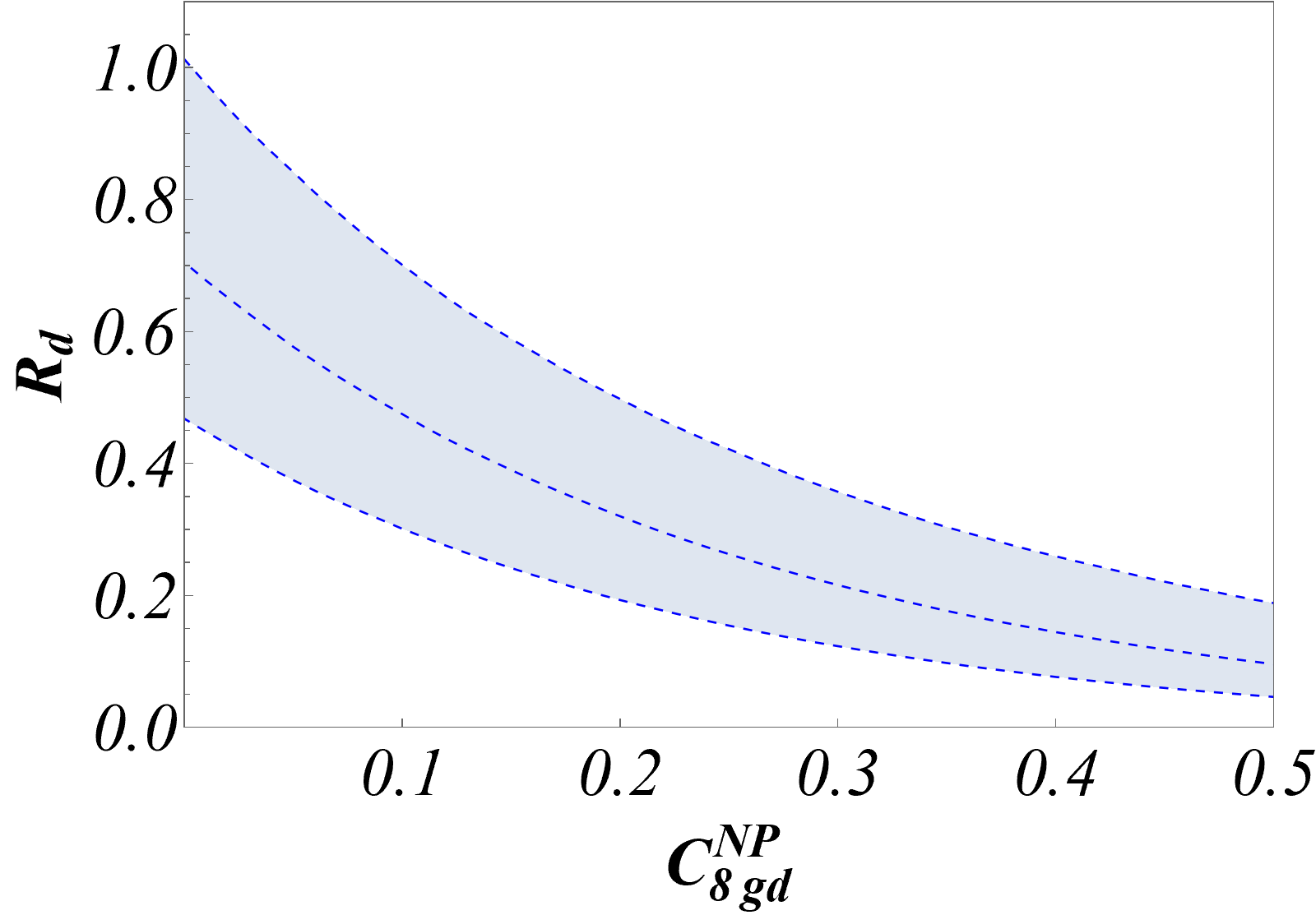}
\caption{$R_d$ versus ${\cal C}_{4d}^{\rm NP}$ (left) and ${\cal C}_{8gd}^{\rm NP}$ (right), in each case within a range of values including  
 those allowed by the individual branching ratios and $L$-observables. }\label{fig:rd1}
\end{figure}
\begin{figure}[h]
\centering
\includegraphics[width=0.59\textwidth,height=0.40\textwidth]{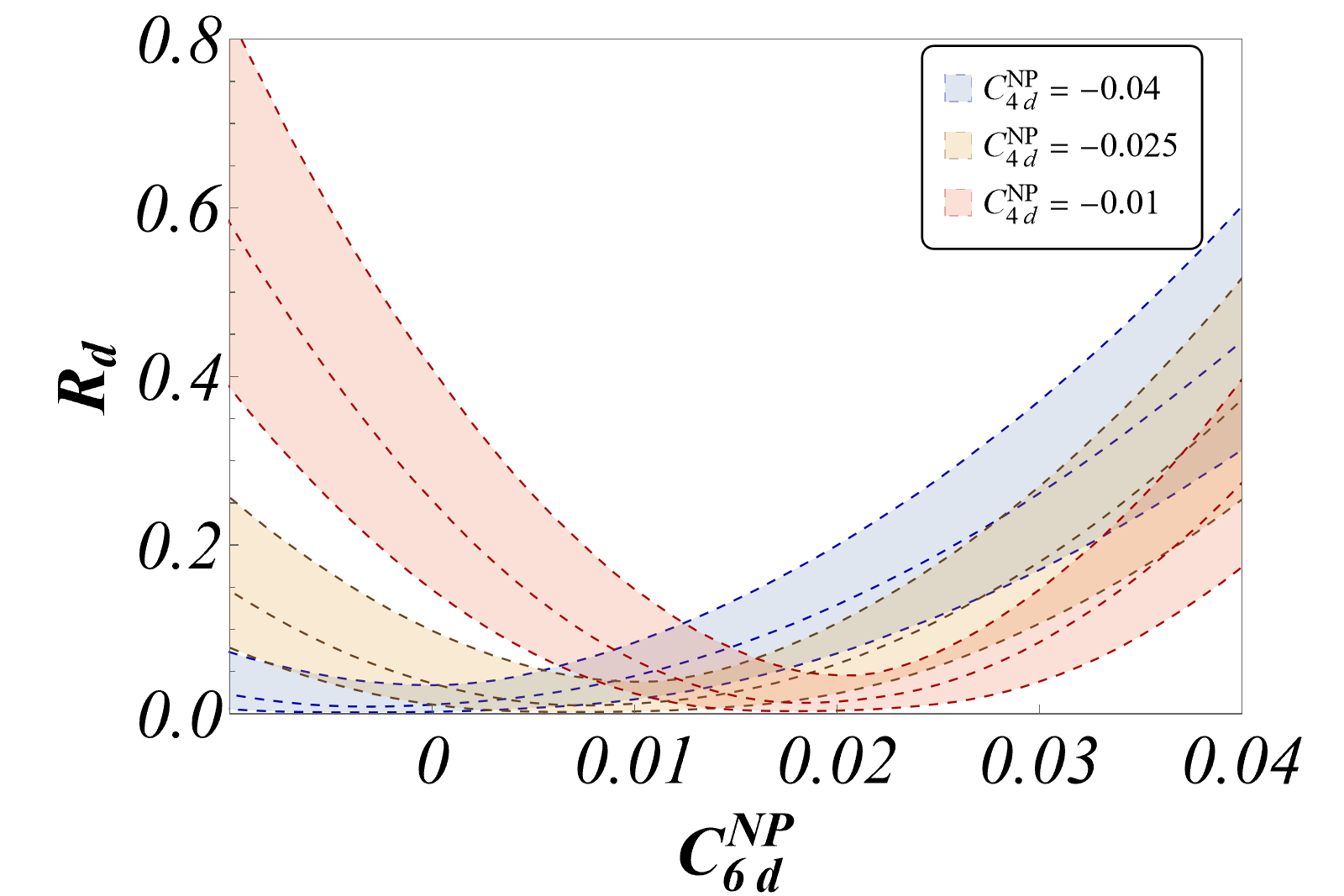}
\caption{$R_d$ versus ${\cal C}_{6d}^{\rm NP}$, within a range of values including  
 those allowed by the individual branching ratios and $L$-observables. We illustrate the impact of ${\cal C}_{4d}^{\rm NP}$ by showing three benchmark values of ${\cal C}_{4d}^{\rm NP}$ that are allowed by the constraints from the branching ratios and the $L$-observables.}\label{fig:rd2}
\end{figure}

 We can use the above expressions (including uncertainties) together with the full expressions for $L_{K\bar{K}}$, $L_{K^*\bar{K}^*}$ allowing for NP 
 in both numerator ($b\to s$) and denominator ($b \to d$) in order to identify
three  representative situations:
\begin{itemize}

\item NP in ${\cal C}_4$: The allowed values in the plane ${\cal C}_{4s}^{\rm NP}$-${\cal C}_{4d}^{\rm NP}$ according to the four individual branching ratios and the two measured $L$ observables is displayed in Fig.~\ref{fig:BR1}, so that
\begin{itemize}
\item[a)] we need NP contributions to ${\cal C}_{4d}$ in the range shown in Fig.~\ref{fig:BR1}
in order to explain the deviations in individual $B_d$ branching ratios,
\item[b)] the region allowed for NP contribution to ${\cal C}_{4s}$ is enlarged, allowing for slightly smaller values up to 0.003 for  ${\cal C}_{4s}^{\rm NP}$ compared  to Fig.~\ref{fig:C4_range} where only NP in $b\to s$ transitions was allowed and where only the $L$ observables were considered (see also the discussion in Sec.~\ref{sec:pattern}).

\end{itemize}

\item NP in both ${\cal C}_4$ and ${\cal C}_6$: 
One may wonder if the allowed region found for ${\cal C}_{4d}$-${\cal C}_{4s}$ is enlarged once NP in ${\cal C}_{6d,6s}$ is introduced. The black dash-dotted contour~\footnote{
The dashed contour in Fig.~\ref{fig:BR1} goes beyond the $L_{K^*\bar{K}^*}$ bound as the latter is shown here taking ${\cal C}_{6d,6s}=0$.}
in Fig.~\ref{fig:BR1} shows that this is indeed the case. However allowing  ${\cal C}_{6d,6s} \neq 0$ does not provide
additional solutions for ${\cal C}_4$ closer to the SM.
For completeness
we also show the allowed region for ${\cal C}_{6d}$-${\cal C}_{6s}$ letting ${\cal C}_{4d}$, ${\cal C}_{4s}$ float within their allowed range in Fig.~\ref{fig:BR2}. None of the points allowed in this plot correspond to ${\cal C}_{4d,s}=0$.

\item NP in ${\cal C}_{8g}$: Like in the previous cases, letting NP enter an additional Wilson coefficient (${\cal C}_{8gd}$) allows ${\cal C}_{8gs}$ to reach values closer to zero, whereas very negative values, previously allowed in Fig.\ref{fig:C8_range}, are now forbidden, as can be seen in Fig.~\ref{fig:BR3}. Non-vanishing values of ${\cal C}_{8gd}$ are required in a narrow range to explain the individual branching ratios.

\end{itemize}

Due to the presence of NP in the ${\cal C}_{id}$ coefficients, $R_d$ cannot be considered SM-like any more. Fig.~\ref{fig:rd1} shows the variation of $R_d$ within the relevant range for ${\cal C}_{4d}$ and ${\cal C}_{8gd}$ (including the region accommodating the individual branching ratios). %$R_d$ turns out to be particularly useful to distinguish pattern 1 (NP in ${\cal C}_4$) from pattern 3 (NP in ${\cal C}_{8g}$). 
In both cases (${\cal C}_{8gd}$ and ${\cal C}_{4d}$), $R_d$ is significantly suppressed by NP contributions. However, this observable requires tagging, which means that it is likely to be measured only at the end of Run 3. In Fig.~\ref{fig:rd2}, we show the sensitivity of $R_d$ to ${\cal C}_{6d}^{\rm NP}$, which is modulated by the value of ${\cal C}_{4d}^{\rm NP}$ (relevant in particular for NP scenarios involving both ${\cal C}_{4d}^{\rm NP}$ and ${\cal C}_{6d}^{\rm NP}$).

We can now reconsider the sensitivity study of the $L$-observables performed in Sec.~\ref{sec:combined}
but allowing for NP also in $b \to d$ transitions.  The discussion of Sec.~\ref{sec:sensitivity} regarding the sensitivity of the $L$-observables on NP in ${\cal C}_{is}^{\rm NP}$ can be easily reinterpreted as a study on the sensitivity of these observables on ${\cal C}_{is}^{\rm NP}-{\cal C}_{id}^{\rm NP}$, simply by changing the $x$-axis from ${\cal C}_{is}^{\rm NP}$ to ${\cal C}_{is}^{\rm NP}-{\cal C}_{id}^{\rm NP}$ in the plots of Sec.~\ref{sec:combined}. This corresponds to a similar reinterpretation of Eqs.~(\ref{eq:LKstLKst-Cis}) to (\ref{eq:Ltotal}).
Indeed, one can perform a Taylor expansion of the ratios of branching ratios involved in $L$ in powers of the NP contribution to the Wilson coefficients (${\cal C}_{id}^{\rm NP}$ and ${\cal C}_{is}^{\rm NP})$: we checked that this reinterpretation is possible as long as the size of the NP contribution is small enough.
More precisely, this reinterpretation will be deemed possible if the error (defined as the difference among the central value of the exact predictions  and the one induced by this approximation) is  smaller   than the size of the uncertainty of the theory prediction at the point. Depending on the NP scenario considered, this approximation can be valid over the whole range allowed by the current measurements, or only to a smaller range:
\begin{itemize}
\item In the case of ${\cal C}_{4s}$, Figs.~\ref{fig:C4_range}, \ref{fig:C46s_pred2}, \ref{fig:C4_pred}, \ref{fig:C4_pred_Ltot}, \ref{fig:dev-patternsI} and \ref{fig:dev-patternsII} can be used upon replacing the $x$-axis ${\cal C}_{4s}^{\rm NP}$ by ${\cal C}_{4s}^{\rm NP}-{\cal C}_{4d}^{\rm NP}$ up to NP values in the range $0 \leq {\cal C}_{4s}^{\rm NP} \leq 0.040$ when the value of ${\cal C}_{4d}^{\rm NP}$ is around  ${\cal C}_{4d}^{\rm NP}\sim -0.006$ (minimal allowed value in absolute value). For $L_{K\bar{K}}$ and $L_{K^*\bar{K}^*}$ the approximation works very well in a wider range for most allowed values for ${\cal C}_{4d}$ according to Fig.~\ref{fig:BR1} and for values of ${\cal C}_{4s}^{\rm NP}$ up to 0.040. $\hat{L}_{K^*}$ is the observable limiting the range of validity of this approximation, which fails when the values of ${\cal C}_{4d}^{\rm NP}$ are not small in absolute value.
\item 
The case of ${\cal C}_{8gd}^{\rm NP}$-${\cal C}_{8gs}^{\rm NP}$ can be discussed similarly. The approximation holds for all observables   in the whole range for ${\cal C}_{8gs}^{\rm NP}$ if ${\cal C}_{8gd}^{\rm NP} \sim +0.13$ (minimal allowed   value). This larger range of validity comes from the smaller factors in front of the Wilson coefficients ${\cal C}_{8gi}$  ($i=d,s$) for the expression of the branching ratios, which increases the convergence of the Taylor expansion required to express the $L$-observables. 
\end{itemize}

In  App.~\ref{app:benchmark}, we illustrate the behaviour of the $L$-observables taking the full expressions allowing for NP in numerator and denominator of the  observables without any Taylor expansion. We consider benchmark points which are allowed according to Figs.~\ref{fig:BR1}, \ref{fig:BR2} and \ref{fig:BR3}, but outside the limited range where the approximation discussed above holds. At these points, the expansions discussed in Sec.~\ref{sec:sensitivity} cannot be used
substituting ${\cal C}_{is}^{\rm NP}$ for ${\cal C}_{is}^{\rm NP}-{\cal C}_{id}^{\rm NP}$ for some of the observables, in particular ${\hat L}_{K^*}$.
The study in App.~\ref{app:benchmark} suggests that an increase in the tensions in the $B_d$ branching ratios would require more negative (positive) values for ${\cal C}_{4d}^{\rm NP}$ (${\cal C}_{8gd}^{\rm NP}$) and a dedicated analysis based  on our full expressions of the observables will be required.
\newpage

\section{Conclusions}\label{sec:conclusions}

Several hints of New Physics have occurred in $b\to s$ transitions mediated by flavour-changing neutral currents, which are expected to exhibit such a sensitivity due to their suppression in the Standard Model. On the one hand, semileptonic transitions $b\to s\mu\mu$ have shown a consistent trend of deviations, even without taking into account the waning signs of Lepton Flavour Universality violation. On the other hand, the non-leptonic transition $\bar{B}_{s}\to K^*\bar{K^*}$ has also showed an interesting deviation from SM expectation, with a very low longitudinal polarisation fraction. In a previous article, we designed an optimised observable $L_{K^*\bar{K}^*}$ based on the ratio of longitudinal branching ratios of $\bar{B}_{s} \to K^{*0} \bar{K}^{*0}$ and $\bar{B}_{d} \to K^{*0} \bar{K}^{*0}$ decay modes. We have a better theoretical control on this observable, as it is protected by $U$-spin symmetry and is computable within QCDF with a reduced sensitivity to $1/m_b$-suppressed contributions. We found that the 2.6 $\sigma$ deviation from the SM could be explained by an NP contribution to two Wilson coefficients of the Weak Effective Theory at the scale of the $b$-quark mass, associated with the QCD penguin operator $Q_{4s}$ and the QCD chromomagnetic operator $Q_{8gs}$. The NP contributions to the Wilson coefficients needed are at most of the same size as the SM contribution for QCD penguin operators and up to three times for the chromomagnetic one, which does not violate the current experimental bounds on these Wilson coefficients.

In the present article, we extend the discussion to related decays with the same quark content, but different spins for the outgoing mesons, i.e. ${\bar B}_{d,s} \to K^{0} \bar{K}^{0}$, ${\bar B}_{d,s} \to K^{0} \bar{K}^{*0}$ and ${\bar B}_{d,s} \to \bar{K}^{0} {K^{*0}}$ together with their CP conjugate partners. We are able to design
optimised observables for these decays, with reduced hadronic uncertainties, mainly coming from form factors and power-suppressed infrared divergences, thanks to $U$-spin symmetry and QCD factorisation. We may then maximize their sensitivity to NP. We found a 2.4 $\sigma$ deviation
in the pseudoscalar mode ${\bar B}_{d,s} \to K^{0} \bar{K}^{0}$ whereas the lack of experimental information prevented us from analysing the pseudoscalar-vector modes
${\bar B}_{d,s} \to K^{0} \bar{K}^{*0}$ and ${\bar B}_{d,s} \to \bar{K}^{0} {K^{*0}}$ in more detail. 

We reconsidered some NP scenarios able to explain the  tensions between the SM prediction and data in $L_{K^* \bar{K}^*}$ and $L_{K\bar{K}}$, focusing first on NP in $b\to s$ transitions only.
 It turns out that the various observables have different but significant sensitivities not only to  ${\cal C}_{4s}$ and ${\cal C}_{8gs}$, but also to ${\cal C}_{6s}$. Indeed, these decays can be analysed within the factorisation approach: some contributions of operators may be  absent, present or enhanced, depending on how the spin and flavour structure of the operators match those of the outgoing mesons, leading to different sensitivities to the Wilson coefficients  based on whether one or two pseudoscalars are involved and  whether they pick up the spectator quark or not.
We identified simple NP scenarios that could accommodate the deviations in $L_{K^* \bar{K}^*}$ and $L_{K\bar{K}}$ within their $1\sigma$ experimental and theoretical ranges in terms of an NP contribution to Wilson coefficients in $b\to s$ transitions. These scenarios would yield very distinctive patterns of deviations that can be probed  in the LHCb and Belle II experiments. In particular the optimised observables associated with the pseudoscalar-vector modes $\bar{B}_{d,s}\to K^{(*)0}\bar{K}^{(*)0}$ provide powerful probes  for various scenarios corresponding to NP contributions in  ${\cal C}_{4s}$, ${\cal C}_{6s}$ and/or ${\cal C}_{8gs}$. Some of the pseudoscalar-vector observables ($\hat{L}_K$ and $\hat{L}_{K^*}$) deviate among them and from SM expectations in a very remarkable way in the presence of NP (sometimes by order of magnitude) but are difficult to measure experimentally as they require $B_d$ tagging. Related observables ($L_K$ and $L_{K^*}$) can be used for similar purposes without relying on $B_d$ tagging and should be accessible more easily experimentally. 

In addition to the ratios of branching ratios such as $L_{K^* \bar{K}^*}$ and $L_{K\bar{K}}$, we also considered individual branching ratios in the same framework. It turns out that deviations occur in ${\cal B}(\bar{B}_s\to K\bar{K})$ but also in ${\cal B}(\bar{B}_d\to K^*\bar{K}^*)$, though at a more modest level than $L_{K^* \bar{K}^*}$ and $L_{K\bar{K}}$. It thus suggests that NP scenarios with contributions to both $b\to d$ and $b\to s$ transitions should be considered. 
We have identified domains of NP contributions to ${\cal C}_{4(d,s)}$, ${\cal C}_{6(d,s)}$, ${\cal C}_{8g(d,s)}$ 
(see Figs.~\ref{fig:BR1}, \ref{fig:BR2} and \ref{fig:BR3})
which could accommodate all the measurements related to $K\bar{K}$ and $K^*\bar{K}^*$ (both $L$-observables and individual branching ratios) within their theoretical and experimental 1$\sigma$ ranges. As in the previous case, significant deviations of branching ratios in the pseudoscalar-vector modes should be observed according to different patterns of deviations associated with different NP scenarios. 
For such scenarios, it is particularly important to measure also the pseudoscalar-vector observables $\hat{L}_K$ and $\hat{L}_{K^*}$ and not only $L_K$ and $L_{K^*}$ which have a more limited sensitivity to NP. We kept
our discussion of these scenarios at a qualitative level without trying to perform a detailed statistical analysis. Since individual branching ratios are rather sensitive to the models used to describe long-distance contributions that are $1/m_b$-suppressed within QCD factorisation, we did not attempt to perform a global fit analysis, which is left for a future work.

A (future) confirmation of such consistent set of deviations in these channels would be extremely valuable. It would point towards a common origin and would provide a possible strong hint of NP in the non-leptonic sector, which would require a more elaborate statistical framework than the simple approach presented here. In any case, we hope that our analysis will thus constitute a strong incentive to study these penguin-mediated modes experimentally in more detail in the coming years.

\section*{Acknowledgements}

J.M. thanks Nicola Serra, Patrick Owen, Julian Garcia Pardinas and Gino Isidori for useful discussions. This project has received support from the European Union’s Horizon 2020 research and innovation programme under the Marie Sklodowska-Curie grant agreement No 860881-HIDDeN [S.D.G.] and the Marie Sklodowska-Curie grant agreement No 945422 [G.T-X.]. This research has been supported by
the Deutsche Forschungsgemeinschaft (DFG, German Research Foundation) under grant
396021762 - TRR 257.
J.M. gratefully acknowledges the financial support from ICREA under the ICREA Academia programme and from the Pauli Center (Zurich) and the Physics Department of University of Zurich.
 J.M. and A.B. also received financial support from Spanish Ministry of Science, Innovation and Universities (project PID2020-112965GB-I00) and from the Research Grant Agency of the Government of Catalonia (project SGR 1069).

\newpage

\appendix

\begin{table}[h]
	\begin{center}
\tabcolsep=1.26cm\begin{tabular}{|c|c|c|}
\hline\multicolumn{3}{|c|}{$B_{d,s}$ Distribution Amplitudes (at $\mu=1$ GeV)~\cite{Khodjamirian:2020hob,Ball:2006nr} } \\ \hline
$\lambda_{B_d} $ [GeV]&$\lambda_{B_s}/\lambda_{B_d}$& $\sigma_B$\\
\hline
$0.383\pm0.153$&$1.19\pm0.14$&$1.4\pm0.4$\\ \hline
\end{tabular}

\vskip 1pt

\tabcolsep=0.712cm\begin{tabular}{|c|c|c|c|}
\hline\multicolumn{4}{|c|}{$K^*$ Distribution Amplitudes (at $\mu=2$ GeV)~\cite{Ball:2007rt}}  \\ \hline
$\alpha_1^{K^*}$&
$\alpha_{1,\perp}^{K^*}$&
$\alpha_2^{K^*}$&
$\alpha_{2,\perp}^{K^*}$\\
\hline
$0.02\pm0.02$&$0.03\pm0.03$&$0.08\pm0.06$&$0.08\pm0.06$\\ \hline
\end{tabular}

\vskip 1pt

\tabcolsep=2.512cm\begin{tabular}{|c|c|}
\hline\multicolumn{2}{|c|}{$K$ Distribution Amplitudes (at $\mu=2$ GeV)~\cite{RQCD:2019osh}}  \\ \hline
$\alpha_1^{K}$&
%$\alpha_{1,\perp}^{K^*}$&
$\alpha_2^{K}$\\
%$\alpha_{2,\perp}^{K^*}$\\
\hline
$0.0525^{+31}_{-33}$&$0.106^{+15}_{-16}$\\ \hline
\end{tabular}

\vskip 1pt

\tabcolsep=2.1cm\begin{tabular}{|c|c|}
\hline\multicolumn{2}{|c|}{Decay Constants for $B$ mesons (at $\mu=2$ GeV)~\cite{Aoki:2019cca}}  \\ \hline
$f_{B_d}$&$f_{B_s}/f_{B_d}$\\
\hline
$0.190\pm0.0013$&$1.209\pm0.005$\\ \hline
\end{tabular}

\vskip 1pt

\tabcolsep=1.0cm\begin{tabular}{|c|c|c|}
\hline\multicolumn{3}{|c|}{Decay Constants for Kaons (at $\mu=2$ GeV)~\cite{Workman:2022ynf,Allton:2008pn,Straub:2015ica}}  \\ \hline
$f_{K}$&$f_{K^*}$&$f^\perp_{K^*}/f_{K^*}$\\
\hline
$0.1557\pm 0.0003$&$0.204\pm0.007$&$0.712\pm0.012$\\ \hline
\end{tabular}
\vskip 1pt

\tabcolsep=0.52cm\begin{tabular}{|c|c|c|c|}
\hline\multicolumn{4}{|c|}{$B_{d,s}\to K^*$ form factors~\cite{Straub:2015ica} and B-meson lifetimes (ps)}  \\ \hline
$A_0^{B_s}(q^2=0)$&$A_0^{B_d}(q^2=0)$ & $\tau_{B_d}$ & $\tau_{B_s}$\\
\hline
$0.314 \pm 0.048$ &$ 0.356 \pm 0.046$ & $1.519\pm0.004$ & $1.515\pm0.004$ \\ \hline
\end{tabular}

\vskip 1pt

\tabcolsep=2.162cm\begin{tabular}{|c|c|}
\hline\multicolumn{2}{|c|}{$B_{d}\to K$~\cite{Parrott:2022rgu} and $B_s\to K$~\cite{Khodjamirian:2017fxg} form factors}  \\ \hline
$f_0^{B_s}(q^2=0)$&$f_0^{B_d}(q^2=0)$ \\
\hline
$0.336 \pm 0.023$ &$ 0.332 \pm 0.012$ \\ \hline
\end{tabular}

\vskip 1pt

\tabcolsep=0.515cm\begin{tabular}{|c|c|c|c|c|}
\hline
\multicolumn{4}{|c|}{Wolfenstein parameters~\cite{Charles:2004jd} }  \\ \hline
$A$ & $\lambda$ & $\bar\rho$ &
$\bar\eta$\\ \hline
$0.8235^{+0.0056}_{-0.0145}$ &$	0.22484^{+0.00025}_{-0.00006}$&$0.1569^{+0.0102}_{-0.0061}$&$0.3499^{+0.0079}_{-0.0065}$\\
\hline
\end{tabular}

\vskip 1pt

\tabcolsep=0.496cm\begin{tabular}{|c|c|c|c|c|c|}
	\hline
	\multicolumn{6}{|c|}{QCD scale and masses [GeV]}\\
		\hline 
 $\bar{m}_b(\bar{m}_b)$  & $m_b/m_c$ & $m_{B_d} $& $m_{B_s} $& $m_{K^*} $ & $\Lambda_{{\rm QCD}}$
      \\ \hline
      $4.2$  & $4.577\pm0.008$  & $5.280$ & $5.367$&$0.892$&$0.225$
      \\ \hline
\end{tabular}
\vskip 1pt
\begin{tabular}{|c|c|c|c|c|c|}
\hline
	\multicolumn{6}{|c|}{SM Wilson Coefficients (at $\mu=4.2$ GeV)}\\
		\hline 
${\cal C}_1$ &  ${\cal C}_2$ & ${\cal C}_3$ & ${\cal C}_4$ &  ${\cal C}_5$ & ${\cal C}_6$
      \\ \hline
 1.082 & -0.191 & 0.013 & -0.036 & 0.009 &  -0.042
      \\ \hline
 ${\cal C}_{7}/\alpha_{em}$ & ${\cal C}_{8}/\alpha_{em}$ & ${\cal C}_{9}/\alpha_{em}$ &  ${\cal C}_{10}/\alpha_{em}$ & ${\cal C}^{\rm eff}_{7\gamma}$ &  ${\cal C}^{\rm eff}_{8g}$
      \\ \hline
 -0.011 & 0.058 & -1.254 & 0.223 & -0.318 & -0.151 
      \\ \hline
		\end{tabular}
		\caption{Input parameters.}
		\label{tab:inputs}
	\end{center}
\end{table}

\section{Elements of QCDF} \label{app:QCDF}

\subsection{Weak Effective Theory}\label{app:WET}

At the scale $m_b$, the relevant effective Hamiltonian separating  small- and large-distances for $b\to q$ transitions in non-leptonic $B$-decays is
\begin{equation}\label{eq:wet}
H_{\rm eff}=\frac{G_F}{\sqrt{2}}\sum_{p=c,u} \lambda_p^{(q)}
 \Big({\cal C}_{1s}^{p} Q_{1s}^p + {\cal C}_{2s}^{p} Q_{2s}^p+\sum_{i=3 \ldots 10} {\cal C}_{is} Q_{is} + {\cal C}_{7\gamma s} Q_{7\gamma s} + {\cal C}_{8gs} Q_{8gs}\Big) \,. \nonumber
\end{equation}
where $\lambda_p^{(q)}=V_{pb}V^*_{pq}$. We follow the conventions and definitions of Ref~\cite{Beneke:2001ev}:
\begin{align}
 Q_{1s}^p &= (\bar p b)_{V-A} (\bar s p)_{V-A} \,,  & Q_{7s} &= (\bar s b)_{V-A} \sum_q\,\frac{3}{2} e_q (\bar q q)_{V+A} \,, \nonumber \\[-2.2mm]
 Q_{2s}^p &= (\bar p_i b_j)_{V-A} (\bar s_j p_i)_{V-A} \,, & Q_{8s} &= (\bar s_i b_j)_{V-A} \sum_q\,\frac{3}{2} e_q (\bar q_j q_i)_{V+A} \,, \nonumber \\[-2.2mm]
 Q_{3s} &= (\bar s b)_{V-A} \sum_q\,(\bar q q)_{V-A} \,, &Q_{9s} &= (\bar s b)_{V-A} \sum_q\,\frac{3}{2} e_q (\bar q q)_{V-A} \,, \nonumber \\[-2.2mm]
 Q_{4s} &= (\bar s_i b_j)_{V-A} \sum_q\,(\bar q_j q_i)_{V-A} \,, & Q_{10s} &= (\bar s_i b_j)_{V-A} \sum_q\,\frac{3}{2} e_q (\bar q_j q_i)_{V-A} \,, \nonumber\\[-2.2mm]
 Q_{5s} &= (\bar s b)_{V-A} \sum_q\,(\bar q q)_{V+A} \,, &Q_{7\gamma s} &= \frac{-e}{8\pi^2}\,m_b\bar s\sigma_{\mu\nu}(1+\gamma_5) F^{\mu\nu} b \,,\nonumber \\[-2.2mm]
 Q_{6s} &= (\bar s_i b_j)_{V-A} \sum_q\,(\bar q_j q_i)_{V+A} \, , &Q_{8gs} &= \frac{-g_s}{8\pi^2}\,m_b\, \bar s\sigma_{\mu\nu}(1+\gamma_5) G^{\mu\nu} b \,, \nonumber
%\label{operators}
\end{align}
 where $(\bar q_1 q_2)_{V\pm A}=\bar q_1\gamma_\mu(1\pm\gamma_5)q_2$, 
$i,j$ are colour indices, $e_q$ are the electric charges of the quarks in units of $|e|$.
$ Q_{1s,2s}^p$ are the left-handed current-current operators,  $ Q_{3s\ldots 6s}$ and
$ Q_{7s\ldots 10s}$ are QCD and electroweak penguin operators, and $Q_{7\gamma s}$ and $Q_{8gs}$ are electromagnetic and chromomagnetic dipole operators.
A summation over $q=u,d,s,c,b$ is implied. 
In the SM, ${\cal C}_1^c$  is the largest coefficient and it corresponds to the colour-allowed tree-level contribution from the W exchange, whereas ${\cal C}_2^c$  is colour suppressed. QCD-penguin operators are numerically suppressed, and the electroweak operators even more so.

For simplicity, in this article, we will focus on NP scenarios involving shifts  in Wilson coefficients associated to operators already present in the SM, and we will thus use the basis presented in Eq.~(\ref{eq:wet}). Similarly to Ref.~\cite{Alguero:2020xca}, we could extend this basis to the chirally-
flipped ones $\tilde{Q}_i$ as defined in Ref.~\cite{Kagan:2004ia} by exchanging left and right-chirality projector in all quark bilinears.
These right-handed currents would modify the amplitudes of the various decays considered in the following way
\begin{equation}
A(B\to PP)[C-\tilde{C}]\qquad
A(B\to VP)[C+\tilde{C}]\qquad
A(B\to (VV)_0)[C-\tilde{C}]
\end{equation}
The discussion of the impact of right-handed currents on the observables under discussion here is left for future work.

QCDF allows one to compute the hadronic matrix elements starting from the heavy-quark limit $m_b\to\infty$ where the naive factorisation approach holds for some classes of decays and providing corrections to this picture. Indeed, the hadronic matrix elements $T_q$ and $P_q$
can be expressed as an expansion of $\alpha_s$ involving form factors and light-cone distribution amplitudes as hadronic inputs, up to  $1/m_b$-suppressed terms that contain long-distance contributions, corresponding to infrared divergences in the factorisation framework. 
In the case of vector modes, a clear hierarchy among polarisations occurs in the limit $m_b\to\infty$, so that only the longitudinal polarisation can actually be computed accurately within QCDF.
The matrix elements for the $B_d$ decay into a pair of longitudinally polarised $K^{*0}$ is actually given as~\cite{Beneke:2006hg,Bartsch:2008ps} 
\begin{eqnarray}\label{eq:TPlongitudinal}
&&\!\!\!T(\bar{B}_d\to \bar{K}^{*0}K^{*0})=A_{\bar{K}^*K^*}
   [\alpha_4^u-\frac{1}{2}\alpha_{4,EW}^u+\beta_3^u+\beta_4^u-\frac{1}{2}\beta_{3,EW}^u-\frac{1}{2}\beta_{4,EW}^u] \nonumber \\
  && \qquad \qquad \qquad \qquad \, \,  +A_{K^*\bar{K}^*}[\beta_4^u-\frac{1}{2}\beta_{4,EW}^u]\,,  \\
&&\!\!\! P(\bar{B}_d\to \bar{K}^{*0}K^{*0})=A_{\bar{K}^*K^*}
   [\alpha_4^c-\frac{1}{2}\alpha_{4,EW}^c+\beta_3^c+\beta_4^c-\frac{1}{2}\beta_{3,EW}^c-\frac{1}{2}\beta_{4,EW}^c] \nonumber \\
   && \qquad \qquad \qquad \qquad \, \, +A_{K^*\bar{K}^*}[\beta_4^c-\frac{1}{2}\beta_{4,EW}^c]\,,
\end{eqnarray}
whereas the $B_s$ decay yields
\begin{eqnarray}   
&&\!\!\!T(\bar{B}_s\to \bar{K}^{*0}K^{*0})=A_{\bar{K}^*K^*}[\beta_4^u-\frac{1}{2}\beta_{4,EW}^u]\\
&& \qquad \qquad \qquad \qquad \, \,  +A_{K^*\bar{K}^*}
   [\alpha_4^u-\frac{1}{2}\alpha_{4,EW}^u+\beta_3^u+\beta_4^u-\frac{1}{2}\beta_{3,EW}^u-\frac{1}{2}\beta_{4,EW}^u] \,,\nonumber\\
&&\!\!\! P(\bar{B}_s\to \bar{K}^{*0}K^{*0})=A_{\bar{K}^*K^*}[\beta_4^c-\frac{1}{2}\beta_{4,EW}^c]\\
&& \qquad \qquad \qquad \qquad \, \,  +A_{K^*\bar{K}^*}[\alpha_4^c-\frac{1}{2}\alpha_{4,EW}^c+\beta_3^c+\beta_4^c-\frac{1}{2}\beta_{3,EW}^c-\frac{1}{2}\beta_{4,EW}^c] \,. \nonumber 
 \end{eqnarray}
  The coefficients $\beta$ collect weak annihilation
 contributions, whereas the coefficients $\alpha$ collect 
 the remaining contributions. The normalisation $A$ is the same for $\bar{K}^*K^*$ and $K^*\bar{K}^*$
\begin{equation}
A_{\bar{K}^*K^*}=A_{K^*\bar{K}^*}=\frac{G_F}{\sqrt{2}}m^2_{B_q}f_{K^*}A_0^{B_q\to K^*}(0)\,,
 \end{equation}
 but the order of $K^*$ and $\bar{K}^*$ in the normalisation $A(M_1 M_2)$ indicates also the argument of the coefficients $\alpha$ and $\beta$: $M_1$ denotes the meson picking up the spectator quark of the decaying $B_q$-meson. 
 These coefficients involve (in general) form factors, decay constants and convolutions of perturbative kernels with light-cone distribution amplitudes multiplied by the short-distance Wilson coefficients of the Weak Effective Theory, as well as
 $1/m_b$-suppressed long-distance contributions related to annihilation and hard-scattering, non-computable (divergent) within the QCD framework and modelled through the quantities $X_{A,H}$. 

 Similar expressions can be obtained for the $VP,PV,PP$ modes, modifying $\bar{K}^*K^*$ accordingly. The expressions for the $\alpha$ and $\beta$ coefficients are similar, but not completely identical for all four types of modes, and can be found in Ref.~\cite{Beneke:2003zv}.

\subsection{Dependence of $VV$ and $PP$ optimised observables on Wilson coefficients}
\label{app:alphacoeffs-PPVV}

We need to introduce the explicit form of the (numerically) dominant $\alpha_4^c$ coefficients in Eq.~(\ref{eq:TPlongitudinal})
to understand the different structure and dependence on Wilson coefficients of the observables. Following the discussion in Sec.~\ref{sec:BVVBPP}, we will aim to provide the explicit origin  of size (and signs) of  the factors accompanying the Wilson coefficients in the observables.   

Focusing on $b\to s$ transitions, we write~\cite{Beneke:2003zv}~\footnote{We make a slight abuse of notation here: under the convention that we denote $B_q\to M_1M_2$ with $M_1$ picking up the spectator quark, we should actually write $B_s\to \bar{K}K$ but $B_d\to K\bar{K}$.}
\begin{eqnarray} \label{eq:alphac4KstKst}
\alpha^c_4(\bar{B}_s\to K^*\bar{K^*})&=&
\left({\cal C}_{4s}+\frac{{\cal C}_{3s}}{3}\right)
\\ \nonumber
&&\quad +\frac{{\cal C}_{4s}}{3}\frac{\alpha_sC_F}{4\pi}
\left[V_4(K^*)+\frac{4\pi^2}{N_c}H_4(K^*\bar{K}^*)\right]+P_4^c(K^*)\\ \nonumber
&&\quad + r_\chi^{K^*}
\left[
\frac{{\cal C}_{5s}}{3}\frac{\alpha_sC_F}{4\pi}
\left[V_6(K^*)+\frac{4\pi^2}{N_c}H_6(K\bar{K^*})\right]+P_6^c(K^*)
\right]
\\ \label{eq:alphac4KK}
\alpha^c_4(\bar{B}_s\to K\bar{K})&=&
\left({\cal C}_{4s}+\frac{{\cal C}_{3s}}{3}\right)
 + r_\chi^K \left({\cal C}_{6s}+\frac{{\cal C}_{5s}}{3}\right)
\\ \nonumber
&&\quad +\frac{{\cal C}_{4s}}{3}\frac{\alpha_sC_F}{4\pi}
\left[V_4(K)+\frac{4\pi^2}{N_c}H_4(K\bar{K})\right]+P_4^c(K)\\ \nonumber
&&\quad + r_\chi^K
\left[
\frac{{\cal C}_{5s}}{3}\frac{\alpha_sC_F}{4\pi}
\left[V_6(K)+\frac{4\pi^2}{N_c}H_6(K\bar{K})\right]+P_6^c(K)
\right]
\end{eqnarray}
The first line of Eqs.~(\ref{eq:alphac4KstKst}) and (\ref{eq:alphac4KK}) corresponds to the naive factorisation term. The second and third lines correspond to $O(\alpha_s)$ corrections from QCDF ($V_i$, $H_i$ and $P_i$ are vertex, hard-scattering and penguin corrections expressed in terms of convolutions of short-distance kernels and light-cone distribution amplitudes).
They also involve the chiral factors
\begin{equation}
r_\chi^K=\frac{2m_K}{m_b(m_s+m_d)}\simeq 1.41
\qquad
r_\chi^{K^*}=\frac{2m_{K^*}}{m_b}\frac{f_K^\perp}{f_{K^*}}\simeq 0.29
\end{equation}
where the chiral enhancement is clearly visible for the pseudoscalar case compared to the vector case.
 
 We  focus first on $L_{K^*\bar{K}^*}$ and  $L_{K\bar{K}}$ and their sensitivity to the NP in Wilson coefficients.
The factor of three for the term linear in ${\cal C}^{\rm NP}_{4s}$ in $L_{K^*\bar{K}^*}$ with respect to  $L_{K\bar{K}}$ in Eqs.~(\ref{eq:LKstLKst-Cis})-(\ref{eq:LKLK-Cis}) can be traced back to the behaviour of the corresponding $\alpha^c_4$ which is dominant in both cases. Since the different contributions inside $\alpha^c_4$ are complex quantities (and thus are defined up to a global arbitrary phase), we will focus on the relative weight of the SM contribution on one side and the factor multiplying ${\cal C}_{4s}^{\rm NP}$ on the other side.

Let us start with ${\cal C}_{4s}^{\rm NP}$ and ${\cal C}_{6s}^{\rm NP}$.
In the case of $L_{K^*\bar{K}^*}$, the  prefactor of  ${\cal C}_{4s}^{\rm NP}$ is 30 times bigger than the SM term inside $\alpha^c_4$, which we can denote as $\alpha_4^c (K^*K^*) \propto 1-30 {\cal C}_{4s}^{\rm NP}$ (the SM normalisation being factored out). In the case of $L_{K\bar{K}}$, the chirally enhanced part in $\alpha^c_4$ includes an additional contribution from ${\cal C}_{6s} +{\cal C}_{5s}/3$. Indeed, as discussed in Ref.~\cite{Beneke:2003zv} and recalled in Eqs.~(\ref{eq:alphac4KstKst})-(\ref{eq:alphac4KK}), there are no such contributions for vector mesons at tree level: the contribution from $Q_{5,6}$ originate from $(V-A)\otimes (V+A)$  penguin operators in the weak Hamiltonian, which must be Fierz-transformed into $(-2)(S-P)\otimes (S+P)$ so that the second current $(S+P)$ can match the flavour and spin structure of the meson $M_2$ to yield a non-vanishing contribution.
 This additional contribution to $L_{K\bar{K}}$ enhances the SM term significantly and reduces the relative size with respect to the coefficient of ${\cal C}_{4s}^{\rm NP}$ to only a mere factor of 10, i.e.
  $\alpha_4^c (KK) \propto 1-10 {\cal C}_{4s}^{\rm NP}$, which is 3 times smaller than for $K^*\bar{K}^*$, leading to the observed sensitivity of $L_{K\bar{K}}$ and $L_{K^*\bar{K}^*}$ in Eqs.~(\ref{eq:LKstLKst-Cis})-(\ref{eq:LKLK-Cis}). Moreover, this contribution induces a dependence on ${\cal C}_{6s}$ for $L_{K\bar{K}}$ which is not present in the vector-vector case.

 For ${\cal C}_{8gs}^{\rm NP}$, we notice that
 the relative size of its coefficient  compared to SM contribution inside $\alpha^c_4$ is different for $KK$ ($\alpha_4^c (KK) \propto 1 + 0.7 {\cal C}_{8gs}^{\rm NP}$) and $K^*K^*$ ($\alpha_4^c (K^*K^*) \propto 1 + 1.6 {\cal C}_{8gs}^{\rm NP}$). This difference comes from the 
  chiral enhanced terms (proportional to $r_\chi^K$) 
  and more specifically the chirally enhanced penguin $P_6^c(K)$ (see Ref.~\cite{Beneke:2003zv} for the definition of the loop function $\hat G_{K}(s_q=m_q^2/m_b^2)$)
\begin{eqnarray}
   P_6^c(K) &=& \frac{C_F\alpha_s}{4\pi N_c}\,\Bigg\{
    {\cal C}_{1s} \!\left[ \frac43\ln\frac{m_b}{\mu}
    + \frac23 - \hat G_{K}(s_p) \right]\!
    + {\cal C}_{3s} \!\left[ \frac83\ln\frac{m_b}{\mu} + \frac43
    - \hat G_{K}(0) - \hat G_{K}(1) \right] \nonumber\\  &&\hspace{-1.1cm}\mbox{}+ ({\cal C}_{4s}+{\cal C}_{6s})\!
    \left[ \frac{4n_f}{3}\ln\frac{m_b}{\mu}
    - (n_f-2)\,\hat G_{K}(0) - \hat G_{K}(s_c) - \hat G_{K}(1)
    \right] - 2 {\cal C}_{8gs}^{\rm eff} \Bigg\} 
\end{eqnarray}
which makes both the SM contribution and the prefactor of the NP contribution ${\cal C}_{8gs}^{\rm NP}$ of comparable size for $K\bar{K}$.

For   $K^*\bar{K}^*$, the contribution to ${\cal C}_{8gs}$ is absent in $P_6^c(K^*)$\cite{Beneke:2003zv}, 
\begin{eqnarray}
   P_6^c(K^*) &=& - \frac{C_F\alpha_s}{4\pi N_c}\,\Bigg\{
    {\cal C}_1\,\hat G_{K^*}(s_p)
    + {\cal C}_3\,\Big[ \hat G_{K^*}(0) + \hat G_{K^*}(1) \Big] \nonumber\\
   &&\mbox{}+ ({\cal C}_4+{\cal C}_6) \left[ (n_f-2)\,\hat G_{K^*}(0)
    + \hat G_{K^*}(s_c) + \hat G_{K^*}(1) \right] \Bigg\}
\end{eqnarray}
    once again because the spin and flavour content of the outgoing meson has to match the structure of the penguin operator. This difference modifies the balance between the two contributions for $K^*K^*$. This translates as a factor of 1.6 between $L_{K\bar{K}}$ and $L_{K^*\bar{K}^*}$ for the relative coefficients of ${\cal C}_{8gs}^{\rm NP}$ with respect to the corresponding SM contribution.

\subsection{Dependence of $PV$ and $VP$ optimised observables on Wilson coefficients}
\label{app:alphacoeffs-PVVP}

We can move now to the $B\to PV,VP$ observables. Here the most relevant difference with respect to all the other observables is the origin of the positive (negative) sign of the prefactor of ${\cal C}_{4s}^{\rm NP}$ (${\cal C}_{8gs}^{\rm NP}$) in $\hat{L}_{K^*}$, see Eqs.~(\ref{eq:hatLKst-Cis})-(\ref{eq:hatLK-Cis}).

To explain this difference, we can consider the general expression for the $\alpha^c_4$ contribution
for the decay $B_q\rightarrow M_1 M_2$:
\begin{eqnarray}
\label{eq:alpha4generic}
\alpha^c_4=\begin{cases}
    a^c_4  + r^{M_2}_{\chi} a^c_6  & \text{if $M_1 M_2 =PP, PV$},\\
    a^c_4  - r^{M_2}_{\chi} a^c_6  & \text{if $M_1 M_2 = VV, VP$},
  \end{cases}
\end{eqnarray}
with the two contributions
\begin{eqnarray}
\label{eq:alpha4}
a^c_4&=& 
\left({\cal C}_{4q}+\frac{{\cal C}_{3q}}{3}\right)
+\frac{{\cal C}_{4q}}{3}\frac{\alpha_s C_F}{4\pi}
\left[V_4(M_2)+\frac{4\pi^2}{N_c}H_4(M_1 M_2)\right]+P_4^c(M_2) \\ 
\label{eq:alpha6}
a^c_6 &=& \left({\cal C}_{6q}+\frac{{\cal C}_{5q}}{3}\right) \delta_{M_2 P} + \frac{{\cal C}_{5s}}{3}\frac{\alpha_sC_F}{4\pi}
V_6(M_2)+P_6^c(M_2)
\end{eqnarray}
where $\delta_{M_2 P}=1$ only if $M_2$ is a pseudoscalar meson and $\delta_{M_2 P}=0$ otherwise.

For $\hat{L}_{K^*}$,
the SM contribution and the prefactor of $\mathcal{C}^{\rm NP}_{4s}$ in $a^c_4(M_1 M_2)$ have opposite signs. However, the additional $a^c_6$ term
 modifies the situation because of two different effects: first, 
   the  chirally enhanced  term $r^{K}_{\chi}(\mathcal{C}_{6q} + {\mathcal{C}_{5 q}}/{3})$ dominates the SM contribution of $a^c_4$, see Eqs.~(\ref{eq:alpha4})-(\ref{eq:alpha6}). Secondly, 
   since $a^c_6$ has to be subtracted in Eq.~(\ref{eq:alpha4generic}) ($VP$ case), it interferes destructively with the SM value in $a^c_4$. Consequently, the SM and the coefficient of $\mathcal{C}^{\rm NP}_{4s}$ in $\alpha^c_4$ have the same sign. The coefficient of $\mathcal{C}^{\rm NP}_{4s}$ in $\hat{L}_{K^*}$ is the product of both terms, and is thus positive. 
   Notice that this situation is unique for $\hat{L}_{K^*}$ 
   compared to the observables for other modes,
   where the two effects (negative sign for $a^c_6$ or presence $(\mathcal{C}_{6q} + {\mathcal{C}_{5 q}}/{3})$) are not combined.

The explanation of the sign for $\mathcal{C}^{\rm NP}_{8gs}$ in $\hat{L}_{K^{*}}$ is analogous, although the SM term and the relevant NP prefactor have the same sign in $a^c_4$ this time. Once $a^c_6$ is included, the two effects discussed previously lead to a flipped signed for the SM term. This yields a relative minus sign in between the SM term and the prefactor of $\mathcal{C}^{\rm NP}_{8gs}$ in $\alpha^c_4$, which translates into a negative coefficient for  $\mathcal{C}^{\rm NP}_{8gs}$ in $\hat{L}_{K^*}$.

We thus understand how the different decay modes considered can exhibit a different sensitivity to NP contributions to the various Wilson coefficients, depending on the spin of the mesons in the final state and in particular the one picking up the spectator quark.

\begin{table}
\begin{center}
\scalebox{0.89}{
\begin{tabular}{|c|c|c|c|}
\hline
& $T$ & $P$ & $\Delta$ \\
\hline
${\cal B}(B_s \to K^* \bar{K}^*)\cdot 10^{7}$ & $4.58^{+1.04}_{-1.12}+2.38^{+0.85}_{-0.85}i$  &$5.28^{+1.13}_{-1.20}+2.20^{+0.81}_{-0.81}i$  & $-0.71^{+0.13}_{-0.13}+0.17^{+0.07}_{-0.07}i$
\\
${\cal B}(B_d \to K^* \bar{K}^*)\cdot 10^{7}$ & $4.92^{+0.93}_{-1.00}+2.61^{+0.74}_{-0.74}i$  &$5.70^{+1.02}_{-1.07}+2.42^{+0.71}_{-0.71} i$ & $-0.78^{+0.13}_{-0.12}+0.18^{+0.07}_{-0.07} i$ 
\\
${\cal B}(B_s \to K \bar{K})\cdot 10^{6}$ & $1.28^{+0.20}_{-0.16}+0.46^{+0.15}_{-0.15} i$  & $1.38^{+0.16}_{-0.20}+0.33^{+0.15}_{-0.15} i$ & $-0.108^{+0.008}_{-0.008}+0.135^{+0.010}_{-0.010} i$ 
\\
${\cal B}(B_d \to K \bar{K})\cdot 10^{6}$ & $1.19^{+0.16}_{-0.12}+0.44^{+0.12}_{-0.12}i$  & $1.30^{+0.16}_{-0.13}+0.31^{+0.12}_{-0.12}i$ & $-0.105^{+0.004}_{-0.004}+0.136^{+0.006}_{-0.006}i$\\
\hline
\end{tabular}}
\end{center}
\caption{$T$ and $P$ hadronic matrix elements as well as their difference $\Delta$ evaluated in the SM using QCDF. We recall that we consider only the longitudinal polarisation in $K^*\bar{K}^*$.}
\label{tab:deltas}
\end{table}

\subsection{Hadronic matrix elements within the SM from QCDF} \label{app:PTDelta}

Within QCDF, we can estimate the tree and penguin hadronic matrix elements, as well as their difference, for the various modes of interest. The results are given in Table \ref{tab:deltas}. The individual values differ slightly from Ref.~\cite{Alguero:2020xca} due to the update of the inputs in Tab.~\ref{tab:inputs}. Nevertheless we can still see that the difference $\Delta=T-P$ is one order of magnitude smaller than the hadronic matrix elements $T$ and $P$ for penguin-mediated modes, following the arguments presented in Refs.~\cite{Descotes-Genon:2006spp,Descotes-Genon:2007iri,Descotes-Genon:2011rgs,Alguero:2020xca} and recalled in Sec.~\ref{sec:theory}.

\section{Computation of the SM distributions} \label{app:nonGaussian}

We have discussed the theoretical predictions for the various quantities of interest within QCDF by assuming Gaussian distributions for the various theoretical inputs. Since the dependence of the various variables on these inputs is not linear, the resulting distributions for the observables are not perfectly Gaussian. This is particularly true for the observables whose prediction involve the form factors $A_0^{B_q\to K^*}$ which are not known to a very high accuracy currently.

We generate a large number of points (typically $10^5-10^6$) in order to determine the probability density function $PDF(L)$ of a given observable $L$ within the SM. A usual construction to quote a central value $L_0$ and the $1\sigma$ interval $L_0{\,}^{+\sigma_+}_{-\sigma_-}$ is based on the
maximal likelihood ratio (MLR) approach, relying on the test statistic $\chi^2(L)=-2\log(PDF(L))$, with
\begin{equation}
LL: \qquad
\chi^2(L_0)=\min_L\chi^2(L),\qquad
\chi^2(L_0-\sigma_-)=\chi^2(L_0+\sigma_+)=\chi^2(L_0)+1
\end{equation}
In the main text, we consider a slightly different construction, based on the specific points of the cumulative density function for $L$:
\begin{equation} CDF: \qquad CDF(L_0)=0.5, \qquad CDF(L_0-\sigma_-)=0.159,\qquad CDF(L_0+\sigma_+)=0.841
\end{equation}
These definitions are equivalent to the confidence intervals based on the MLR approach in the Gaussian case.
Given the non-Gaussianities present in some of the distributions that we consider here, one can wonder if these two definitions are in good agreement even in this case. We provide the comparison in Tab.~\ref{tab:compnonGauss}, where we can see that the two versions agree well.

\begin{table}
\begin{center}
\begin{tabular}{|c|c|c|}
\hline
& $MLR$ & $CDF$\\
\hline
$L_{K^*\bar{K}^*}$ & $17.2^{+8.3}_{-5.9}$  & $19.5^{+9.1}_{-6.7}$  \\
$L_{K\bar{K}}$ & $25.5^{+4.0}_{-3.3}$  & $26.0^{+3.9}_{-3.6}$  \\
$\hat{L}_{K^*}$ & $20.5^{+6.8}_{-6.2}$  & $21.3^{+7.2}_{-6.3}$  \\
$\hat{L}_{K}$ & $25.3^{+3.7}_{-4.5}$  & $25.0^{+4.2}_{-4.1}$  \\
$L_{K^*}$ & $16.6^{+6.9}_{-6.0}$  & $17.4^{+6.6}_{-5.8}$  \\
$L_{K}$ & $28.8^{+5.2}_{-4.6}$ & $29.2^{+5.5}_{-5.3}$  \\
$L_{\rm total}$ & $23.5^{+3.8}_{-4.0}$ & $23.5^{+4.0}_{-3.8}$  \\
$R_d$ & $0.67^{+0.23}_{-0.24}$  & $0.70^{+0.30}_{-0.22}$  \\
\hline
${\cal B}(B_d\to K^{*0}\bar{K}^{*0})\times 10^6$ & $ 0.22^{+0.08}_{-0.08}$ & $ 0.23^{+0.10}_{-0.08}$\\
${\cal B}(B_s\to K^{*0}\bar{K}^{*0})\times 10^6$ & $ 3.95^{+1.88}_{-1.54}$ & $ 4.36^{+2.23}_{-1.65}$\\
${\cal B}(B_d\to K^{0}\bar{K}^{0})\times 10^6$ 
& $1.01^{+0.24}_{-0.16}$ & $1.09^{+0.29}_{-0.20}$\\
${\cal B}(B_s\to K^{0}\bar{K}^{0})\times 10^6$ & $ 25.6^{+7.5}_{-5.2}$ & $ 28.0^{+8.9}_{-6.2}$\\
\hline
\end{tabular}
\end{center}
\caption{Confidence intervals extracted from the probability density function of various observables, using either the $CDF$ or the $MLR$ definitions. We recall that the longitudinal polarisation is implied for the $K^*\bar{K}^*$ final state.}
\label{tab:compnonGauss}
\end{table}

One can also determine the pull between the experimental and SM theoretical values, when the former are available. Given the limited impact of non-Gaussianities, we can take it as the PDF associated with the difference $X$ between the experimental value (taken as a Gaussian) and the SM theoretical value (generated as before), leading to
\begin{eqnarray}
L_{K^*\bar{K}^*}: {\rm Pull}_{\rm SM}=2.6\sigma&\qquad&
\qquad\qquad\ \  L_{K\bar{K}}: {\rm Pull}_{\rm SM}=2.4 \sigma \\
{\cal B}(B_s\to K^*\bar{K}^*): {\rm Pull}_{\rm SM}=0.9\sigma&\qquad&
{\cal B}(B_d\to K^*\bar{K}^*): {\rm Pull}_{\rm SM}=1.8\sigma\\
{\cal B}(B_s\to K\bar{K}): {\rm Pull}_{\rm SM}=1.6\sigma&\qquad&
\quad {\cal B}(B_d\to K\bar{K}): {\rm Pull}_{\rm SM}=0.4\sigma
\end{eqnarray}

\begin{figure}[h]
\centering
\includegraphics[width=0.75\textwidth]{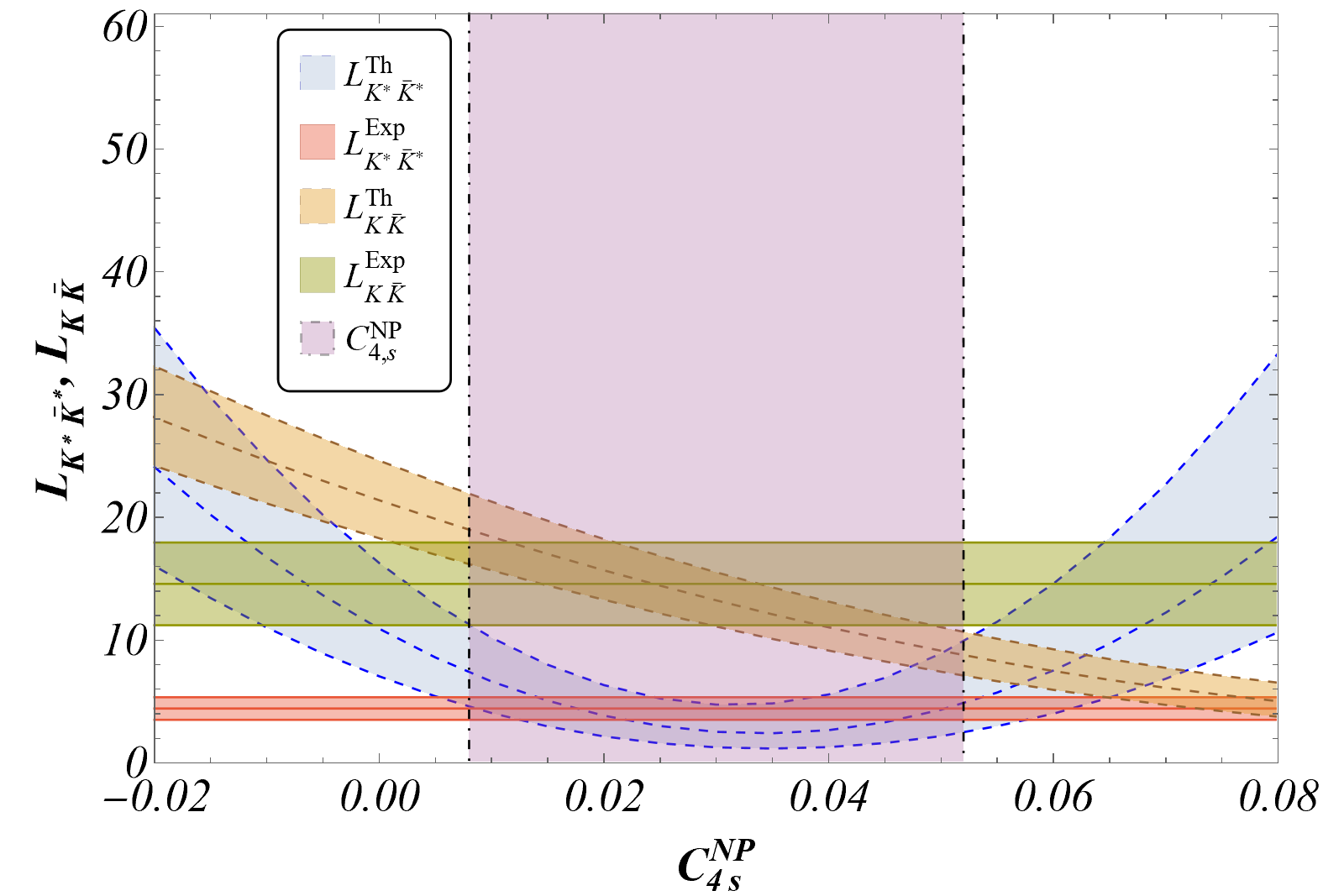}
\caption{Variation of $L_{K^*\bar{K}^*}$ and $L_{K\bar{K}}$ w.r.t ${\cal C}^{\rm NP}_{4s}$ for ${\cal C}^{\rm NP}_{4d} = -0.01$ and ${\cal C}^{\rm NP}_{6,8g d,s} = 0$. The magenta region ($[0.008, 0.052]$) represents the range of ${\cal C}^{\rm NP}_{4s}$ 
 where both observables are
compatible (for this particular value of ${\cal C}^{\rm NP}_{4d}$) theoretically and experimentally within 1$\sigma$.}
\label{fig:C4s_pred_C4d_fxd_LKK_LKstKst}
\end{figure}
\begin{figure}[h]
\centering
\includegraphics[width=0.47\textwidth]{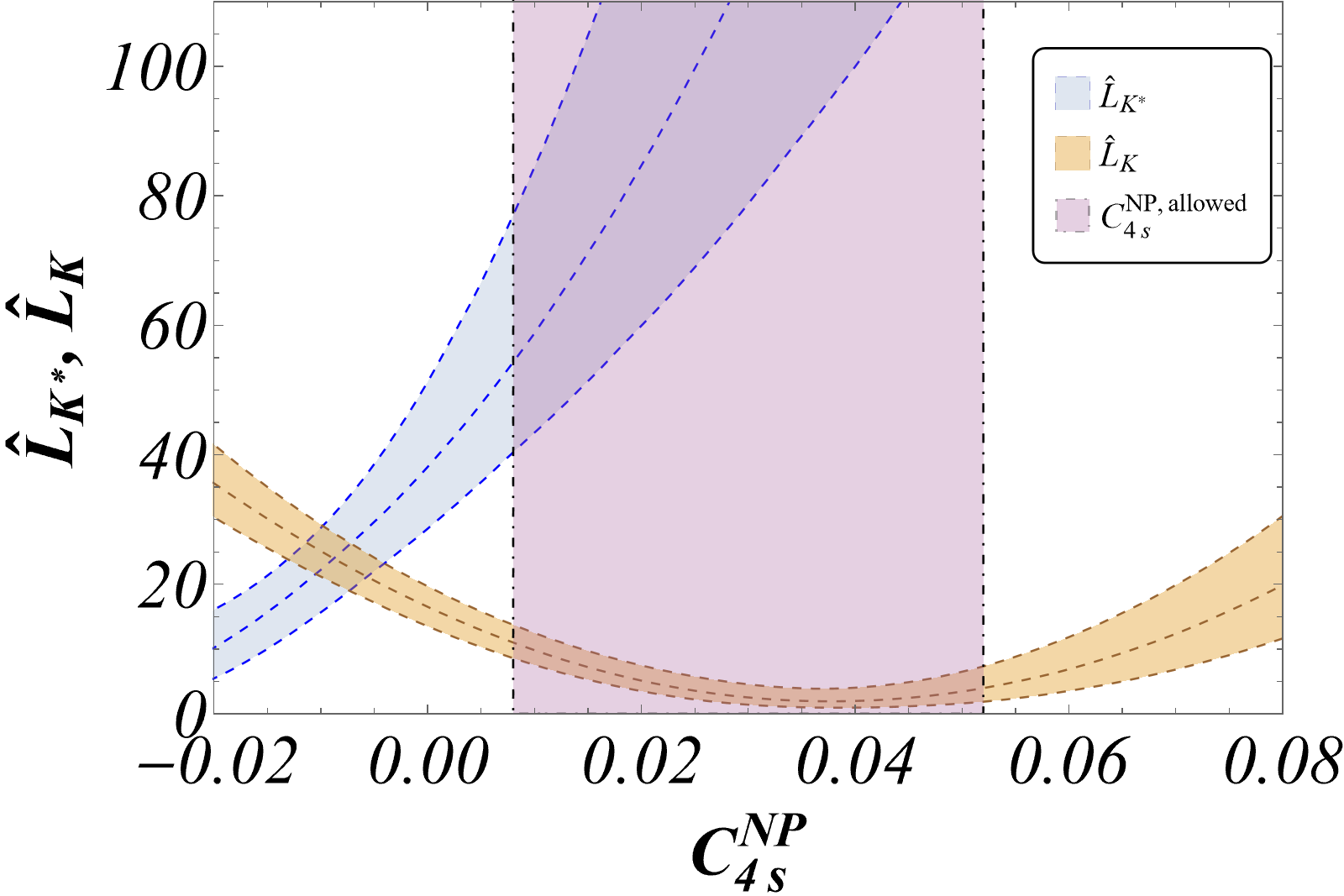}
\includegraphics[width=0.47\textwidth]{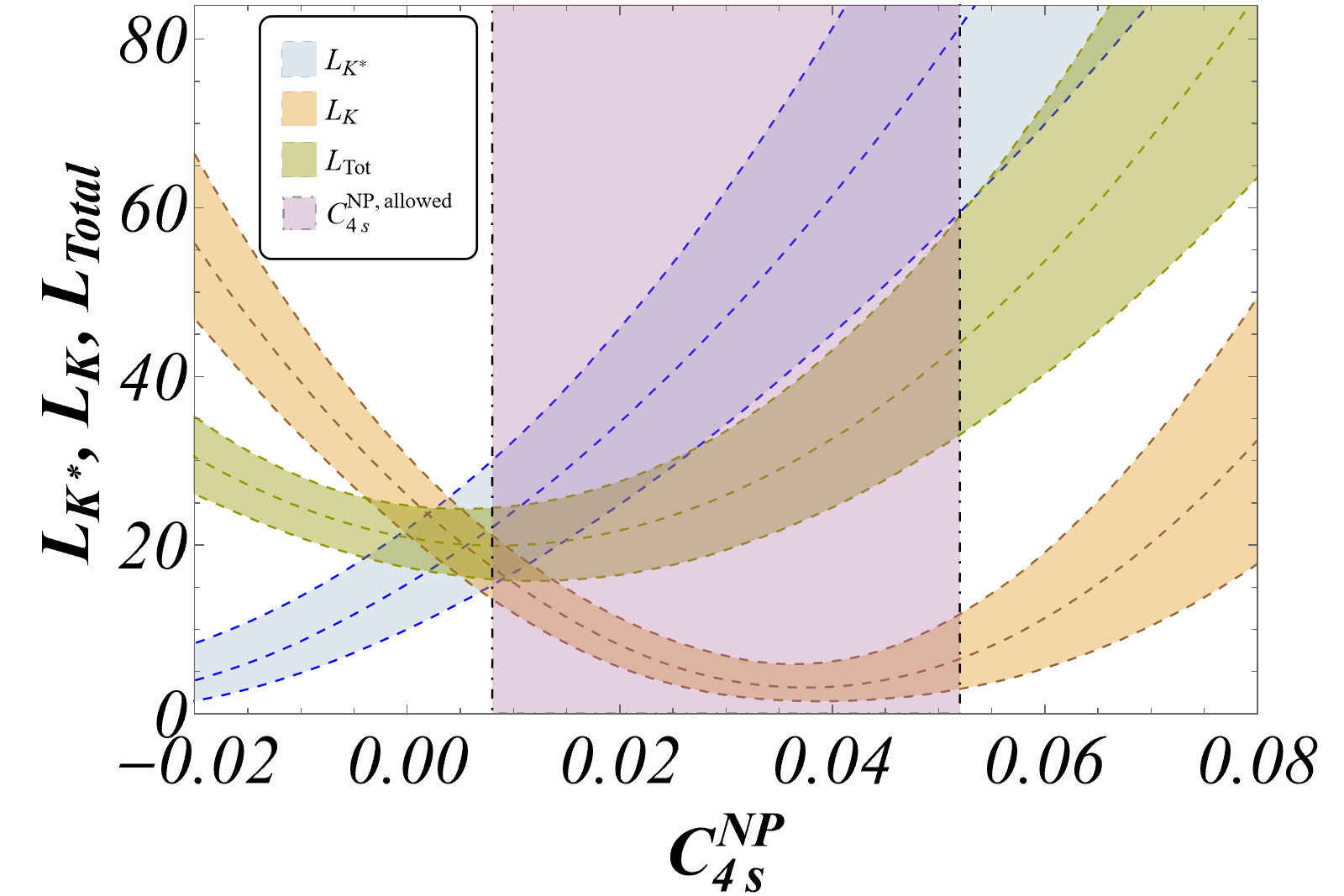}
\caption{Variation of $\hat{L}_{K^*}$ and $\hat{L}_K$ (left) and ${L}_{K^*}$, ${L}_K$ and $L_{\rm total}$ (right) w.r.t ${\cal C}^{\rm NP}_{4s}$ for ${\cal C}^{\rm NP}_{4d} = -0.01$ and ${\cal C}^{\rm NP}_{6,8g d,s} = 0$. The magenta region ($[0.008, 0.052]$) represents the range of ${\cal C}^{\rm NP}_{4s}$ (with ${\cal C}^{\rm NP}_{4d} = -0.01$) where both observables are
compatible theoretically and experimentally within 1$\sigma$ (it is the same as shown in fig.~\ref{fig:C4s_pred_C4d_fxd_LKK_LKstKst}).}
\label{fig:C4s_pred_C4d_fxd_LK_LKst}
\end{figure}

\begin{figure}[h]
\centering
\includegraphics[width=0.75\textwidth,height=0.5\textwidth]{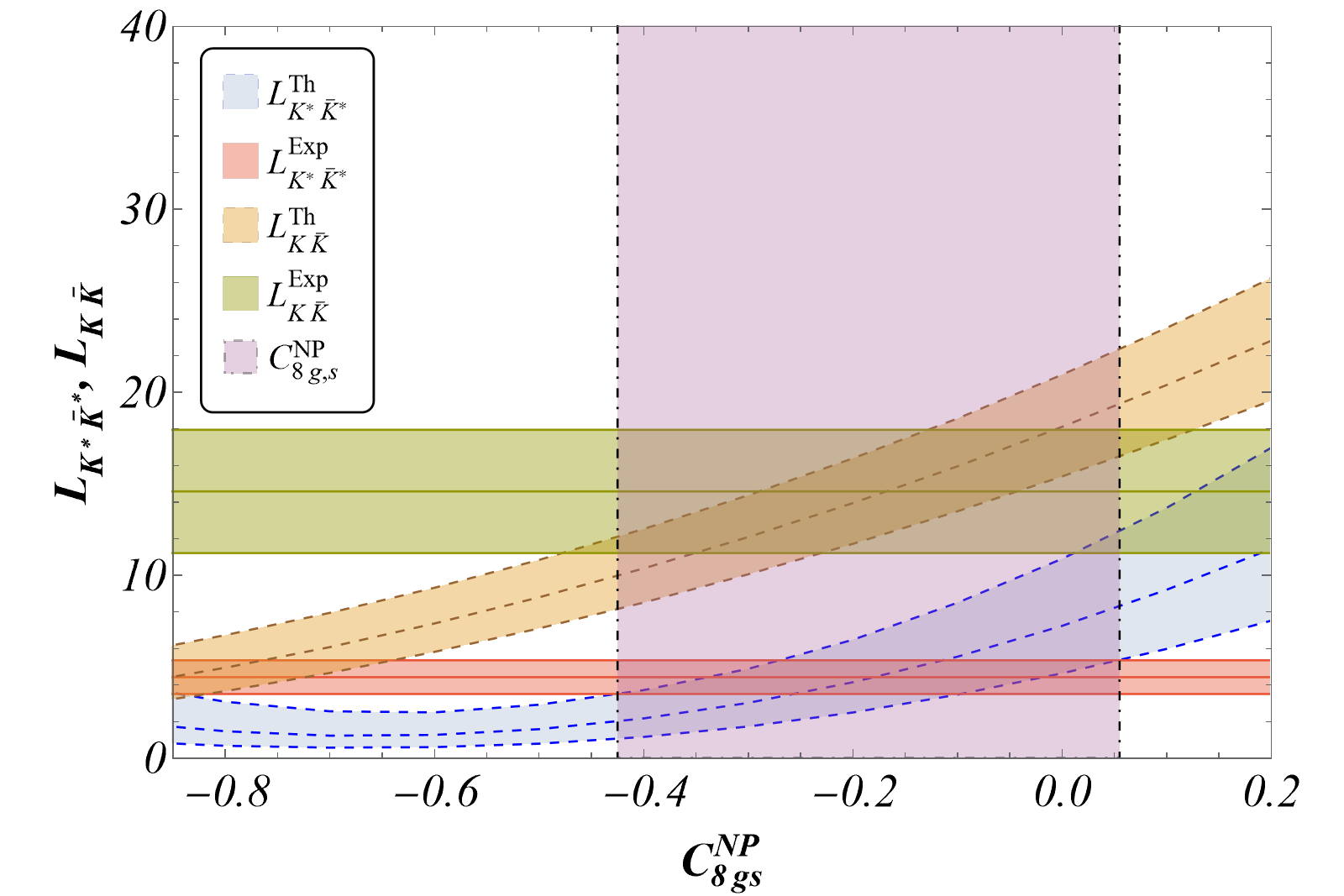}
\caption{Variation of $L_{K^*\bar{K}^*}$ and $L_{K\bar{K}}$ w.r.t ${\cal C}^{\rm NP}_{8gs}$ for ${\cal C}^{\rm NP}_{8gd} = 0.3$ and ${\cal C}^{\rm NP}_{4,6 d,s} = 0$. The magenta region ($[-0.425, 0.055]$) represents the range of ${\cal C}^{\rm NP}_{8gs}$ 
 where both observables are
compatible (for this particular value of ${\cal C}^{\rm NP}_{8gd}$) theoretically and experimentally within 1$\sigma$.}
\label{fig:C8s_pred_C8d_fxd_LKK_LKstKst}
\end{figure}
\begin{figure}[h]
\centering
\includegraphics[width=0.47\textwidth,height=0.32\textwidth]{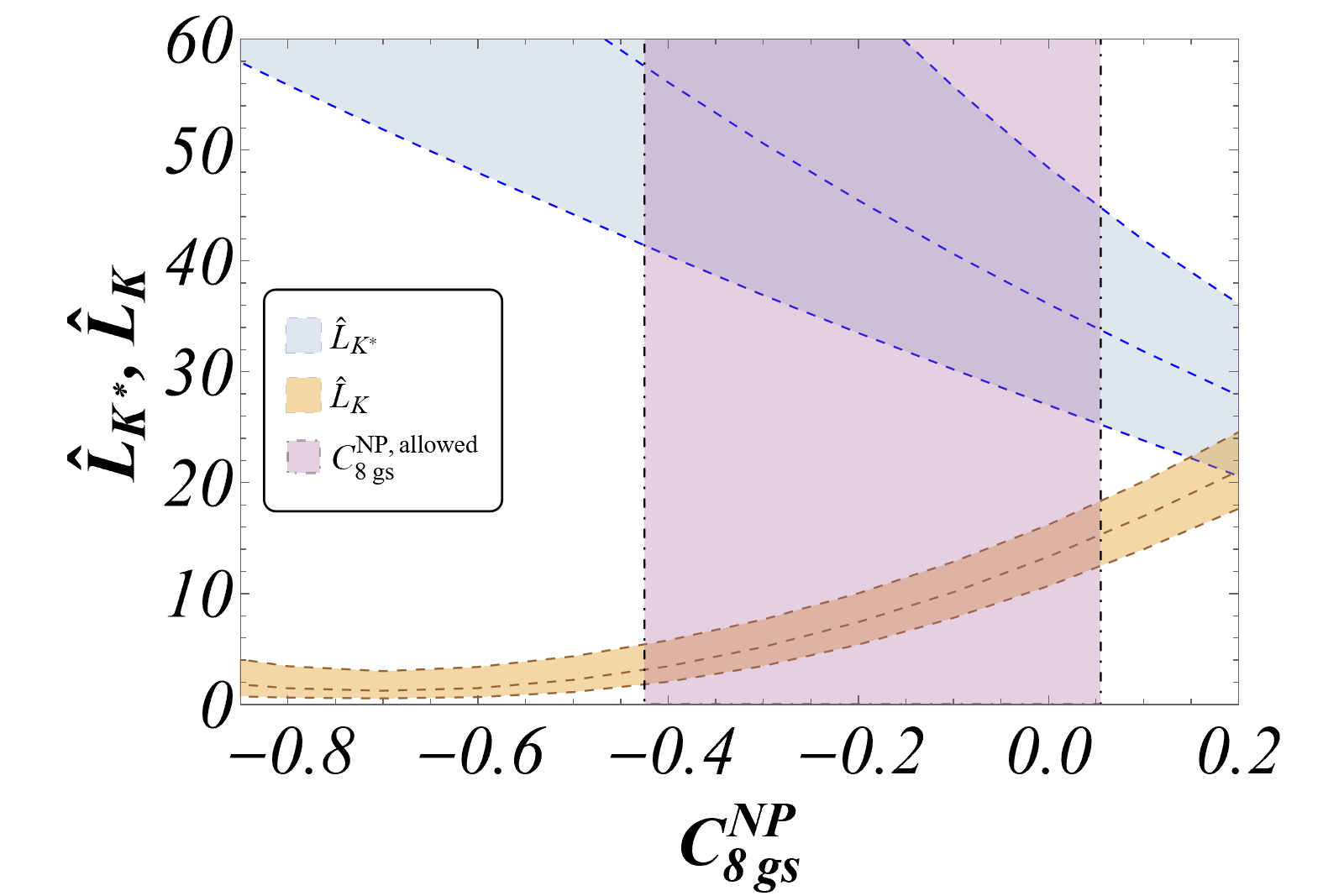}
\includegraphics[width=0.47\textwidth,height=0.32\textwidth]{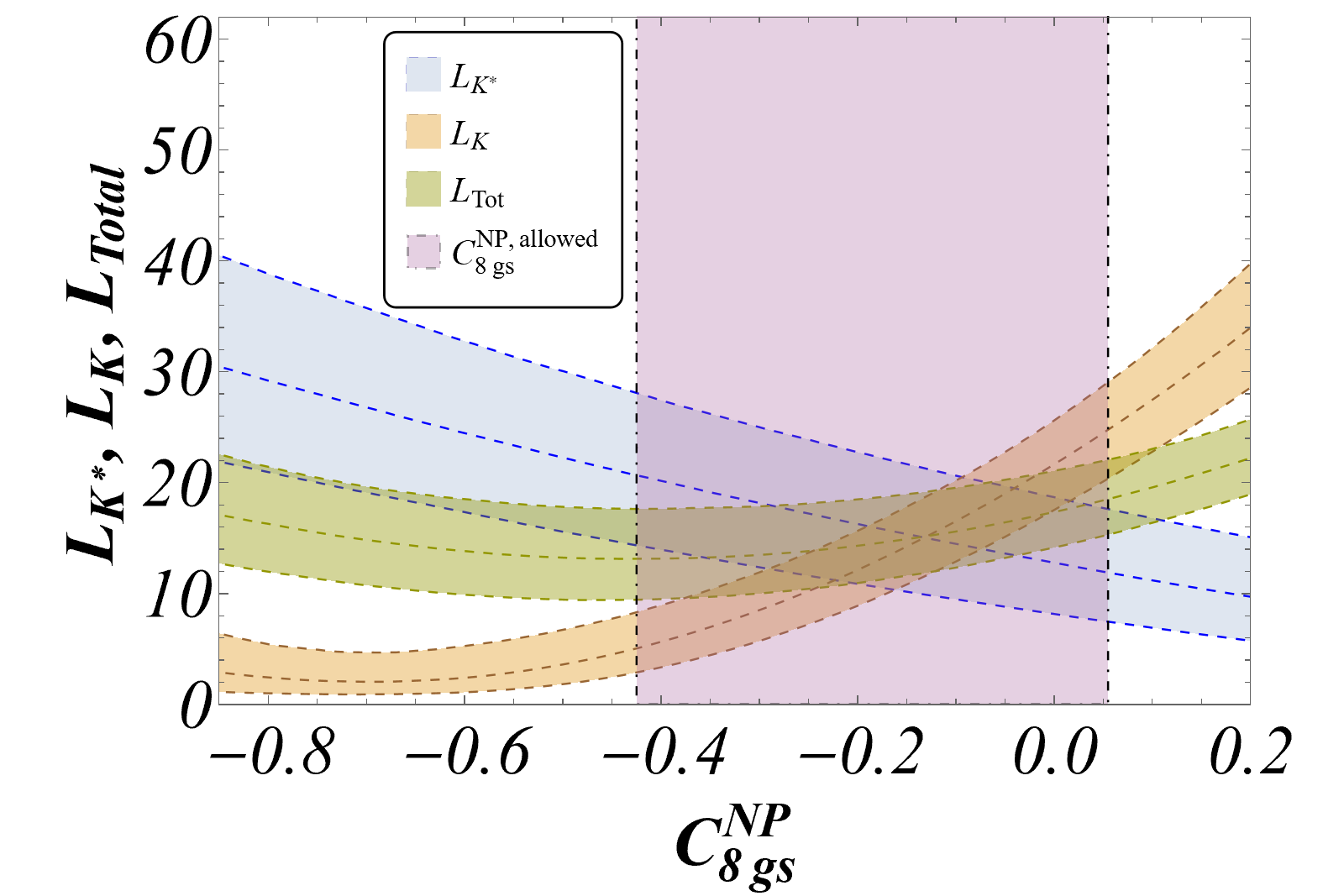}
\caption{Variation of $\hat{L}_{K^*}$ and $\hat{L}_K$ (left) and ${L}_{K^*}$, ${L}_K$ and $L_{\rm total}$ (right)  w.r.t ${\cal C}^{\rm NP}_{8gs}$ for ${\cal C}^{\rm NP}_{8gd} = 0.3$ and ${\cal C}^{\rm NP}_{4,6 d,s} = 0$. The magenta region ($[-0.425, 0.055]$) represents the range of ${\cal C}^{\rm NP}_{8gs}$ (with ${\cal C}^{\rm NP}_{8gd} = 0.3$)  where both observables are
compatible theoretically and experimentally within 1$\sigma$ (it is 
 the same as shown in Fig.~\ref{fig:C8s_pred_C8d_fxd_LKK_LKstKst}).}
\label{fig:C8s_pred_C8d_fxd_LK_LKst}
\end{figure}
\begin{figure}[t]
\centering
\includegraphics[width=1.0\textwidth]{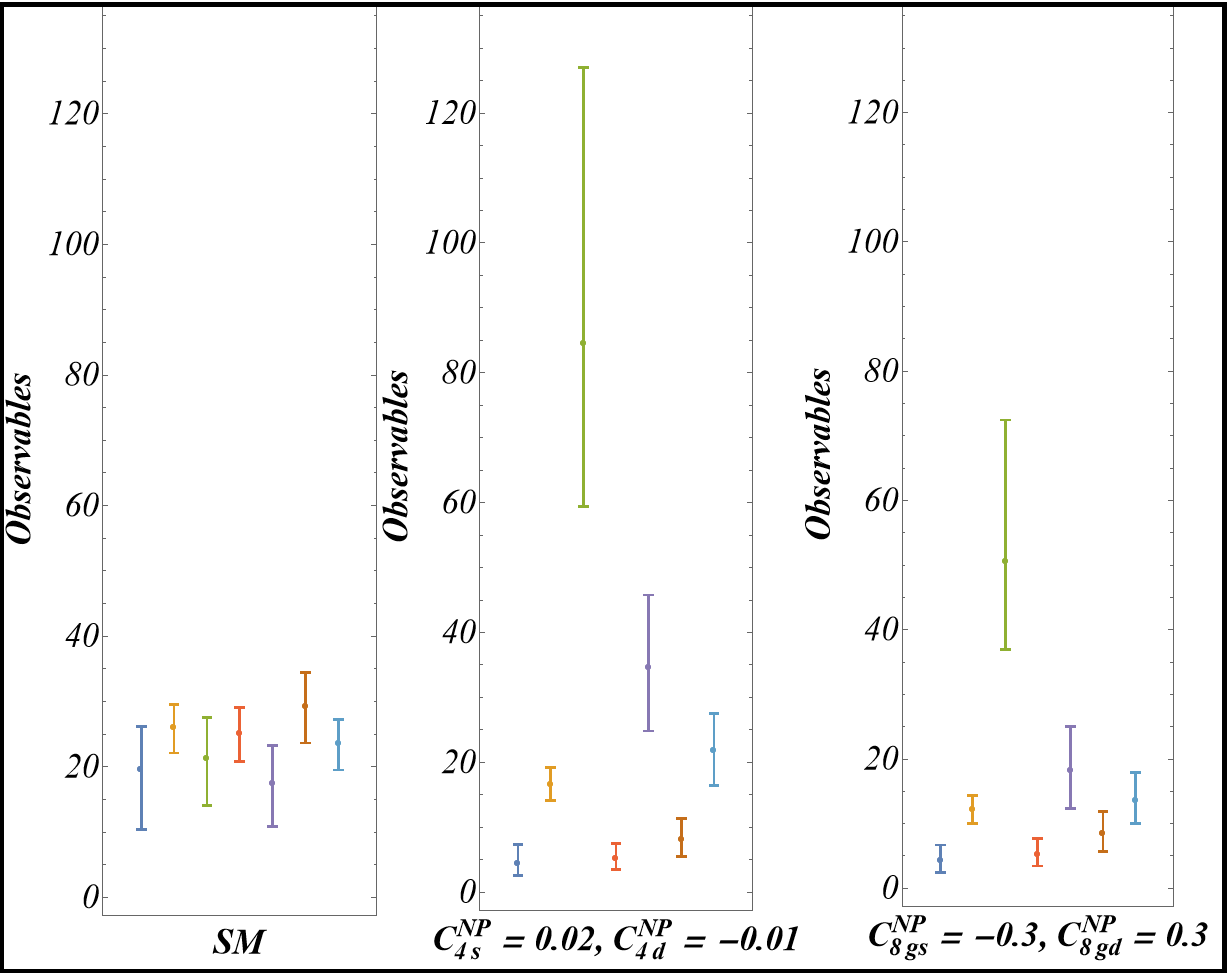}\qquad\\
\centering
\includegraphics[width=0.5\textwidth]{lgndpatt.png}
\caption{Predictions within the SM and different scenarios at specific NP points illustrating the patterns to be expected in each case, 
assuming NP enters both $b\to s$ and $b\to d$ transitions. The specific benchmark values for the NP Wilson Coefficients are taken in agreement with the magenta regions shown in figs.~\ref{fig:BR1} and \ref{fig:BR3}. }
\label{fig:dev-patternsIII}
\end{figure}

\section{Benchmark scenarios for New Physics} \label{app:benchmark}

In this appendix,
we illustrate the behaviour of the observables using their full expressions in terms of Wilson coefficients, going beyond the approximation discussed in 
Sec.~\ref{sec:indivBR} that consisted in performing the substitution
${\cal C}_{is}^{\rm NP}$ for ${\cal C}_{is}^{\rm NP}-{\cal C}_{id}^{\rm NP}$ to infer the dependence on both $b\to d$ and $b\to s$ Wilson coefficients.
We consider benchmark points which are allowed according to Figs.~\ref{fig:BR1}, \ref{fig:BR2} and \ref{fig:BR3} (using also exact expressions) but are outside the limited range where the approximation is valid.

We consider first the  benchmark point ${\cal C}_{4d}^{\rm NP}=-0.01$. 
The plots of Sec.~\ref{sec:combined} are modified in this case and we present
Figs.~\ref{fig:C4s_pred_C4d_fxd_LKK_LKstKst} and \ref{fig:C4s_pred_C4d_fxd_LK_LKst}  for $L_{K\bar{K}}$, $L_{K^*\bar{K}^*}$, $\hat{L}_K$, $\hat{L}_{K^*}$, 
$L_K$, $L_{K^*}$ and $L_{total}$ as a function of ${\cal C}_{4s}^{\rm NP}$. This benchmark point for ${\cal C}_{4d}^{\rm NP}$ illustrates the exact impact (no approximation used here) of allowing NP in QCD penguins for the $b\to d$ transition.
 As can be seen from Fig.~\ref{fig:C4s_pred_C4d_fxd_LKK_LKstKst} allowing  for a non-zero NP contribution to ${\cal C}_{4d}^{\rm NP}$ and considering only the $L$ observables enlarges the domain for 
  ${\cal C}_{4s}^{\rm NP}$  to values closer to zero. This is, as expected, in excellent agreement with Fig.~\ref{fig:BR1}. Moreover,  Figs.~\ref{fig:C4s_pred_C4d_fxd_LK_LKst}  and \ref{fig:dev-patternsIII}  show
   that the size of the deviations between $\hat{L}_{K^*}$ and $\hat{L}_{K}$ may become substantially enhanced (up to two orders of magnitude), due to the presence of a negative contribution to ${\cal C}_{4d}^{\rm NP}$, compared to the case where NP does not affect $b\to d$ transitions.

 Notice that the behaviour  for the $L_{K^*}$ and $L_{K}$ is opposite:  their splitting decreases in the case of NP in ${\cal C}_{4d}$. This comes from the substantial reduction of $R_d$ if ${\cal C}_{4d}=-0.01$  as can be seen in Fig.\ref{fig:rd1}.
 The contribution of $\hat{L}_{K^*}$ to $L_{K^*}$ is then suppressed by a third and the contribution of  $\hat{L}_K$ to $L_K$ gets enhanced by a factor of around 5/3. Both effects reduce the difference among the two observables $L_{K^*}$ and $L_{K}$, which are less sensitive to NP than $\hat{L}_{K^*}$ and $\hat{L}_{K}$.

Similarly we consider the benchmark point  ${\cal C}_{8gd}^{\rm NP}= +0.30$, and show how the plots of Sec.~\ref{sec:combined} are modified to yield
Figs.~\ref{fig:C8s_pred_C8d_fxd_LKK_LKstKst} and \ref{fig:C8s_pred_C8d_fxd_LK_LKst}.  
  Here we observe that  when only the $L$ observables are considered, the extension of the domain for ${\cal C}_{8gs}^{\rm NP}$ towards the SM is even more important, and only the addition of the individual branching ratios cut part of this enhanced region as shown in Fig.~\ref{fig:BR3}. The pattern of deviations between $\hat{L}_K$ and $\hat{L}_{K^*}$ is very significantly enhanced as can be seen by comparing Fig.~\ref{fig:C8s_pred_C8d_fxd_LK_LKst} with Fig.~\ref{fig:C8_pred} and gets reduced for $L_{K^*}$ and $L_{K}$.

In  summary, if the tension in the $B_d$ individual branching ratios are confirmed, it will be necessary to analyse the observables using the full expressions in terms of the Wilson coefficients, as done in this section. On the contrary, if the tension in the $b\to d$ transitions vanishes or gets reduced, the approximation discussed in the main text can be safely used.

\bibliographystyle{JHEP}

\bibliography{main.bib}

\end{document}